\documentclass[prb,aps,reprint,superbib,superscriptaddress,twocolumn]{revtex4-1}
\usepackage{amsmath,bm,bbm}
\usepackage{amstext}
\usepackage{epsfig}
\usepackage{xcolor}
\usepackage{subfig}
\usepackage{graphicx}
\usepackage{multirow}
\usepackage{array}
\usepackage{tikz,pgfplots}
\usepackage{MnSymbol}
\usepackage{dsfont}
\usepackage{bbold}
\usepackage{physics}
\usepackage{mathtools}
\usepackage{hyperref}
\usepackage[english]{babel}
\usepackage{ulem}
\hypersetup{
    colorlinks=true,
    citecolor=gray,
    filecolor=blue,
    linkcolor=blue,
    urlcolor=red
}
\usepackage{mathrsfs}

\definecolor{dgreen}{rgb}{0,.5,0}
\definecolor{dred}{rgb}{.7,.0,.0}
\newcommand{\be}{\begin{eqnarray}}
\newcommand{\ee}{\end{eqnarray}}

\usepackage{xr}
\begin{document}

\title{Recursive relations and quantum eigensolver algorithms within modified Schrieffer--Wolff transformations for the Hubbard dimer}
\author{Quentin Mar\'ecat}
\affiliation{ICGM, Université de Montpellier, CNRS, ENSCM, Montpellier, France}
\author{Bruno Senjean}
\affiliation{ICGM, Université de Montpellier, CNRS, ENSCM, Montpellier, France}
%\author{Gustavo M.\ Pastor}
%\affiliation{Institut fur Physik, Universitat Kassel}
\author{Matthieu Sauban\`ere}
\email{matthieu.saubanere@umontpellier.fr}
\affiliation{ICGM, Université de Montpellier, CNRS, ENSCM, Montpellier, France}

%%%%% Abstract %%%%%

\begin{abstract}
We derive recursive relations for the Schrieffer--Wolff (SW) transformation applied to the half-filled Hubbard dimer. 
While the standard SW transformation is set to block-diagonalize the transformed Hamiltonian solely at the first order of perturbation, 
we infer from  recursive relations two types of modifications, variational or iterative, that approximate, or even enforce for the homogeneous case, the desired block-diagonalization at infinite order of perturbation.
The modified SW unitary transformations are then used to design an test quantum algorithms adapted to the noisy and fault-tolerant era.
This work paves the way toward the design of alternative  quantum algorithms for the general Hubbard Hamiltonian. 
\end{abstract}

\maketitle

\section{Introduction}

By describing the competition between kinetically induced electron delocalization and electron localization due to the Coulomb repulsion, the non-trivial Hubbard model remains one of the most challenging systems in condensed matter physics~\cite{hubbard_electron_1963}. 
Indeed, despite its simplicity, no general and analytic solution exists.
Besides exact results at certain limits such as the Nagaoka theorem~\cite{nagaoka_ground_1965} close to half-band filling or the Bethe Ansatz~\cite{lieb_absence_1968} in one dimension, different approximations, strategies and numerical algorithms have been designed to solve this cornerstone problem on classical computers. 
More precisely, one could mention density functional~\cite{lopez-sandoval_density-matrix_2002,lima_density_2003} or  Green's functions~\cite{georges_dynamical_1996,senechal_spectral_2000,potthoff_self-energy-functional_2003} based theories, renormalization methods~\cite{white_density_1992} or more recently divide and conquer strategies~\cite{knizia_density_2012,sekaran_householder_2021}, to cite but a few.

In that context, the emergence of quantum computers has revived the hope of obtaining accurate physically relevant quantities for any dimension, size, regime and filling. 
Indeed, a growing interest on developing quantum algorithms to solve the Hubbard model emerges from the literature~\cite{wecker_solving_2015,wecker_towards_2015,kivlichan_quantum_2018,reiner_finding_2019,montanaro_compressed_2020,cai_resource_2020,cade_strategies_2020,mineh_solving_2022,martin_simulating_2022,stanisic_observing_2022,dallaire-demers_low-depth_2018,dallaire-demers_application_2020,suchsland_simulating_2022,gard_classically_2022,kivlichan_improved_2020,campbell_early_2022,clinton_hamiltonian_2021}. 
On the one hand, most of the proposed algorithms targets Noisy Intermediate Scale Quantum (NISQ) devices and  relies mainly on hybrid classical/quantum strategies such as the Variational Quantum Eigensolver (VQE)~\cite{peruzzo_variational_2014,bharti_noisy_2022}. Roughly speaking, it consists in applying a parameterized unitary transformation on an
easy-to-prepare initial state, generally the Hartree--Fock state, on the quantum device while the variational parameters are  optimized on a classical computer. 
Several type of Ansatz have been proposed to design this unitary transformation, either physically motivated such as the variational Hamiltonian Ansatz~\cite{wecker_towards_2015,kivlichan_quantum_2018,reiner_finding_2019,montanaro_compressed_2020,cai_resource_2020,cade_strategies_2020,mineh_solving_2022,martin_simulating_2022,stanisic_observing_2022} and the unitary coupled cluster Ansatz~\cite{dallaire-demers_low-depth_2018}, or hardware efficient ones~\cite{dallaire-demers_application_2020,suchsland_simulating_2022,gard_classically_2022}. Most of these approaches, as they are based on an initial Hartree--Fock state, are particularly relevant for the weakly correlated regime.~\cite{cade_strategies_2020,martin_simulating_2022}  In any case, a compromise between the desired accuracy and the computational cost has to be reached.
It depends in particular on the Ansatz circuit depth, the number of CNOT gates and the number of variational parameters, for which the development of improved or new types of Ansatz is needed.
On the second hand, some algorithms target long-term expected fault-tolerant devices~\cite{kivlichan_improved_2020,campbell_early_2022,clinton_hamiltonian_2021},
and rely for instance on Hamiltonian propagation for which the associated quantum circuits are much deeper than those devoted to the NISQ era.

Concerning the application of a unitary transformation onto a easy-to-prepare known state, the unitary Van--Vleck (VV) similarity transformation, developed in the framework of many-body perturbation theory~\cite{van_vleck__1929,jordahl_effect_1934,foldy_on_1950,primas_generalized_1963,brandow_formal_1979,shavitt_quasidegenerate_1980,bravyi_schriefferwolff_2011}, appears relevant to serve as a basis for new quantum algorithms.
In few words, given an Hamiltonian $\hat{H} = \hat{H}_0 + \hat{V}$ where $\hat{H}_0$ is called the unperturbed Hamiltonian whose eigenstates are known, and $\hat{V}$ is a perturbation, the VV similarity transformation aims to design perturbatively a unitary transformation $\hat{U} = e^{\hat{S}^{\rm VV}}$, where $\hat{S}^{\rm VV}$ is called the generator, that leads to an effective Hamiltonian $\bar{H}_{\rm eff}$ in the low-energy subspace of $\hat{H}_0$. 
Ultimately, at infinite order of perturbation, the transformed Hamiltonian  $\bar{H}=\hat{U}\hat{H}\hat{U}^{\dagger}$ is block-diagonal
and is reduced to $\bar{H}_{\rm eff}$ in the low-energy subspace of $\hat{H}_0$, such that the eigenvalues of $\bar{H}_{\rm eff}$ strictly match the lowest eigenvalues of $\hat{H}$. 
It follows a straightforward quantum algorithm for which the ground state (or excited states) of a given Hamiltonian can be prepared on a quantum computer by applying $e^{\hat{S}^{\rm VV}}$ on the known ground state (or excited states) of the unperturbed Hamiltonian $\hat{H}_0$.
However, an explicit expression for $\hat{S}^{\rm VV}$ is in general unknown
and truncation of the perturbative order or approximations are mandatory.
Considering the non-interacting Hamiltonian as $\hat{H}_0$ and the electron-electron Coulomb repulsion as the perturbation, the VV similarity transformation is closely related to the unitary coupled cluster Ansatz~\cite{shavitt_many-body_2009}.
On the other limit where $\hat{H}_0$ is the Coulomb repulsion operator and $\hat{V}$ is the non-interacting Hamiltonian, Schrieffer and Wolff (SW) derived an analytic form of $\hat{S}^{\rm VV}$, such that $\bar{H}$ is block-diagonal at the first-order of perturbation~\cite{schrieffer_relation_1966}.
Moreover, they showed that at the limit of small perturbation, the Kondo model corresponds to the effective low-energy approximation of the Anderson model.
Following the work of SW, the Heisenberg model was also shown to be the effective Hamiltonian of the Hubbard model at half-band filling for large Coulomb repulsion strength~\cite{harris_single-particle_nodate,chao_kinetic_1977}.
Yet, improvements of the SW approximation can fairly serve as a basis to approximate $\hat{S}^{\rm VV}$ and construct an efficient and hopefully accurate quantum algorithm for the Hubbard model.
In that context, Zhang {\it et al.} proposed two quantum algorithms devoted to finding the VV unitary transformation in the context of spin chains.~\cite{zhang_quantum_2022} 
The first one is a quantum phase estimation based algorithm that provides the exact transformation,
but which is only realizable in the fault-tolerant era.
The second one, more adapted to the NISQ era, is an hybrid quantum-classical algorithm
based on a variational approach where the unitary transformation (Ansatz) is built from the exponentiation of the commutator $[\hat{H}_0, \hat{V}]$, expressed as a linear
combination of Pauli operators.
%Recently, Zhang {\it et al.} proposed quantum algorithms to blindly determine the VV unitary transformation through a variational Ansatz applied on spin chains~\cite{zhang_quantum_2022}.

In this contribution, we derive recursive relations to the perturbative expansion of $\bar{H}$ within the standard SW generator for the Hubbard dimer. 
Following these relations, we propose two modifications of this generator, one variational with a single parameter thanks to the recursive relations, and the other iterative in the spirit of the Foldy--Wouthuysen transformation~\cite{foldy_on_1950}.
Both modified SW transformations are shown to approximate, or even perform for the homogeneous case, the desired block-diagonalization at infinite order of perturbation, as the VV generator would provide.
As a proof of concept, we introduce two quantum algorithms associated to the modified SW transformations on the Hubbard dimer. 
Finally, in light of our findings, we discuss the perspective of generalizing our approach to larger Hubbard systems that is left for future investigations.
In particular, we show that in contrast to most of the currently proposed Ansatz, our strategies are relevant close to the strongly interacting regime.

\section{Van--Vleck similarity  and standard Schrieffer--Wolff transformations}

Let us first recall the Van--Vleck
canonical perturbation theory following Shavitt and Redmon~\cite{shavitt_quasidegenerate_1980}. 
Consider a Hamiltonian $\hat{H}$ with (unknown) orthonormal eigenvectors $\{\ket{\Psi_i}\}$ such that
\begin{equation}
  \hat{H} |\Psi_i\rangle = E_i |\Psi_i\rangle,
  \label{eq:eigenvaluepb}
\end{equation}
%We note the density matrix operator  $\hat{\Gamma} = \sum_{i,j} |\Psi_i\rangle \langle \Psi_j |$.\\
and another orthonormal basis set $\{\ket{\Phi_i}\}$, eigenvectors of another Hamiltonian $\hat{H}^0$ with the same dimension than $\hat{H}$, that is related 
to $\{\ket{\Psi_i}\}$ by a unitary transformation,
\begin{equation}
  | \Psi_i \rangle = \hat{U}^{\dagger} | \Phi_i \rangle = \sum_j  | \Phi_j \rangle \langle \Phi_j | \hat{U}^{\dagger}  | \Phi_i \rangle = \sum_j  | \Phi_j \rangle U_{ij}^{\dagger}.
\end{equation}
The eigenvalues of $\hat{H}$
can be inferred as the elements of the diagonal representation of the similar Hamiltonian,
\begin{equation}\label{eq:UHU}
\bar{H}^{\rm VV} = \hat{U} \hat{H}\hat{U}^{\dagger},
\end{equation}
in the orthonormal basis $\{\ket{\Phi_i}\}$.
Thus, solving the eigenvalue problem in Eq.~(\ref{eq:eigenvaluepb}) is equivalent to searching for a unitary transformation $\hat{U}$ such that $\bar{H}^{\rm VV} = \hat{U} \hat{H}\hat{U}^{\dagger}$ is diagonal in a given basis set $\{\ket{\Phi_i}\}$. 
The reasoning remains equivalent, though less restrictive, if solely a block-diagonalization in a target subspace is desired. 
In other words, we are looking for an unknown Hamiltonian $\bar{H}^{\rm VV}$ with eigenvectors $ | \Phi_i \rangle$ that shares the same eigenvalues than $\hat{H}$. 
If one focuses on the ground state $\ket{\Psi_0}$, it is enough to only block-diagonalize $\bar{H}$,
\begin{align}
  \langle \Phi_i | \hat{U} \hat{H} \hat{U}^{\dagger} | \Phi_0\rangle & =  \langle \Phi_0 | \hat{U}^{\dagger} \hat{H} \hat{U} | \Phi_i\rangle = 0\quad \forall i\neq 0, \\
  \langle \Phi_0 | \hat{U} \hat{H} \hat{U}^{\dagger} | \Phi_0\rangle &= E_0.
\end{align}
Many $\hat{U}$ fulfill these conditions up to a unitary transformation acting only on the subspace of $\lbrace \ket{\Phi_i}\rbrace$ with $i \neq 0 $.\\
Let us now consider the following decomposition of the Hamiltonian,
\begin{equation}\label{eq:H0_V}
  \hat{H} = \hat{H}^0 + \hat{V},
\end{equation}
where $ \hat{H}^0 $ is diagonal in the $\{\ket{\Phi_i}\}$ basis set, i.e. $\langle \Phi_i | \hat{H}^0 |  \Phi_j \rangle = E_i^{0} \delta_{ij}$. 
%We define another operator $\bar{V}$ corresponding to the correction added to $\hat{H}^0$ to obtain $\bar{H}$,
%\begin{equation}
%  \bar{H}^{\rm VV} = \hat{H}^0  + \bar{V}.
%\end{equation}
If one wants to block-diagonalize $\bar{H}$ with respect to a  given subspace $\Omega$,  for instance the one that contains all degenerate ground states of $\hat{H}^0$, one can define the operator
\begin{equation}
  \hat{P} = \sum_{i \in \Omega} |\Phi_i \rangle\langle \Phi_i |
\end{equation}
that projects onto $\Omega$, and its complementary projector
\begin{equation}
  \hat{Q} = \hat{1} - \hat{P} =  \sum_{i \notin \Omega} |\Phi_i \rangle\langle \Phi_i |.
\end{equation}
We note $\hat{O}_D =  \hat{P}\hat{O}\hat{P} + \hat{Q}\hat{O}\hat{Q}$ the block-diagonal projection of an operator $\hat{O}$ and its complementary off-block-diagonal part $\hat{O}_X =  \hat{P}\hat{O}\hat{Q} + \hat{Q}\hat{O}\hat{P}$.
Adopting the exponential form of the unitary transformation $\hat{U} = e^{\hat{G}}$, $\hat{G}$ being an anti-Hermitian generator with $\hat{G} = \hat{G}_X$ and  $\hat{G}_D = 0$, we seek conditions for $\hat{G}$ such that $\bar{H}^{\rm VV} $ is block-diagonal, i.e. $\bar{H}_X^{\rm VV}  = \hat{0}$. 
Within the super-operator formalism~\cite{primas_generalized_1963}, 
% \bru{(this sentence is necessary ? I don't find it useful, especially as you use mathcal(G) and then you talk about a mathcal(S)): $\hat{X}$ being an operator and $\mathcal{S}(\hat{X})$ a super-operator that transforms $\hat{X}$ into another operator acting in the same Hilbert space~\cite{primas_generalized_1963},}
$\bar{H}^{\rm VV}$ reads:
  \begin{align}
    \bar{H}^{\rm VV} & = e^{\hat{G}}\hat{H} e^{-\hat{G}} = \hat{H} + [\hat{G}, \hat{H}] + \frac{1}{2}[\hat{G}, [\hat{G}, \hat{H}] ] + \dots \nonumber \\
     & = \sum_{n=0}^{\infty} \frac{1}{n!} \mathcal{G}^n(\hat{H}) = e^{\mathcal{G}}(\hat{H}),
    \end{align}
    where $\mathcal{G}(\hat{X}) = [\hat{G},\hat{X}]$.
    By decomposing $e^{\mathcal{G}}(\hat{H}) = {\rm cosh}\mathcal{G}(\hat{H}) + \rm{ sinh}\mathcal{G}(\hat{H})$, it follows that the condition $\bar{H}_X^{\rm VV}  = \hat{0}$ is fulfilled for
    \begin{equation}
      [\hat{G}, \hat{H}^0] = -[\hat{G}, \hat{V}_D] - \sum_{n=0}^{\infty}c_n \mathcal{G}^{2n}(\hat{V}_X), \label{eq:VV_MBPT}
    \end{equation}
    where $c_n = 2^{2n}B_{2n}/(2n)!$ are functions of Bernoulli numbers $B_{2n}$. Eq.~(\ref{eq:VV_MBPT}) is the central equation of the VV canonical perturbation theory that defines the generator $\hat{G}$ such that $\bar{H}^{\rm VV} $ is block-diagonal, 
    thus expressing the eigenstates of $\hat{H}$ in terms of eigenstates of $\hat{H}^0$ through $\hat{G}$.
    Using an order by order expansion of $\hat{G}$, i.e.  $\hat{G}  =  \sum_{n = 1} \hat{G}^{(n)}$, conditions to cancel $\bar{H}_X$ can be obtained at each order as,
  \begin{align} 
    & [\hat{G}^{(1)}, \hat{H}^0] = -\hat{V}_X, \label{eq:1storder} \\
    & [\hat{G}^{(2)}, \hat{H}^0] = -[\hat{G}^{(1)}, \hat{V}_D],\\
    & [\hat{G}^{(3)}, \hat{H}^0] = -[\hat{G}^{(2)}, \hat{V}_D] - \frac{1}{3}[\hat{G}^{(1)}, [\hat{G}^{(1)}, \hat{V}_X]], \\
    & \dots \nonumber 
  \end{align}
  It follows that $\bar{H}^{\rm VV} $ can also be expressed order by order as
\begin{equation}
  \bar{H}^{\rm VV} = \hat{H}^0 + \hat{V}_D + \sum_{n=0}^{\infty}t_n\mathcal{G}^{2n+1}(\hat{V}_X),\label{eq:VV_barV}
\end{equation}
with $t_n = 2(2^{2n+2} -1) B_{2n+2}/(2n + 2)!$. 
As mentioned in Ref.~[\onlinecite{shavitt_quasidegenerate_1980}], the Van--Vleck perturbation theory equations (\ref{eq:VV_MBPT}) and (\ref{eq:VV_barV}) are expressed in the domain of a Lie algebra, thus allowing an equivalent diagrammatic expansion. 
Note that the convergence of perturbative series and the diagrammatic expansion has been thoroughly investigated in Ref.~[\onlinecite{bravyi_schriefferwolff_2011}], which also provides recursive relations to obtain the $n$-th order term 
$\hat{G}^{(n)}$ of the VV generator $\hat{G}$ as a function of the previous $n-1$ terms.

In practice, finding both an analytic and a numerical form of $\hat{G}$ for a given $\hat{H}^0$ and $\hat{V}$ remains challenging, at least equivalent as the explicit diagonalisation of $\hat{H}$.
From the perspective of developing quantum algorithms based on the VV formalism, one realizes that the number of terms in the generator drastically increases order by order,
thus leading to deeper circuits
and, consequently, to an increase in complexity and
sensibility to noise of quantum algorithms.
To overcome this issue, we explore an alternative approach
which consists in using a truncated generator, the Schrieffer--Wolff generator, that is later modified by adding a variational parameter or by using an iterative process to compensate the resulting truncation error.

First of all, following Ref.~[\onlinecite{schrieffer_relation_1966}], let us recall the Schrieffer--Wolff transformation in the context of
the half-filled Hubbard model that we decompose as in Eq.~(\ref{eq:H0_V})
into a local part,
\begin{eqnarray} 
  \hat{H}^0 &=& \sum_{i \sigma}\mu_i \hat{n}_{i\sigma} + \sum_{i}U_i \hat{n}_{i\uparrow}\hat{n}_{i\downarrow},
\end{eqnarray}
and a non-local (kinetic) part,
\begin{eqnarray}
  \hat{V} &=& -\dfrac{1}{2} \sum_{ i \neq j,\sigma} t_{ij} \left( \hat{\gamma}_{ij\sigma} + \hat{\gamma}_{ji\sigma}\right),
\end{eqnarray}
with $\hat{n}_{i\sigma} = \hat{c}_{i\sigma}^{\dagger}\hat{c}_{i\sigma}$ and $ \hat{\gamma}_{ij\sigma} = \hat{c}_{i\sigma}^{\dagger}\hat{c}_{j\sigma}$,
and $\hat{c}^\dagger_{i\sigma}$ ($\hat{c}_{i\sigma}$) the
creation (annihilation) operator of an electron of spin $\sigma = \lbrace \uparrow, \downarrow \rbrace$ in site $i$.
This decomposition contrasts with the usual decomposition between the non-interacting part for which the solution is easily accessible and the non-trivial canonical (interacting) part. 
Indeed, the ground state of $\hat{H}^0$ is degenerate at half filling for $U>0$, and consists in a superposition of all states having no double occupation (spanning the so-called Heisenberg subspace in this paper). 
Starting from the atomic limit ($U/t \rightarrow \infty$), Schrieffer and Wolff have proposed, in the original context of an Anderson Hamiltonian, to use the unitary transformation $\hat{U} = e^{\hat{S}}$
such that
\begin{equation}
  \bar{H}^{\rm SW}  = e^{\hat{S}}He^{-\hat{S}},
\end{equation} 
that we denote simply  $\bar{H}$ in the following to simplify notations, is block-diagonalized at first order of perturbation, i.e. 
%\matthieu{Je pense qu'il était un peu confus d'utiliser $\bar{H}$ pur VV and SW sachant qu'ils sont {\it in fine} differents. J'ai pris partit de renomer $\bar{H}^{\rm VV}$ dans la partie précédente. Pensez vous qu'il est plus clair de mettre aussi $\bar{H}^{\rm SW}$ ici, ou simplement laisse $\bar{H}$ pour SW est suffisant pour distinger les deux car je trouve que le $^{\rm SW}$ alourdie beaucoup les equations  ?}
%\bru{Je pense que c'est bon, mais peut-être serait-il utile d'écrire "$\bar{H}^{\rm SW}$, that we denote $\bar{H}$ in the following to simplify notations".}
%is block-diagonalized at first order of perturbation,
%i.e. the anti-hermitian generator $\hat{S}$ is defined \ma{ as the first order truncated VV generator $\hat{S} = \hat{G}^{(1)}$,} i.e. to fulfill Eq.~(\ref{eq:1storder}),
\begin{equation} 
  [\hat{S}, \hat{H}^0 ] = -\hat{V}.\label{eq:1storderS}
\end{equation}
We highlight that the above equation corresponds to the first order of perturbation of the VV relations, Eq.~(\ref{eq:1storder}), i.e that the SW generator $\hat{S}$ block-diagonalizes the Hamiltonian only at first order, contrary to the VV generator $\hat{G}$.\\
Note that $\hat{V}_X = \hat{V}$ when the operator $\hat{P}$ projects onto the Heisenberg subspace.
It can be shown that under the SW condition (\ref{eq:1storderS}), $\hat{S}$ takes the following form,
\begin{equation}
  \label{eq:SWS}
  \hat{S} = \dfrac{1}{2}\sum_{i\neq j, \sigma}\hat{p}_{ij\bar{\sigma}}\left(\hat{\gamma}_{ij\sigma} - \hat{\gamma}_{ji\sigma}\right), 
\end{equation}
 with $\hat{p}_{ij\sigma}$ defined as 
\begin{align}
  \hat{p}_{ij \sigma} = \sum_{x=0}^3 \lambda_{ij\sigma,x}\,\hat{p}_{ij \sigma,x},
\end{align}
where $\hat{p}_{ij \sigma,0} = \left(1-\hat{n}_{i\sigma}\right)\left(1-\hat{n}_{j\sigma}\right) $, $\hat{p}_{ij \sigma,1} = \hat{n}_{i\sigma}\left(1-\hat{n}_{j\sigma}\right)$, $\hat{p}_{ij \sigma,2} = \left(1-\hat{n}_{i\sigma}\right)\hat{n}_{j\sigma}$, $\hat{p}_{ij \sigma,3} = \hat{n}_{i\sigma}\hat{n}_{j\sigma}$ are projectors, i.e. $\sum_{x=0}^3 \hat{p}_{ij \sigma,x} = \hat{1}$,
and
\begin{align}
  \lambda_{ij\sigma,0} & = -\frac{t_{ij}}{\Delta \mu_{ij}}\;\; {\rm if}\;  \Delta \mu_{ij}\neq 0;\;\; \lambda_{ij\sigma,0} = 0 \;\; {\rm else}, \label{eq:lamb0}\\
  \lambda_{ij \sigma,1} & = -\frac{t_{ij}}{\Delta \mu_{ij}+U_i}\;\; {\rm if}\;   \Delta \mu_{ij} + U_i\neq 0;\;\; \lambda_{ij\sigma,1} = 0 \;\; {\rm else}, \label{eq:lamb1}\\
  \lambda_{ij \sigma,2} & = -\frac{t_{ij}}{\Delta \mu_{ij}-U_j}\;\; {\rm if}\;  \Delta \mu_{ij} - U_j \neq 0;\;\; \lambda_{ij\sigma,2} = 0 \;\; {\rm else}, \label{eq:lamb2}\\
   \lambda_{ij \sigma,3} & =-\frac{t_{ij}}{\Delta \mu_{ij}  + \Delta U_{ij}}\;\; {\rm if}\;   \Delta \mu_{ij}  + \Delta U_{ij}\neq 0;\;\; \lambda_{ij\sigma,3} = 0 \;\; {\rm else}, \label{eq:lamb3}
\end{align}
with $\Delta \mu_{ij} = \mu_i - \mu_j$ and $\Delta U_{ij} = U_i - U_j$. 
Within the SW transformation, we obtain
\begin{equation}
  \bar{H}  =  \hat{H}^0 + \sum_{n=2}^{\infty} \frac{n-1}{n!} \mathcal{S}^{n-1}(\hat{V}) =   \hat{H}^0 +  \sum_{n=2}^{\infty}  \bar{H}^{(n)} , \label{eq:barH_SW}
\end{equation}
where $\mathcal{S}$ is the super-operator defined as  $\mathcal{S}(\hat{X}) = [\hat{S},\hat{X}]$, and $\bar{H}^{(n)} = \dfrac{n-1}{n!}\mathcal{S}^{n-1}(\hat{V})$.

Since $\hat{S}$ consists in a truncated form of $\hat{G}$, $\bar{H}$ is not expected to be block-diagonal anymore, i.e. $\bar{H}_X \neq \hat{0}$. 
In the following, we propose recursive relations between $\mathcal{S}^{n}(\hat{V})$ and $\mathcal{S}^{n-1}(\hat{V})$,
derived for the Hubbard dimer, that provide an explicit expression for $\bar{H}$ and in particular for $\bar{H}_X$ in terms of two- and three-body operators. 
These relations are further exploited to develop two modifications of $\hat{S}$, one variational and the other iterative, designed to minimize or even cancel $\bar{H}_X$ while conserving the same complexity as $\hat{S}$.

\section{Modified Schrieffer--Wolff transformations}

\subsection{Recursive relations}

We establish recursive relations for each order of Eq.~(\ref{eq:barH_SW}), which details are provided in  Appendix~\ref{app:Recurence}. More precisely, we find that even orders are block-diagonal, i.e. $\bar{H}^{(2n)}_D = \bar{H}^{(2n)}$ and reads
\begin{eqnarray}
    \bar{H}^{(2n)}  &= &\dfrac{2n - 1}{2(2n)!}\sum_{\substack{i \neq j \\ \sigma}}\sum_{x=0}^3K^{(2n-1)}_{ij\sigma,x}\hat{p}_{ij \bar{\sigma},x}\left(\hat{n}_{i\sigma} - \hat{n}_{j\sigma}\right) \nonumber \\
  & &+ \dfrac{2n - 1}{2(2n)!}\sum_{\substack{i \neq j\\\sigma}}J^{(2n-1)}_{ij\sigma}\left(\hat{\gamma}_{ij\sigma}\hat{\gamma}_{ji\bar{\sigma}} + \hat{\gamma}_{ji\sigma}\hat{\gamma}_{ij\bar{\sigma}}\right) \nonumber \\
 & &+ \dfrac{2n - 1}{2(2n)!}\sum_{\substack{i \neq j\\\sigma}}L^{(2n-1)}_{ij\sigma}\left(\hat{\gamma}_{ij\sigma}\hat{\gamma}_{ij\bar{\sigma}} + \hat{\gamma}_{ji\sigma}\hat{\gamma}_{ji\bar{\sigma}}\right),  \label{eq:MainS2n+1}
\end{eqnarray}
while odd orders are found to be off-block-diagonal, i.e. $\bar{H}^{(2n+1)}_X = \bar{H}^{(2n+1)}$ and take the following expression,
\begin{equation}
  \bar{H}^{(2n + 1)} = \dfrac{2n}{2(2n + 1)!}\sum_{i\neq j\sigma} \sum_{x=0}^3T_{ij\sigma,x}^{(2n)} \hat{p}_{ij \bar{\sigma},x}\left(\hat{\gamma}_{ij\sigma} + \hat{\gamma}_{ji\sigma}\right).   \label{eq:MainS2n}
\end{equation}
and where only the expression of interaction integrals  $I_{ij\sigma,x}^{(k)}$ ($I = J, K, L$ or $T$)  depend on the order $k$. 
Explicit formulas for the interaction integrals $I_{ij\sigma,x}^{(k)}$ are given in Appendix~\ref{app:Recurence}. 
By summing over all orders, $\bar{H}$ takes exactly the following form,
\begin{align}
  \label{eq:Htrans}
  % \bar{H}  & = H^0 + \sum_{n=2}^{\infty}  \bar{H}^{(n)}= \hat{H}^0 +  \bar{H}^{\rm dia} + \bar{H}^{\rm ex} + \bar{H}^{\rm de} +  \bar{H}^{\rm cpl},
               \bar{H}  & = H^0 + \sum_{n=2}^{\infty}  \bar{H}^{(n)} = \bar{H}_D + \bar{H}_X,
\end{align}
with $\bar{H}_D = \hat{H}^0 +  \sum_{n = 1}^{\infty}\bar{H}^{(2n)}$ and $\bar{H}_X =   \sum_{n = 1}^{\infty}   \bar{H}^{(2n +1)}$. Specifically, the non block-diagonal contribution $\bar{H}_X$ couples states from the Heisenberg subspace to states from its complementary subspace, and reads explicitely
\begin{align}
  \label{eq:Hcouple}
 \bar{H}_X =\dfrac{1}{2}\sum_{\substack{i \neq j \\ \sigma}}\sum_{x=0}^3T_{ij\sigma,x}\hat{p}_{ij \bar{\sigma},x}\left(\hat{\gamma}_{ij\sigma} + \hat{\gamma}_{ji\sigma}\right),
\end{align}
where the interaction integrals $T_{ij\sigma,x}$ are obtained by summing over all orders as
\begin{equation}
T_{ij\sigma,x} = \sum_{n=1}^{\infty}  \dfrac{2n}{(2n + 1)!}  T_{ij\sigma,x}^{(2n)}.
\end{equation}
For the homogeneous case, the associated integrals are simply given by
\begin{align}
  \label{eq:Jcpl}
  T_{ij\sigma,0} = T_{ij\sigma,3} = 0,
  \end{align}
  and
\begin{align}
  T_{ij\sigma,1} = T_{ij\sigma,2} = -t_{ij}\left(\cos{(4t/U)} - {\rm sinc}\,(4t/U)\right).
\end{align}
Explicit form of the block-diagonal contributions and relations for the inhomogeneous Hubbard dimer are derived in  Appendix~\ref{app:Recurence}.
\begin{figure}[h]
\vspace{0.2cm}
\resizebox{\columnwidth}{!}{
\includegraphics[scale=1]{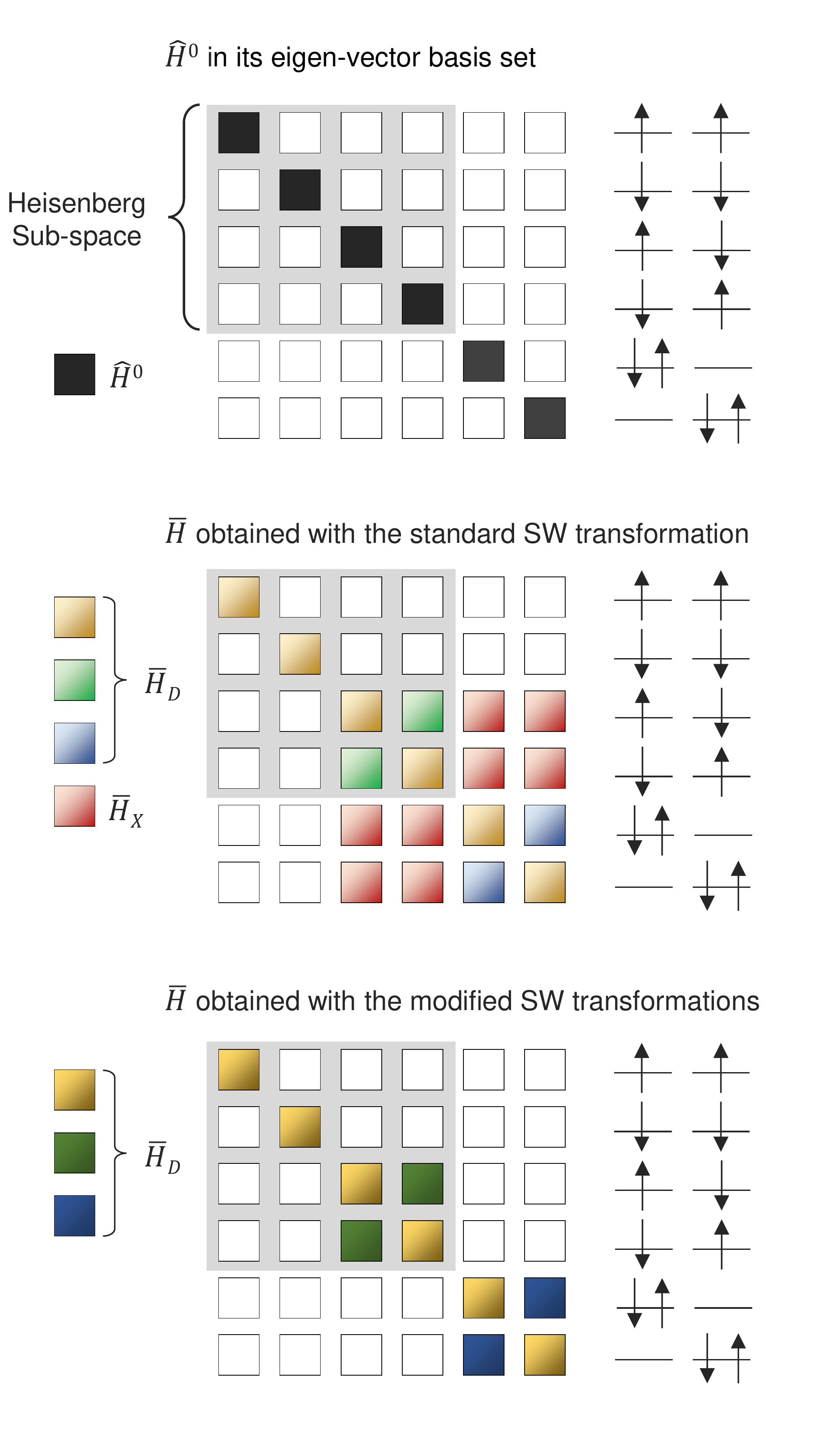}
}
\caption{Schematic representation of the action of the different operators $\bar{H}$ in the Hilbert space of the half-filled Hubbard dimer. }
\label{fig:fig1}
\end{figure}

At this stage, we have established recursive relations to obtain the similar Hamiltonian $\bar{H}$ within the standard SW transformation (SWT) at infinite order of perturbation.
However, given the definition of  $\hat{S}$ in Eq.~(\ref{eq:1storderS}), the standard SW transformation at infinite order does not lead to a block-diagonal representation of $\bar{H}$ with respect to the Heisenberg subspace since  $\bar{H}_X \neq \hat{0}$, see Fig~\ref{fig:fig1}.  
Based on the previous recursive relations, in the following subsections we present two strategies, denoted as modified SW (MSW) transformations, one variational and the other iterative, to fully perform the desired block-diagonalization.

% with $\bar{H}_D = \hat{H}^0 +  \bar{H}^{\rm dia} + \bar{H}^{\rm ex} + \bar{H}^{\rm de}$ and $\bar{H}_X = \bar{H}^{\rm cpl}$. More precisely, the block diagonal part contains diagonal contributions
% \begin{align}
%   \bar{H}^{\rm dia} = \dfrac{1}{2}\sum_{\substack{i \neq j \\ \sigma}}\sum_{x=0}^3K_{ij\sigma,x}\hat{p}_{ij \bar{\sigma},x}\left(\hat{n}_{i\sigma} - \hat{n}_{j\sigma}\right),
% \end{align}
% contributions  that couples spins of different sites,
% \begin{align}
%   \bar{H}^{\rm ex} = \dfrac{1}{2}\sum_{\substack{i \neq j\\\sigma}}J_{ij\sigma}\left(\hat{\gamma}_{ij\sigma}\hat{\gamma}_{ji\bar{\sigma}} + \hat{\gamma}_{ji\sigma}\hat{\gamma}_{ij\bar{\sigma}}\right),
% \end{align}
% and contributions that acts on doubly occupied or empty sites,
% \begin{align}
%   \bar{H}^{\rm de} = \dfrac{1}{2}\sum_{\substack{i \neq j\\\sigma}}L_{ij\sigma}\left(\hat{\gamma}_{ij\sigma}\hat{\gamma}_{ij\bar{\sigma}} + \hat{\gamma}_{ji\sigma}\hat{\gamma}_{ji\bar{\sigma}}\right).
% \end{align}

\subsection{Variational Schrieffer--Wolff transformation }
\label{sec:variational}

We propose to introduce a variational scaling parameter $\theta$ to the unitary transformation, 
\begin{equation}
  \label{eq:Var_SWT_U}
  \hat{U}(\theta) = e^{\theta \hat{S}},
\end{equation}
such that for $\theta = 0$, $\hat{U} = \hat{1}$, and for $\theta = 1$ one recovers the standard SW transformation $\hat{U} = e^{\hat{S}}$. 
Within this unitary transformation, the similar Hamiltonian $\bar{H}(\theta)$ reads
\begin{align}
  \bar{H}(\theta) &= e^{\theta \hat{S}}\hat{H} e^{-\theta \hat{S}}, \nonumber \\
 & = \hat{H}^0 +\hat{V} \left( 1 - \theta \right) + \sum_{n = 2}^{\infty}\frac{\theta^{n-1}(n - \theta)}{n!}\mathcal{S}^{n-1}(\hat{V}).
\end{align}
Using the previously established recursive relations and after summation till the  infinite order, see Appendix~\ref{app:Var}, it can be decomposed
as follows, similarly as in Eq.~(\ref{eq:Htrans}),
%\begin{eqnarray}
%  \label{eq:Htheta}
%  \bar{H}(\theta)  = \hat{H}^0(\theta) +  \bar{H}^{\rm dia}(\theta) + \bar{H}^{\rm ex}(\theta) + \bar{H}^{\rm de}(\theta)  +  \bar{H}^{\rm cpl}(\theta), \nonumber \\
   %  \end{eqnarray}
\begin{eqnarray}
  \label{eq:Htheta}
  \bar{H}(\theta)  =  \bar{H}_D(\theta)  +  \bar{H}_X(\theta),
     \end{eqnarray}
where the $\theta$-dependence lies in the renormalized interaction integrals
that read for the non block-diagonal contribution in the homogeneous case,
\begin{align}
  T_{ij\sigma,0}(\theta) = T_{ij\sigma,3}(\theta) = 0,
\end{align}
and
\begin{align}
  T_{ij\sigma,1}(\theta) = T_{ij\sigma,2}(\theta) = -t\left(\cos{(4t\theta/U)} - \theta{\rm sinc}\,(4t\theta/U)\right),
\end{align}
%and
%\begin{align}
%  K_{ij\sigma,1}(\theta) &= - K_{ij\sigma,2}(\theta) = J_{ij\sigma}(\theta) = - L_{ij\sigma}(\theta) \nonumber\\
%  &=  -t\left(\sin{(4t\theta/U)} - (2t\theta/U){\rm sinc}^2\,(2t\theta/U)\right)/2.
%\end{align}
The scaling parameter $\theta$ can be optimized to minimize the contributions from the coupling operator $\bar{H}_X(\theta)$, which is shown to cancel out for the homogeneous case at 
\begin{align}\label{eq:theta_analytic}
  \theta = (U/4t)\tan^{-1}{(4t/U)},
\end{align}
leading to an exact block-diagonalization of $\bar{H}(\theta)$. 
More precisely, we demonstrate in  Appendix~\ref{app:Var} that in this case and at the saddle point, the variational generator fulfill the VV condition in Eq.~(\ref{eq:VV_MBPT}).  
Relations become more complex for the inhomogeneous cases and the variational process has to be done numerically, see  Appendix~\ref{app:Var}.
In this case, the cancellation of $\bar{H}_X(\theta)$ cannot always be reached.
Alternatively, one can minimize the energy of $\bar{H}(\theta)$ restricted to the Heisenberg subspace, which is equivalent to maximize the overlap between the minimizing state
and the exact ground state $\ket{\Psi_0}$.
The difference between the two optimization schemes is discussed in Appendix~\ref{app:min}.

\subsection{Iterative Schrieffer--Wolff transformation}
\label{sec:iterative}

Alternatively to the variational approach, one can take advantage of the similarity between the coupling terms in Eq.~(\ref{eq:Jcpl}) and 
the perturbation $\hat{V}$
in Eq.~(\ref{eq:Vdecomp}). 
In the spirit of the Foldy--Wouthuysen transformation~\cite{foldy_on_1950}, we propose the following iterative scheme:
\begin{enumerate}
\item Initialize the iterative process by applying the standard SW transformation on the original problem to obtain $\bar{H}^{(s=0)}$.
\item At the iteration $s = s+1$, define the new problem $\bar{H}^{0(s)} = \bar{H}^{(s)} - \bar{V}^{(s)} $ and $\bar{V}^{(s)} = \bar{H}_X^{(s-1)}$.
\item Find the corresponding generator $\hat{S}^{(s)}$ such that $[\bar{H}^{0(s)}, \hat{S}^{(s)}] =  \bar{V}^{(s)}$.
\item Use the recursive relations derived in Appendix~\ref{app:Itgen} to obtain the new  $\bar{H}^{(s)}$.
\item Repeat steps 2 to 4 until convergence is reached, i.e. $\bar{H}_X^{(s)} \rightarrow \hat{0}$.\\
\end{enumerate}
After $N_s$ iterations, the iterative unitary transformation
and the similar Hamiltonian are given by
\begin{equation}
  \label{eq:Ite_SWT_U}
  \hat{U}^{(N_s)} = \left(\prod_{s = 0}^{N_s-1}e^{\hat{S}^{(s)}}\right),
\end{equation}
and
\begin{equation}
  \bar{H}^{(N_s)} = \hat{U}^{(N_s)} \hat{H}\hat{U}^{(N_s)^{\dagger}},
  \end{equation}
  respectively.
The amplitudes of the resulting coupling terms for large $U/t$ behave asymptotically in $(t^2/U)^{N_s}$ for $N_s$ iterations,
such that the iterative algorithm converges exponentially to a precise decoupling.
%\ma{Indeed, as shown in Appendix~\ref{app:Itgen}, after only four iterations the norm of $\bar{H}_X$ is closed to the numerical errors even at $U/t =1$.}

%\subsection{Proof of concept on classical computers}
%\subsection{Perspectives for larger Hubbard systems}
\subsection{Perspectives for larger Hubbard rings}

The iterative and variational MSW transformations are shown to perform exact (quasi) block-diagonalization for the homogeneous (inhomogeneous) Hubbard dimer, respectively, thanks to the recursive properties in Eqs.~(\ref{eq:MainS2n+1}) and (\ref{eq:MainS2n}).
Before investigating the quantum algorithms associated to the presented MSW transformation applied to the Hubbard dimer, we discuss possible extensions to larger systems.
%Indeed, one quickly realize that for larger systems the truncation of the VV perturbative devlopment cannot be anymore fully compensated by the addition of a single variational parameter or by an iterative process. Consequently a resulting truncation error is expected that we propose to evaluate on a classical computer on homogeneous nearest neighbor (NN) Hubbard rings up to $N = 10$ sites. 
First of all, note that the recursive relations obtained in Eqs.~(\ref{eq:MainS2n+1}) and (\ref{eq:MainS2n}) are not valid for larger systems, where additional terms giving rise to interactions among more than two sites emerge. 
Nonetheless, for the purposes of this study, we neglect these terms, meaning that VV perturbation condition in Eq.~(\ref{eq:VV_MBPT}) is not satisfied, but the SW condition in Eq.~(\ref{eq:1storderS}) (i.e first order) still is. 
In this section, the truncation error is assessed on a classical computer for homogeneous nearest neighbour (NN) Hubbard rings of up to $N=10$ sites.
To do so we apply the unitary transformation $\hat{U}^\dagger(\theta) = e^{-\theta \hat{S}}$ 
to the ground state $\ket{\Phi_{\rm Heis}}$ of the NN antiferromagnetic Heisenberg model $J\sum_{ij} \hat{s}_i\hat{s}_j$, where $\hat{s}_i$ denotes the spin operator on site $i$ and $J>0$ is the spin-coupling element, which corresponds to the strongly correlated limit of the NN Hubbard model~\cite{schrieffer_relation_1966}.
%Beyond this limit, the rotated Heisenberg state $U^\dagger(\theta)\ket{\Phi_{\rm Heis}}$ is expected to remain a good approximation of the ground state of the Hubbard Hamiltonian. 
We follow the variational scheme presented in Sec.~\ref{sec:variational}, where $\theta$ is optimized to minimize the expectation value $E(\theta) = \langle \Phi_{\rm Heis}|\bar{H}(\theta)|\Phi_{\rm Heis}\rangle$ with $\bar{H}(\theta)=U(\theta)\hat{H}U^\dagger(\theta)$,
which is equivalent to maximize the overlap of $U^\dagger(\theta)|\Phi_{\rm Heis}\rangle$ with the exact ground state $|\Psi_{0}\rangle$.

\begin{figure}
\resizebox{\columnwidth}{!}{
\includegraphics[scale=1]{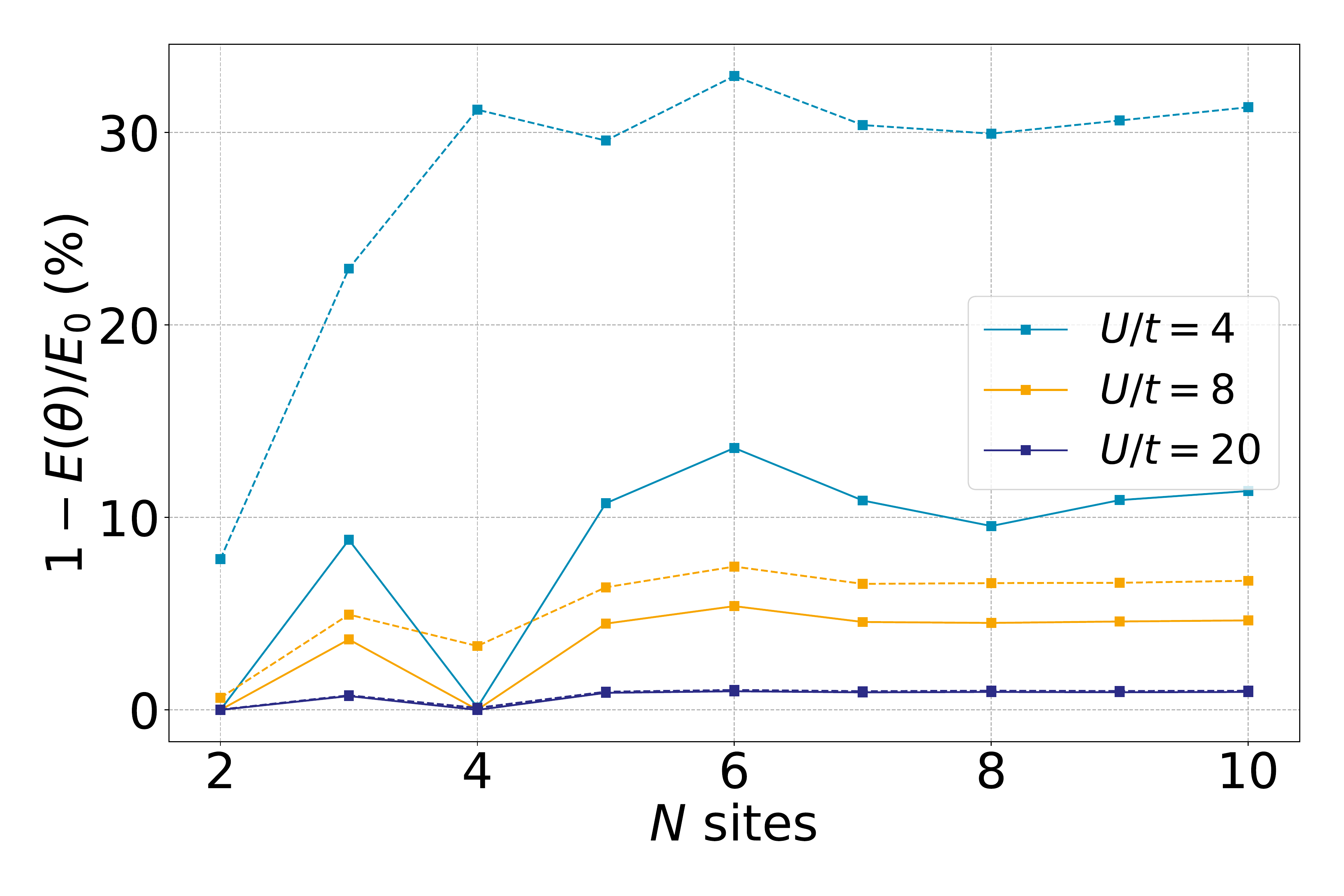}
}
\caption{Relative errors in the ground-state energy calculated for homogeneous half-filled Hubbard rings with respect to the number of sites. Results are given for the variational MSW transformation (solid lines) and the standard SWT ($\theta=1$, dashed lines). Lines are guide for the eye.}
\label{fig:swt_nsites}
\end{figure}

In Fig.~\ref{fig:swt_nsites} we show, as a function of the number of sites $N$ and for different values of the Coulomb repulsion strenght $U/t$, the relative error (in $\%$) of $E(\theta)$ with respect to the exact ground-state energy. 
Results are provided for $\theta = 1$, corresponding to the standard SWT, and for the optimal value $\theta^*$. 
As $N$ increases the relative error increases and appears to converge to what would correspond to the truncation error. 
As expected, the truncation error increases as $U/t$ decreases, i.e. $\sim 1\%\, (1\%), \sim 5\%\, (7\%)$ and $\sim 11\%\, (30\%)$ for $U/t = 20$, 8, 4 and $\theta = \theta^*$ ($\theta = 1)$, respectively. 
The introduction of a single and variational parameter systematically and drastically improves over the standard SWT. 
Consequently, the variational extension to the SW approximation, that is exact for the homogeneous half-filled Hubbard dimer, remains a good approximation for
larger system sizes in the intermediate to strongly correlated regime.
Straightforward improvements can be envisioned by  considering higher-order contributions from the generator,
following the recursive relations proposed in Ref.~[\onlinecite{bravyi_schriefferwolff_2011}], for instance.   
However, the implementation of the MSW transformations
on classical computers is computationally intractable for systems beyond $\sim 16$ orbitals, in analogy with the unitary coupled cluster ansatz~\cite{romero2018strategies}.
This is also the case for the construction of the trial state $|\Phi_{\rm Heis} \rangle$ for large system's size, which we disregard in the following by investigating quantum algorithms applied to the Hubbard dimer, for which $|\Phi_{\rm Heis} \rangle$ is easy to prepare.

\section{Modified SW transformations applied on quantum computers}

At this stage,
we investigate the relevance of the aforementioned MSW transformations, $\bar{H}^{\rm MSW}=\bar{H}(\theta)$ and $\bar{H}^{\rm MSW}=\bar{H}^{(N_s)}$ in Secs.~\ref{sec:variational} and \ref{sec:iterative}, respectively, for the design of new quantum algorithms.

In both cases, $\bar{H}^{\rm MSW}$ is  block-diagonal for the homogeneous case, so that ground or excited states can easily be constructed as linear combination of two basis vectors for the Hubbard dimer, see Fig.~\ref{fig:fig1}.
For the homogeneous half-filled Hubbard dimer,
relevant trial eigenstates of $\bar{H}^{\rm MSW}$
consist in the Heisenberg state $|\Phi_{\rm Heis} \rangle = (1/\sqrt{2})\left(|\uparrow \; \downarrow \,\rangle + | \downarrow \;  \uparrow \,\rangle \right)$
and the ionic state $|\Phi_{\rm Ionic}^\alpha \rangle = \cos(\alpha)|\uparrow \downarrow\; \cdot \,\rangle + \sin(\alpha)|\,  \cdot  \;  \uparrow \downarrow\,\rangle$.
Indeed, $e^{\hat{S}}$ preserving the spin symmetry,  triplet states $|\uparrow \; \uparrow \,\rangle$ and $| \downarrow \;  \downarrow \,\rangle$ are discarded.
The eigenstates of $\hat{H}$ can then be constructed from 
the trial eigenstates
of $\bar{H}^{\rm MSW}$ by applying the transformation $\hat{U}^{\rm MSW}$, which refers to the variational [see Eq.~(\ref{eq:Var_SWT_U})] or to the iterative [see Eq.~(\ref{eq:Ite_SWT_U})] MSW transformation.
It appears clear that both the variational or iterative MSW approaches are adapted to the design of quantum algorithms, as they are both formulated as a unitary transformation applied to an easy-to-prepare initial state.

\begin{figure*}
\resizebox{2\columnwidth}{!}{
\includegraphics[scale=1]{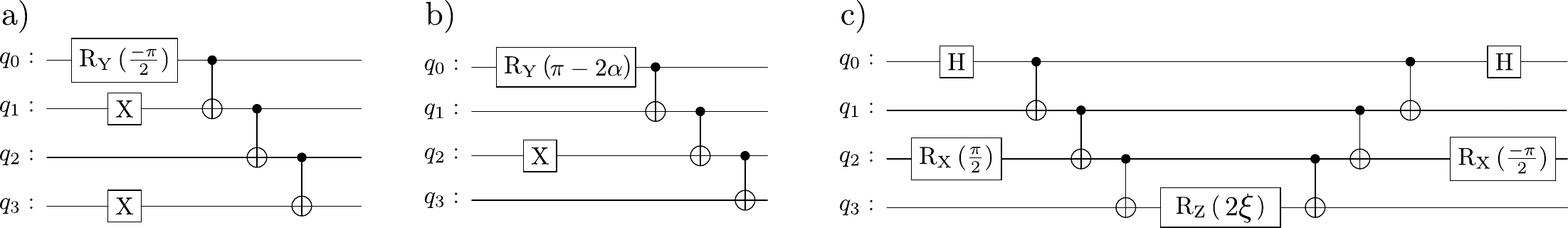}
}
\caption{a) Quantum circuit corresponding to the Heisenberg state $\ket{\Phi_{\rm Heis}} = \left(\ket{\,\uparrow \; \downarrow\,} - \ket{\,\downarrow \; \uparrow\,}\right)/\sqrt{2} =\left(\ket{1001} - \ket{0110}\right)/\sqrt{2} $, b) Quantum circuit corresponding to the linear combination of the ionic states $\ket{\Phi_{\rm Ionic}^\alpha} = \cos(\alpha) \ket{\,\uparrow \downarrow \; \cdot\,} + \sin(\alpha) \ket{\,\cdot \; \uparrow \downarrow\,} = \cos(\alpha) \ket{1100} + \sin(\alpha)\ket{0011}$, c) Quantum circuit corresponding to the implementation of $e^{\xi X_0 Z_1 Y_2 Z_3}$.}
\label{fig:circuit}
\end{figure*}

As a proof of concept, we have implemented both quantum algorithms to treat the homogeneous and inhomogeneous Hubbard dimer, using Qiskit~\cite{Qiskit} to construct the quantum circuits.
We use the one-to-one correspondence between the states of the qubits and the occupation of the spin-orbitals of the Hubbard dimer to map our states onto qubits, with even-numbered qubits corresponding to spin-up orbitals and odd-numbered qubits to spin-down orbitals.
The fermionic creation and annihilation operators are
mapped onto Pauli strings $\hat{\mathcal{P}}_i$
using the Jordan--Wigner (JW) transformation~\cite{jordan1928p}.
To implement the unitary transformation on quantum circuits, the first-order Trotter--Suzuki approximation is used, i.e. the exponential of the sum of Pauli strings is decomposed into
a product of exponential of a single Pauli string,
\begin{eqnarray}\label{eq:trotter}
e^{\theta \hat{S}} \xrightarrow{\rm JW} 
e^{\theta \sum_i \xi_i \hat{\mathcal{P}}_i} \approx \prod_i e^{\theta  \xi_i \hat{\mathcal{P}}_i},
\end{eqnarray}
for which the associated circuit is known (see panel (c) of Fig.~\ref{fig:circuit}),
and $\lbrace\xi_i\rbrace$ are the coefficients that are functions of the SW generator parameters $\{\lambda\}$ obtained after the JW transformation.
The trial Heisenberg and ionic states can be easily prepared on the quantum computer, as shown in panels (a) and (b) of Fig.~\ref{fig:circuit}.
Finally, we simulate our variational MSW transformation
using a noise model built on Qiskit.
This noise model consists in a depolarizing quantum error channel applied on every one- and two-qubit gates, with depolarizing error parameters of $\lambda_1 = 0.0001$ and $\lambda_2 = 0.001$, respectively.
Note that the 4-qubit circuit resulting from the variational MSW transformation is composed of 32 one-qubit gates and 35 CNOT gates, and that no readout error is considered.
Sampling noise is also added to this noise model by considering $n_{\rm shots} = 8192$ for the estimation of the expectation value of each Pauli string.
The variational parameter was optimized by using the SPSA optimizer with a maximum of 1000 iterations.
Then, the optimal parameter $\theta^*$ is calculated as the mean of the last 25 iterations, and the expectation values of $\bar{H}(\theta^*)$
with respect to the Heisenberg and ionic states are estimated as the mean of another 100 noisy simulations (with fixed parameter $\theta^*$).
The noisy results are then compared to the exact references
obtained by exact diagonalization, as well as to the noiseless state-vector simulation, without considering any quantum or sampling noise and for which the L-BFGS-B optimizer was used to update the variational parameter.
For the iterative MSW transformation, only state-vector simulations is performed.

\subsection{Variational approach}

The variational approach described in Sec.~\ref{sec:variational} consists in finding the optimal parameter $\theta_X$ of the unitary in Eq.~(\ref{eq:Var_SWT_U}) such that the couplings $\bar{H}_X(\theta_X)$ are minimized, thus enforcing the block-diagonalization of $\bar{H}(\theta_X)$. 
Compared to the strategy of Zhang and coworkers~\cite{zhang_quantum_2022},
our minimization process implies only a single variational parameter, rather than a number of parameters that would correspond to the number of Pauli strings composing the generator [i.e. each $\xi_i$ in Eq.~(\ref{eq:trotter})].
Minimizing the couplings $\bar{H}_X(\theta_X)$ requires the estimation of the off-diagonal matrix elements of $\bar{H}$ such as done in Ref.~[\onlinecite{zhang_quantum_2022}].
However, this is not straightforward on quantum computers in contrast to the measurement of expectation values, although
one can note some improvements in the literature~\cite{huggins2020non,stair2021simulating}.
Consequently in this work, in analogy with the VQE algorithm, we minimize the energy $\langle \Phi \hat{U}(\theta)  | \hat{H} | \hat{U}^{\dagger}(\theta) \Phi \rangle $ measured on the quantum device, rather than minimizing the norm of $\bar{H}_X$. Note that the two strategies are equivalent when the initial trial state $|\Phi\rangle$ is indeed the ground state of $\bar{H}(\theta)$. Otherwise, it does not lead to the expected block-diagonalization of the Hamiltonian, as discussed in more details in Appendix~\ref{app:min}.

\begin{figure}
\resizebox{\columnwidth}{!}{
\includegraphics[scale=1]{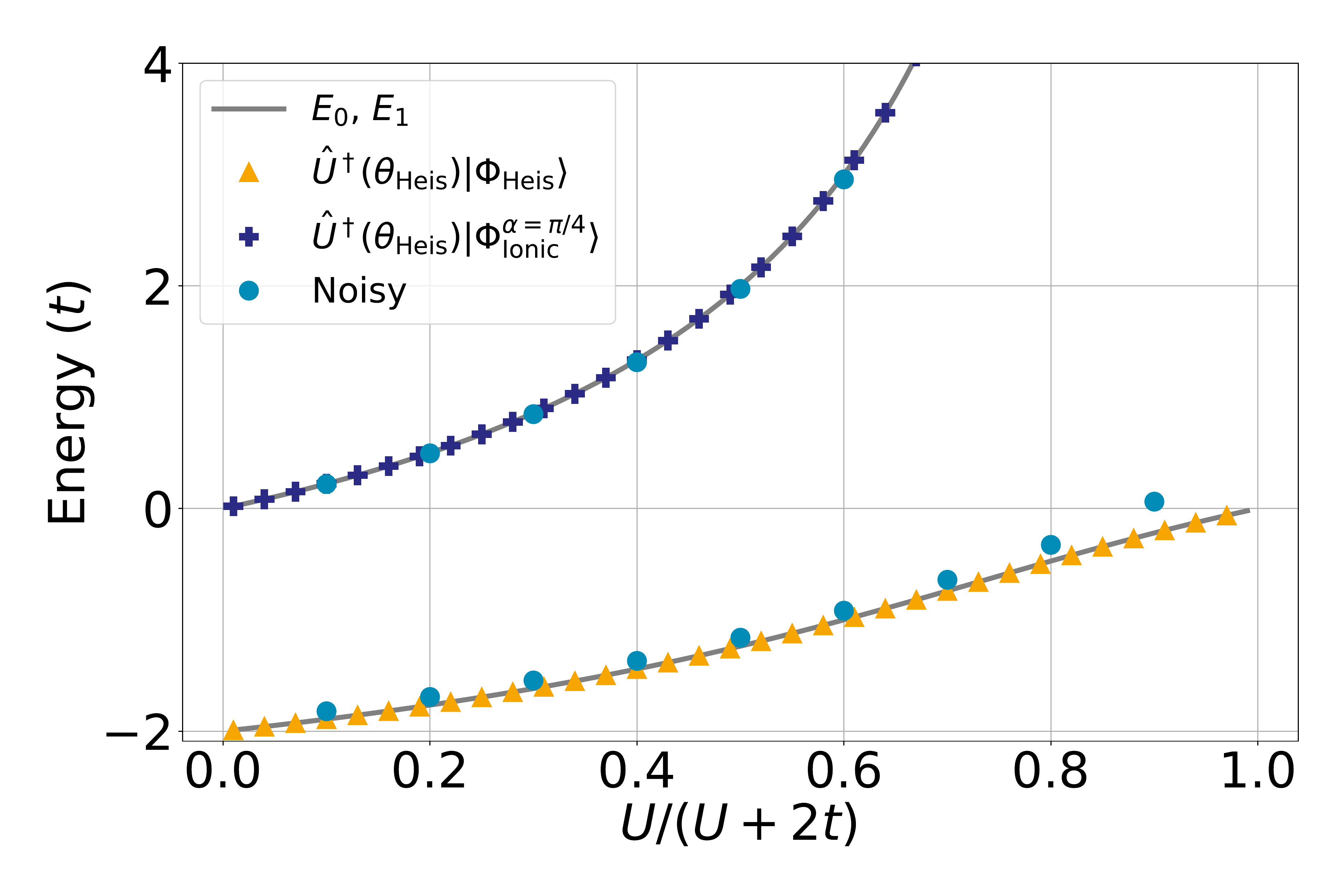}
}
\caption{Energies of the half-filled Hubbard dimer for $\Delta \mu = 0$ with respect to the repulsion strength, using the variational SW method
to minimize the energy $\bra{\Phi_{\rm Heis}}\bar{H}(\theta)\ket{\Phi_{\rm Heis}}$.}
\label{fig:var_deltav0}
\end{figure}

Let us start with the homogeneous Hubbard dimer ($\Delta \mu = 0$).
In this case, analytical expressions for the variational SW transformation can be derived
as shown in Sec.~\ref{sec:variational}.
As readily seen in Fig.~\ref{fig:var_deltav0},
the energy obtained by minimizing
$\bra{\Phi_{\rm Heis}}\bar{H}(\theta)\ket{\Phi_{\rm Heis}}$
matches exactly the ground-state energy of $\hat{H}$, and the minimizing parameter, denoted by $\theta_{\rm Heis}$, is exactly the same as the analytical expression in Eq.~(\ref{eq:theta_analytic}) (not shown).
In addition, using the exact same unitary $\hat{U}(\theta_{\rm Heis})$ but on the
equi-weighted ionic state $\vert \Phi_{\rm Ionic}^{\alpha = \pi/4} \rangle$, one recovers the first-excited singlet energy of $\hat{H}$.
Thus, our variationally optimized SW transformation has indeed block-diagonalized $\hat{H}$ exactly for any repulsion strength $U/t$, 
with the Heisenberg subspace containing the singlet ground state and the triplet states.
Looking at the energies obtained from the noisy simulation,
they follow closely the noiseless results, especially for the first-excited state energy for which the relative error does not exceed 1.5\%. 
We note also an increase of around 0.03 in the expectation value of the spin operator $\hat{S}^2$ due to the noise, showing that the
final state is not a pure singlet state anymore.

\begin{figure}
\resizebox{\columnwidth}{!}{
\includegraphics[scale=0.2]{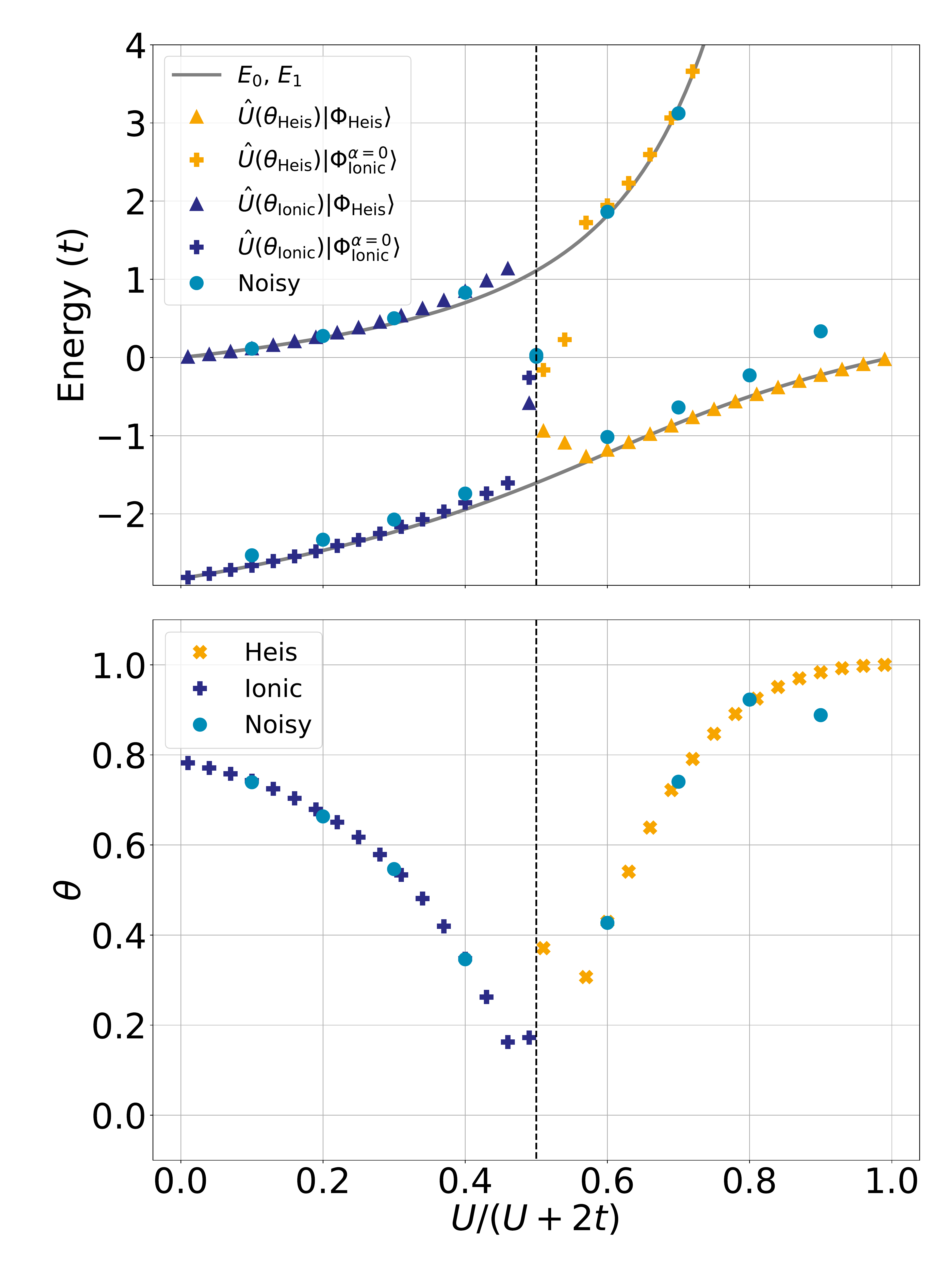}
}
\caption{Energies of the half-filled Hubbard dimer for $\Delta \mu/t = 2$ (top panel) with respect to the repulsion strength, using the variational SW method
to minimize the energies $\bra{\Phi_{\rm Heis}}\bar{H}(\theta)\ket{\Phi_{\rm Heis}}$ (orange markers, shown for $U > \Delta \mu$) and $\bra{\Phi_{\rm Ionic}^{\alpha=0}}\bar{H}(\theta)\ket{\Phi_{\rm Ionic}^{\alpha = 0}}$ (blue markers, shown for $U < \Delta \mu$).
The associated minimizing parameters $\theta_{\rm Heis}$ and
$\theta_{\rm Ionic}$ are shown on the bottom panel, respectively.
The vertical dotted line corresponds to $U = \Delta \mu$.}
\label{fig:var_deltav2}
\end{figure}

Turning to the inhomogeneous Hubbard dimer
with $\Delta \mu /t = 2$,
no analytical expressions are known
for the optimal parameter $\theta$.
In contrast to the homogeneous case,
minimizing the energy $\bra{\Phi_{\rm Heis}}\bar{H}(\theta)\ket{\Phi_{\rm Heis}}$ doesn't lead to a block-diagonal
$\bar{H}(\theta_{\rm Heis})$ in the entire range of interaction,
but only for $U \gg \Delta \mu$ as shown in Fig.~\ref{fig:var_deltav2}.
In the other case,
the ground state doesn't belong to the Heisenberg
subspace such that
$\bar{H}(\theta_{\rm Heis})$ is not block-diagonal.
Hence, the Heisenberg state is not an eigenstate of $\bar{H}(\theta_{\rm Heis})$, neither is the ionic state (see Appendix~\ref{app:min} for more details).
However, rather than minimizing the energy with respect to the Heisenberg state, one can prepare a different initial trial state corresponding to the ground state (or a good approximation of it)
that belongs to the other subspace.
In the case of the Hubbard dimer, this is the ionic subspace which ground state is
a linear combination of the ionic states (see Fig.~\ref{fig:fig1} and panel (b) of Fig.~\ref{fig:circuit}).
As $\Delta \mu/t = 2$, the optimal $\alpha$ value
of $\ket{\Phi_{\rm Ionic}^\alpha}$ is not trivial and
we approximate it as 0, i.e. $\vert \Phi_{\rm Ionic}^{\alpha=0}\rangle = \ket{\, \uparrow \downarrow \; \cdot \,}$.
Minimizing $\langle\Phi_{\rm Ionic}^{\alpha=0}\vert\bar{H}(\theta)\vert\Phi_{\rm Ionic}^{\alpha=0}\rangle$
now leads to an optimal $\theta_{\rm Ionic}$ that approximately block-diagonalizes $\bar{H}(\theta_{\rm Ionic})$.
More precisely, one recovers the correct ground-state and first-excited-state singlet energies
for $U \ll \Delta \mu$ by measuring the expectation values
$\langle\Phi_{\rm Ionic}^{\alpha=0}\vert\bar{H}(\theta_{\rm Ionic})\vert\Phi_{\rm Ionic}^{\alpha=0}\rangle$
and
$\bra{\Phi_{\rm Heis}}\bar{H}(\theta_{\rm Ionic})\ket{\Phi_{\rm Heis}}$, where $\theta_{\rm Ionic}$ is defined as the optimal parameter that minimizes $\langle\Phi_{\rm Ionic}^{\alpha=0}\vert\bar{H}(\theta)\vert\Phi_{\rm Ionic}^{\alpha=0}\rangle$.
Interestingly, the Heisenberg state now belongs to the subspace which contains the first-excited singlet state, as opposed to the correlation regime $U \gg \Delta \mu$.
Note that in the strictly correlated (or atomic) limit $U/t \rightarrow \infty$, $\theta_{\rm Heis}$ tends to 1
(see bottom panel of Fig.~\ref{fig:var_deltav2}), which is expected as the variational SW transformation tends to the standard SW transformation that is exact in this limit.
Moving from this limit, the value of the optimal parameter $\theta_{\rm Heis}$ decreases to compensate the error from applying the MSW transformation in the non-atomic limit.
Finally, the noisy simulations show a relatively good agreement with the noiseless results.
In analogy with the homogeneous model on Fig.~\ref{fig:var_deltav0}, the expectation value of $\hat{S}^2$ also increases from 0 to around 0.05, 
and the deviation in energy is more significant on the ground-state energy and when $U/t$ increases.
According to the bottom panel of Fig.~\ref{fig:var_deltav2}, it seems that the optimized parameter obtained from the noisy simulation deviates significantly from the exact one for large $U/t$ values (last blue circle on the curve), thus indicating that the classical optimization for large $U/t$ values is more challenging
in the noisy environment. 
This could be mitigated by employing error mitigation strategies that are outside of the scope of this manuscript~\cite{cai2022quantum}.

\subsection{Iterative approach}

\begin{figure}
\resizebox{\columnwidth}{!}{
\includegraphics[scale=1]{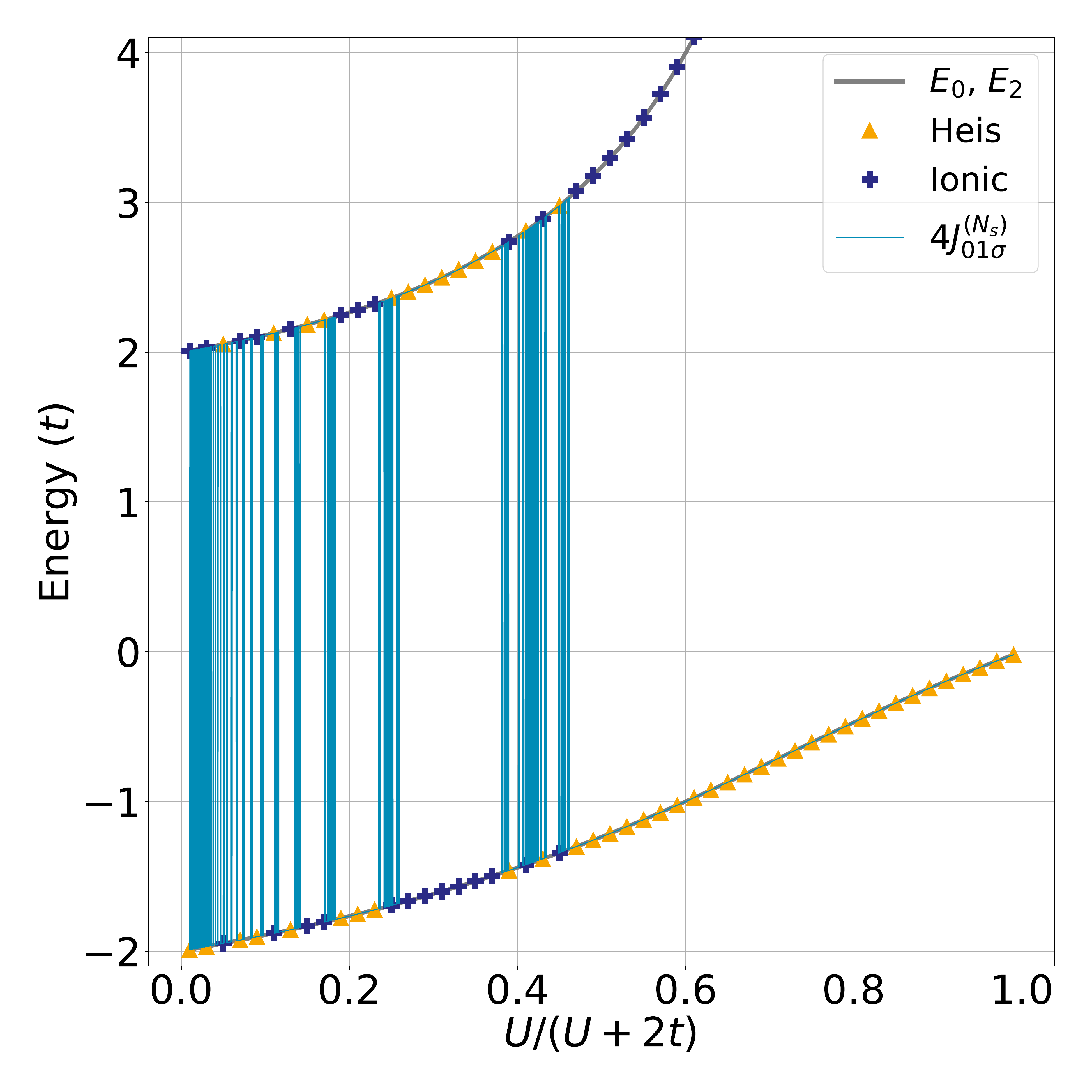}
}
\caption{Energies of the half-filled Hubbard dimer for $\Delta \mu = 0$ with respect to the repulsion strength, using the iterative SW transformation applied on the Heisenberg state (orange triangles) and the equi-weighted ionic state (blue crosses). The exchange integrals $J_{01\sigma}^{(N_s)}$ obtained at convergence are also represented.}
\label{fig:it_deltav0}
\end{figure}

The steps 1 to 5 of the iterative approach described in Sec.~\ref{sec:iterative}
can all be performed on a classical computer by using the recursive relations
derived in Appendix~\ref{app:Var}, such that only the preparation of the final state $U^{(N_s)\dagger}\ket{\Phi}$
and the measurement of $\hat{H}$ are done on the quantum device.
Let us start with the homogeneous dimer in Fig.~\ref{fig:it_deltav0}.
Interestingly, and in contrast to the variational approach,
the ground state doesn't always belong to the Heisenberg subspace.
Indeed, the ground and second-excited singlet states of $\bar{H}^{(N_s)}$
oscillate between the Heisenberg and the equi-weighted ($\alpha = \pi/4$) ionic states.
This can be rationalized by analyzing the behaviour of the exchange integrals.
Indeed, the analytical function in Eq.~(\ref{eq:Jana}) at iteration 0 (corresponding to the standard SW transformation) shows that the exchange integrals oscillate and change sign for different correlation strength (not shown).
The iterative process strongly sharpens these oscillations (though the function remains continuous and infinitely differentiable for all $U/t > 0$), as shown by the solid blue lines in Fig.~\ref{fig:it_deltav0}.
The change of sign of the exchange integrals indicates a change in the ground state
of $\bar{H}^{(N_s)}$.
If they are negative, the ground state belongs to the Heisenberg subspace, while it belongs to the ionic subspace if they are positive.
One can also verify that the energy required to go from the Heisenberg state to the equi-weighted ionic state is of $4J_{01\sigma}^{(N_s)}$, where $J_{01\sigma}^{(N_s)}$ are the couplings terms obtained after $N_s$ iterations.
Finally, note that the first-excited singlet state energy
of $\hat{H}$
is actually exactly recovered from the $\vert\Phi_{\rm Ionic}^{\alpha = - \pi/4}\rangle$ state that is an eigenstate of $\bar{H}^{(N_s)}$ (not shown).

\begin{figure}
\resizebox{\columnwidth}{!}{
\includegraphics[scale=0.2]{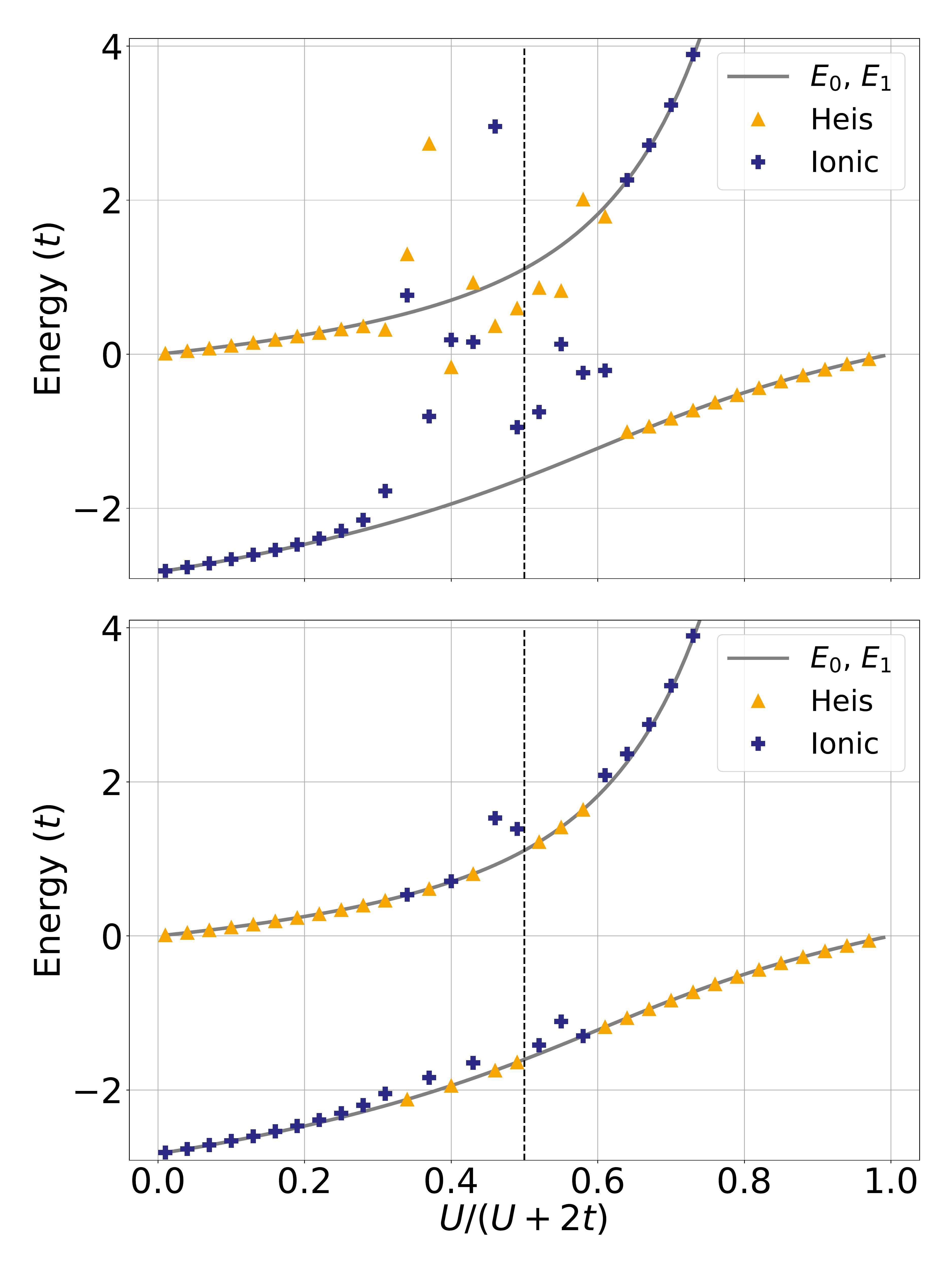}
}
\caption{Energies of the half-filled Hubbard dimer for $\Delta \mu/t = 2$ with respect to the repulsion strength.
The iterative SW transformation is applied on the Heisenberg state (orange triangles) and on the pure ionic state (blue crosses), with (top panel) and without (bottom panel) trotterization error.
The vertical dotted line corresponds to $U = \Delta \mu$.}
\label{fig:it_deltav2}
\end{figure}

Turning to the inhomogeneous dimer with $\Delta \mu/t = 2$
in Fig.~\ref{fig:it_deltav2},
one observes a similar behaviour than for the variational approach in Fig.~\ref{fig:var_deltav2},
i.e. the ground state belongs to the Heisenberg subspace for $U \gg \Delta \mu$ and to the ionic subspace for $U \ll \Delta \mu$.
However, the values around the transition $U \sim \Delta \mu$ (top panel of Fig.~\ref{fig:it_deltav2}) are much less accurate than for the variational method.
This can be rationalized by comparing the energies obtained with and without trotterizing the SW transformation (top and bottom panels of Fig.~\ref{fig:it_deltav2}, respectively).
Indeed, $\bar{H}^{(N_s)}$ obtained without trotterization is block-diagonal
(appart from some small deviation for a few points),
although there are some interchanges between the nature of the ground state around the transition $U \sim \Delta \mu$ in contrast to the variational approach.
Trotterizing the iterative SW transformation does lead to significant errors and to a non-block-diagonalized $\bar{H}^{(N_s)}$, such that the Heisenberg and the ionic states are not eigenstates of $\bar{H}^{(N_s)}$ anymore.
Such trotterization errors are much more pronounced within the
iterative method than the variational one for two reasons.
On the one hand, the successive applications of more than one (most of the time, 3 iterations in this work) unitary transformation does multiply the number of operators that have to be trotterized.
On the other hand,
it is known that the variational optimization of the parameters in VQE-based algorithms does compensate the
trotter errors~\cite{grimsley2019trotterized}.\\

\subsection{Variational versus iterative approach: numerical efficiency}

In contrast to the variational approach, the iterative approach leads to a fully quantum (parameter-free) algorithm as it simply consists in applying the unitary transformation of Eq.~(\ref{eq:Ite_SWT_U}) on a prepared eigenstate of $\bar{H}^{\rm MSW} = \bar{H}^{(N_s)}$.
However, the associated quantum circuit is much deeper than for the variational approach, which applies a single unitary transformation only.
In terms of gate complexity, the number of CNOT required to implement a single SW unitary transformation scales
with the number of Pauli terms in the SW generator, as well as with the number of qubits as shown by the cascade of CNOT in panel (c) of Fig.~\ref{fig:circuit}.
To evaluate the relevance of our approach to more complex systems, we extrapolate the computational scaling for a $N$-sites Hubbard model.
As the operators of the SW generator only act on nearest-neighbor sites, the number of Pauli terms scales as $\mathcal{O}(N)$.
For the iterative approach, one has to multiply by the number of iterations $N_s$.
Hence, the number of CNOT scales as $\mathcal{O}(N^2)$ and $\mathcal{O}(N_s N^2)$ for the variational and the iterative approach, respectively.
Although the variational approach is more attractive in the NISQ era due to its shallower circuit depth,
it is at the expense of much more measurements
as it has to be multiplied at least by the number of iterations 
dictated by the type of cost function and the method used for the classical optimization  of the circuit parameters.
Note that while the iterative approach appears less adapted to NISQ computers, its associated circuit depth still remains far shallower than quantum phase estimation based approaches, such as the fault-tolerant one proposed in Ref.~[\onlinecite{zhang_quantum_2022}].
Which method is the most efficient will depend on the ability of the considered quantum computer to afford deep quantum circuit.
Within noisy quantum computer, the variational approach appears more adapted, while the iterative method can be used on fault tolerant devices.

\section{Conclusions and perspectives}

In this paper, we derived recursive relations for the Schrieffer--Wolff transformation applied on the half-filled Hubbard dimer. 
Based on these findings, we proposed a variational and an iterative modification of the standard SW transformation to approach, 
or even to perform for homogeneous case, a block-diagonalization of the Hamiltonian.
These modified Schrieffer--Wolff transformations have been used to design two quantum algorithms that have been implemented and compared on the half-filled Hubbard dimer.
Regarding the extension of this work to design efficient and alternative quantum algorithms for the general Hubbard model, or even for other models or {\it ab-initio} Hamiltonian, several challenges have to be addressed.

At this stage, one could directly, and without modification, use the variational SW Ansatz [Eq.~(\ref{eq:Var_SWT_U})], the iterative Ansatz [Eq.~(\ref{eq:Ite_SWT_U})] or a combination of both 
to evaluate the ground-state energy of a given Hubbard model.
Beside the fact that it consists in a serious approximation, as additional terms in the perturbative expansion will implicitly be neglected for Hubbard models larger than two sites, it also requires to prepare a relevant trial state $\ket{\Phi}$ that generalizes the Heisenberg state used for the Hubbard dimer.
If no trivial and easy-to-prepare trial eigenstates of $\hat{H}^0$ are known, this step could be performed variationally
using the VQE algorithm, for instance.
Within this strategy, one can expect valuable results for the regime of large $U/t$ values and close to half-filling.
Alternatively, one could apply the modified SW transformations
on a few relevant and easy-to-prepare states that 
we know belong to the
low-energy subspace we are interested in,
thus forming a basis on which all the
Hamiltonian matrix elements are measured on the quantum computer,
followed by a classically diagonalization in the same spirit of the quantum subspace diagonalization methods~\cite{mcclean2017hybrid,
motta2020determining,stair2020multireference}.

Finally,
the generalization of this work to any filled Hubbard model or to the Quantum Chemistry Hamiltonian probably requires to improve the generator. 
This could be done for instance, in the spirit of coupled cluster approaches, by introducing more complex terms or more variational parameters.
All the aforementioned developments are beyond the scope of this manuscript and are left for future work.

\begin{acknowledgments}
The authors would like to thank the ANR (Grant No. ANR-19-CE29-0002 DESCARTES
project) for funding.
\end{acknowledgments}

\appendix

\section{Recursive relations for the standard Schrieffer--Wolff transformation}
\label{app:Recurence}
As originally presented by SW, defining $\hat{S}$ with the SW condition (\ref{eq:1storderS}) is relevant to construct low energy effective Hamiltonian close to the atomic limit $U/t \rightarrow \infty$.  However, as shown in the following, the use of the generator $\hat{S}$ rapidly leads to off-block-diagonal terms in the perturbative expansion that are far from negligible, already at intermediate value of $U/t$. 
More precisely, second-order ($n=2$) contributions read
 \begin{align}
   \bar{H}^{(2)} =\dfrac{1}{2} \mathcal{S}(\hat{V})  =    \dfrac{1}{2}\left[\hat{S}, \hat{V} \right] =  \bar{H}^{(2),{\rm dia}} + \bar{H}^{(2),{\rm ex}} + \bar{H}^{(2),{\rm de}},
 \end{align}
 where
 \begin{align}
   &  \bar{H}^{(2),{\rm dia}} =\dfrac{1}{2}\sum_{\substack{i \neq j \\ \sigma}}\sum_{x=0}^3K^{(1)}_{ij\sigma,x}\hat{p}_{ij \bar{\sigma},x}\left(\hat{n}_{i\sigma} - \hat{n}_{j\sigma}\right)
 \end{align}
 corresponds to diagonal contributions with $K^{(1)}_{ij\sigma,x} = -2t_{ij}\lambda_{ij\sigma,x}$, and
 \begin{align}
   & \bar{H}^{(2),{\rm ex}} =  \dfrac{1}{2}\sum_{\substack{i \neq j\\\sigma}}J^{(1)}_{ij\sigma}\left(\hat{\gamma}_{ij\sigma}\hat{\gamma}_{ji\bar{\sigma}} + \hat{\gamma}_{ji\sigma}\hat{\gamma}_{ij\bar{\sigma}}\right)
 \end{align}
 corresponds to exchange terms, couples spins of different sites and acts solely in the Heisenberg subspace with $J^{(1)}_{ij\sigma} = t_{ij}(\lambda_{ij\sigma,1} - \lambda_{ij\sigma,2})$.
 Finally,
 \begin{align}
   &\bar{H}^{(2),{\rm de}} =  \dfrac{1}{2}\sum_{\substack{i \neq j\\\sigma}}L^{(1)}_{ij\sigma}\left(\hat{\gamma}_{ij\sigma}\hat{\gamma}_{ij\bar{\sigma}} + \hat{\gamma}_{ji\sigma}\hat{\gamma}_{ji\bar{\sigma}}\right)
 \end{align}
creates and annihilates double occupations,  with $L^{(1)}_{ij\sigma} = -t_{ij}(\lambda_{ij\sigma,1} - \lambda_{ij\sigma,2})$, and thus does not act on the Heisenberg subspace at half band filling. At this stage, the perturbed Hamiltonian is stable through the Heisenberg space.
The third-order contributions lead to
\begin{align}
  &  \bar{H}^{(3)} = \dfrac{1}{3}\mathcal{S}^2(\hat{V})  =  \dfrac{1}{3}\left[\hat{S}, \left[\hat{S}, \hat{V} \right]\right]  = \bar{H}^{(3),\rm cpl} ,
\end{align}
where
\begin{align}\label{eq:H3cpl}
 &\bar{H}^{(3),{\rm cpl}} =\dfrac{1}{2}\sum_{\substack{i \neq j \\ \sigma}}\sum_{x=0}^3T^{(2)}_{ij\sigma,x}\hat{p}_{ij \bar{\sigma},x}\left(\hat{\gamma}_{ij\sigma} + \hat{\gamma}_{ji\sigma}\right),
\end{align}
with
\begin{eqnarray}
T^{(2)}_{ij\sigma,0} &=& 4t_{ij}\lambda^2_{ij\sigma,0}, \nonumber \\
T^{(2)}_{ij\sigma,1} &=& 2t_{ij}[\lambda_{ij\sigma,1}(2\lambda_{ij\sigma,1} + (\lambda_{ij\sigma,1} - \lambda_{ij\sigma,2})) \nonumber \\
& & + (\lambda_{ij\sigma,1} - \lambda_{ij\sigma,2})^2], \nonumber \\
T^{(2)}_{ij\sigma,2} &=& 2t_{ij}[\lambda_{ij\sigma,2}(2\lambda_{ij\sigma,2} + (\lambda_{ij\sigma,2} - \lambda_{ij\sigma,1})) \nonumber \\
                     & & + (\lambda_{ij\sigma,2} - \lambda_{ij\sigma,1})^2],
\end{eqnarray}
and
\begin{eqnarray}
  T^{(2)}_{ij\sigma,3} &=& 4t_{ij}\lambda^2_{ij\sigma,3}.
\end{eqnarray}
Obviously, $\bar{H}^{(3),\rm cpl}$ couples states from the Heisenberg subspace to the other states belonging to the complementary subspace. 
Hence, already by truncating at the third order, the similar Hamiltonian $\bar{H} =  H^0 + \sum_{n=2}^{3} \bar{H}^{(n)} $ is not block-diagonal (i.e stable)
anymore with respect to the Heisenberg subspace, as one can expect since the SW generator $\hat{S}$ is designed to keep $\bar{H}$ block-diagonal at first order only, see Fig~\ref{fig:fig1}.\\
Interestingly, the form of $\bar{H}^{(3),\rm cpl}$ 
in Eq.~(\ref{eq:H3cpl}) is analogous to the original $\hat{V}$ which can be rewritten as
\begin{equation}
  \label{eq:Vdecomp}
\hat{V} =   \dfrac{1}{2} \sum_{ij}\sum_{x=0}^3t_{ij\sigma,x}\hat{p}_{ij \bar{\sigma},x}\left(\hat{\gamma}_{ij\sigma} + \hat{\gamma}_{ji\sigma}\right),
\end{equation}
with $t_{ij\sigma,x} = -t_{ij}$, $\forall 0 \leq x\leq 3$.
Given the SW generator, see in Eq.~(\ref{eq:SWS}),
\begin{equation}
  \label{SSW_app}
   \hat{S} = \dfrac{1}{2}\sum_{i\neq j,\sigma}\sum_{x=0}^3 \lambda_{ij\sigma,x}\hat{p}_{ij\bar{\sigma},x}\left(\hat{\gamma}_{ij\sigma} - \hat{\gamma}_{ji\sigma}\right), 
 \end{equation}
 where the coefficients $\lambda_{ij\sigma,x}$ are defined in Eqs.~(\ref{eq:lamb0})--(\ref{eq:lamb3}),
the definition $\hat{H} = \hat{H}^0 + \hat{V}$ and the unitary transformation $e^{\hat{S}}$,
one obtains the transformed Hamiltonian
\begin{align}
  \bar{H} & = e^{\hat{S}} \hat{H} e^{-\hat{S}} \nonumber \\
  &= \left(\sum_n \frac{\hat{S}^{n}}{n!} \right) \hat{H}  \left(\sum_n \frac{(-1)^n \hat{S}^{n}}{n!}\right) \nonumber \\
  & =  \hat{H}^0 + \sum_{n=2}^{\infty} \frac{n-1}{n!} \mathcal{S}^{n-1}(\hat{V}), \label{eq:Hbarapp}
\end{align}
where $\mathcal{S}(\hat{X}) = [\hat{S},\hat{X}]$ is  a super-operator that transforms $\hat{X}$ into another operator acting in the same Hilbert space~\cite{primas_generalized_1963}.
Within the previous definition of $\hat{S} $ and $\hat{V}$, it can be shown that even orders of Eq.~(\ref{eq:Hbarapp}) take the following form,
\begin{equation}
  \mathcal{S}^{2n}(\hat{V}) = \dfrac{1}{2}\sum_{i\neq j\sigma}\sum_{x=0}^3T_{ij\sigma,x}^{(2n)}\hat{p}_{ij \bar{\sigma},x}\left(\hat{\gamma}_{ij\sigma} + \hat{\gamma}_{ji\sigma}\right),  \label{eq:S2n}
\end{equation}
while odd orders are written as
\begin{eqnarray}
  \mathcal{S}^{2n+1}(\hat{V}) &= &\dfrac{1}{2}\sum_{\substack{i \neq j \\ \sigma}}\sum_{x=0}^3K^{(2n+1)}_{ij\sigma,x}\hat{p}_{ij \bar{\sigma},x}\left(\hat{n}_{i\sigma} - \hat{n}_{j\sigma}\right) \nonumber \\
  & &+ \frac{1}{2}\sum_{\substack{i \neq j\\\sigma}}J^{(2n+1)}_{ij\sigma}\left(\hat{\gamma}_{ij\sigma}\hat{\gamma}_{ji\bar{\sigma}} + \hat{\gamma}_{ji\sigma}\hat{\gamma}_{ij\bar{\sigma}}\right) \nonumber \\
 & &+ \dfrac{1}{2}\sum_{\substack{i \neq j\\\sigma}}L^{(2n+1)}_{ij\sigma}\left(\hat{\gamma}_{ij\sigma}\hat{\gamma}_{ij\bar{\sigma}} + \hat{\gamma}_{ji\sigma}\hat{\gamma}_{ji\bar{\sigma}}\right).  \label{eq:S2n+1}
\end{eqnarray}
Moreover, interaction integrals in Eqs.~(\ref{eq:S2n}) and (\ref{eq:S2n+1}) can be recursively obtained as follows,
\begin{align}
  K^{(2n+1)}_{ij\sigma,x} = & 2\lambda_{ij\sigma,x}T_{ij\sigma,x}^{(2n)},
\end{align}
\begin{align}
  J^{(2n+1)}_{ij\sigma} = &\dfrac{1}{2}\left(K^{2n+1}_{ij\sigma,2} - K^{(2n+1)}_{ij\sigma,1}\right),
\end{align}
\begin{align}
  L^{(2n+1)}_{ij\sigma} = &\dfrac{1}{2}\left(K^{(2n+1)}_{ij\sigma,2}\beta - K^{(2n+1)}_{ij\sigma,1}/\beta\right),
\end{align}
\begin{align}
  T_{ij\sigma,0}^{(2n)} = & -2\lambda_{ij\sigma,0} K^{(2n-1)}_{ij\sigma,0} = -4\lambda_{ij\sigma,0}^2T_{ij\sigma,0}^{(2n-2)} \nonumber \\
   T_{ij\sigma,0}^{(2n)} = & (-1)^nt_{ij\sigma,0} (2\lambda_{ij\sigma,0})^{2n}, \label{eq:tau_ij_0}
\end{align}
\begin{align}
  T_{ij\sigma,1}^{(2n)} = & -\lambda_{ij\sigma,1}\left(3K^{(2n-1)}_{ij\sigma,1} - K^{(2n-1)}_{ij\sigma,2} - 2J^{(2n-1)}_{ij\sigma}\right)\nonumber\\
                        & + 2\lambda_{ij\sigma,2}L^{(2n-1)}_{ij\sigma},  \label{eq:tau_ij_1} 
\end{align}
\begin{align}
  T_{ij\sigma,2}^{(2n)} = & -\lambda_{ij\sigma,2}\left(3K^{(2n-1)}_{ij\sigma,2} - K^{(2n-1)}_{ij\sigma,1} + 2J^{(2n-1)}_{ij\sigma}\right)  \nonumber\\
                        & - 2\lambda_{ij\sigma,1}L^{(2n-1)}_{ij\sigma},\label{eq:tau_ij_2}
\end{align}
and, finally, 
\begin{align}
  T_{ij\sigma,3}^{(2n)} = & -2\lambda_{ij\sigma,3} K^{(2n-1)}_{ij\sigma,3} = -4\lambda_{ij\sigma,3}^2T_{ij\sigma,3}^{(2n-2)} \label{eq:tau_ij_3}\nonumber \\
  = & (-1)^nt_{ij\sigma,3} (2\lambda_{ij\sigma,3})^{2n},
\end{align}
with  $\beta_{ij\sigma} = \lambda_{ij\sigma,1}/\lambda_{ij\sigma,2}$ and the initial condition $T_{ij\sigma,x}^{(0)} = t_{ij\sigma,x}$.\\
We uncouple Eqs.~(\ref{eq:tau_ij_1}) and (\ref{eq:tau_ij_2}) by introducing the following geometric series,
\begin{eqnarray}
   W_1^{(2n)} &=& \lambda_{ij\sigma,2}T_{ij\sigma,1}^{(2n)} + \lambda_{ij\sigma,1}T_{ij\sigma,2}^{(2n)}\nonumber \\
  &=& W_1^{(0)}(-1)^n(2\alpha)^{2n},
\end{eqnarray}
and
\begin{eqnarray}
W_2^{(2n)} &=& \lambda_{ij\sigma,1}T_{ij\sigma,1}^{(2n)} - \lambda_{ij\sigma,2}T_{ij\sigma,2}^{(2n)}\nonumber \\
 &=& W_2^{(0)}(-1)^n(4\alpha)^{2n},
\end{eqnarray}
with $\alpha^2 =  \left( \lambda_{ij\sigma,1}^2 +  \lambda_{ij\sigma,2}^2 \right)/2$ and
the initial conditions $W_1^{(0)} = \lambda_{ij\sigma,2}t_{ij\sigma,1} + \lambda_{ij\sigma,1}t_{ij\sigma,2} $ and $ W_2^{(0)} = \lambda_{ij\sigma,1}t_{ij\sigma,1} - \lambda_{ij\sigma,2}t_{ij\sigma,2}$,
thus leading to
\begin{align}
T_{ij\sigma,1}^{(2n)} =  \frac{(-1)^n}{\lambda_{ij\sigma,2}\left(1 +  \beta^2 \right)} \left(W_1^{(0)}(2\alpha)^{2n} + W_2^{(0)}(4\alpha)^{2n}\beta \right), \label{eq:tau_ij_1-2} 
\end{align}
and
\begin{align}
  T_{ij\sigma,2}^{(2n)} = \frac{(-1)^n}{\lambda_{ij\sigma,2}\left(1 +  1/\beta^2 \right)} \left(W_1^{(0)}(2\alpha)^{2n} - W_2^{(0)}(4\alpha)^{2n}/\beta \right).\label{eq:tau_ij_2-2} 
\end{align}
Summing all odd and even contributions at the infinite order allows to  recover Eq.~(\ref{eq:Htrans}) to Eq.~(\ref{eq:Hcouple}) of the manuscript.
The values of the different interaction integrals at infinite order can
be obtained using the recursive relations,
for instance for $T_{ij\sigma,0}$,
\begin{align}
  \label{eq:Jcpl_app}
  T_{ij\sigma,0} & = \sum_{n=0}^{\infty}\frac{2n}{(2n + 1)!}T^{(2n)}_{ij\sigma,0} \nonumber \\
                           & = t_{ij\sigma,0}\sum_{n=0}^{\infty}\frac{2n (-1)^n}{(2n + 1)!}(2\lambda_{ij\sigma,0})^{2n} \nonumber \\
                           & = t_{ij\sigma,0}\sum_{n=0}^{\infty}\frac{ (-1)^n}{(2n)!}(2\lambda_{ij\sigma,0})^{2n} \nonumber \\
  & \quad - \frac{t_{ij\sigma,0}}{\lambda_{ij\sigma,0}}\sum_{n=0}^{\infty}\frac{ (-1)^n}{(2n + 1)!}(2\lambda_{ij\sigma,0})^{2n+1}\nonumber  \\
                           & =t_{ij\sigma,0}\left( \cos(2\lambda_{ij\sigma,0}) - {\rm sinc}(2\lambda_{ij\sigma,0})\right).
\end{align}
The other integrals are similarly obtained and reads, for the electronic integrals $T$ corresponding to the coupling between the Heisenberg subspace and its complementary subspace,
\begin{align}
  T_{ij\sigma,1}  = & \frac{W_1^{(0)}}{\lambda_{ij\sigma,2}(1+\beta_{ij\sigma}^2)}\left(\cos(2\alpha)- {\rm sinc}(2\alpha)\right) \nonumber\\
  &  + \frac{W_2^{(0)}}{\lambda_{ij\sigma,1}(1+1/\beta_{ij\sigma}^2)}\left(\cos(4\alpha) - {\rm sinc}(4\alpha)\right),
\end{align}
\begin{align}
  T_{ij\sigma,2}  = & \frac{W_1^{(0)}}{\lambda_{ij\sigma,1}(1+1/\beta_{ij\sigma}^2)}\left(\cos(2\alpha)- {\rm sinc}(2\alpha)\right) \nonumber\\
  &  - \frac{W_2^{(0)}}{\lambda_{ij\sigma,2}(1+\beta_{ij\sigma}^2)}\left(\cos(4\alpha) - {\rm sinc}(4\alpha)\right),
\end{align}
\begin{equation}
T_{ij\sigma,3}  =  t_{ij\sigma,3}\left( \cos(2\lambda_{ij\sigma,3}) - {\rm sinc}(2\lambda_{ij\sigma,3})\right),
\end{equation}
for the electronic integrals $K$ associated to operators that acts diagonally on each subspace,
\begin{equation}
K_{ij\sigma,0}  =  t_{ij\sigma,0}\left( \sin(2\lambda_{ij\sigma,0}) - \lambda_{ij\sigma,0}\,{\rm sinc}^2(\lambda_{ij\sigma,0})\right),
\end{equation}
\begin{align}
  K_{ij\sigma,1} = & \frac{W_1^{(0)}}{\beta_{ij\sigma} + 1/\beta_{ij\sigma}} \left( 2\,{\rm sinc}(2\alpha) -  {\rm sinc}^2(\alpha)\right) \nonumber\\
                   &\frac{W_2^{(0)}}{1 + 1/\beta_{ij\sigma}^2}\left(2\,{\rm sinc}(4\alpha) -  {\rm sinc}^2(2\alpha)\right),
\end{align}
  \begin{align}
  K_{ij\sigma,2} = & \frac{W_1^{(0)}}{\beta_{ij\sigma} + 1/\beta_{ij\sigma}} \left( 2\,{\rm sinc}(2\alpha) -  {\rm sinc}^2(\alpha)\right) \nonumber\\
  &-\frac{W_2^{(0)}}{1 + \beta_{ij\sigma}^2}\left(2\,{\rm sinc}(4\alpha) -  {\rm sinc}^2(2\alpha)\right),
\end{align}
\begin{equation}
K_{ij\sigma,3}  =  t_{ij\sigma,3}\left( \sin(2\lambda_{ij\sigma,3}) - \lambda_{ij\sigma,3}\,{\rm sinc}^2(\lambda_{ij\sigma,3})\right),
\end{equation}
and finally, for the electronic integrals $J$ and $L$ associated to spin-spin operators and doubly occupied-  empty sites operators,
\begin{equation}
J_{ij\sigma}  =  (1/2)\left(K_{ij\sigma,1}  - K_{ij\sigma,2} \right),
\end{equation}
and
\begin{equation}
L_{ij\sigma}  =  (1/2)\left(K_{ij\sigma,1}/\beta_{ij\sigma}  - \beta_{ij\sigma} K_{ij\sigma,2} \right),
\end{equation}
respectively.
For the homogeneous case,
the associated integrals
are simply given by
\begin{align}
  \label{eq:Jcpl}
  T_{ij\sigma,0} = T_{ij\sigma,3} = K_{ij\sigma,0} = K_{ij\sigma,3} = 0,
  \end{align}
\begin{align}
  T_{ij\sigma,1} = T_{ij\sigma,2} = -t\left(\cos{(4t/U)} - {\rm sinc}\,(4t/U)\right),
\end{align}
and
\begin{align}\label{eq:Jana}
  J_{ij\sigma} = K_{ij\sigma,1} &= - K_{ij\sigma,2} = - L_{ij\sigma} \nonumber\\
  &=  -t\left(\sin{(4t/U)} - (2t/U){\rm sinc}^2\,(2t/U)\right)/2.
   \end{align} 
\section{Recursive relations for the variational Schrieffer--Wolff transformation}
\label{app:Var}

The variational SW transformation given in Eq.~(\ref{eq:Var_SWT_U}) leads to the following transformed Hamiltonian,
\begin{align}
  \bar{H}(\theta) &= e^{\theta \hat{S}}\hat{H} e^{-\theta \hat{S}} \nonumber \\
 & = \hat{H}^0 +\hat{V} \left( 1 - \theta \right) + \sum_{n = 2}^{\infty}\frac{\theta^{n-1}(n - \theta)}{n!}\mathcal{S}^{n-1}(\hat{V}).
\end{align}
Interestingly, Eqs.~(\ref{eq:S2n}) and (\ref{eq:S2n+1}) still hold in this case, such that a strategy similar to the one introduced in Appendix~\ref{app:Recurence} can be used to obtain interaction integrals at each order, recursively. 
The summation to the infinite order is then possible, thus leading to Eq.~(\ref{eq:Htheta}) with the electronic integrals $T(\theta)$ corresponding to the coupling between the Heisenberg subspace and its complementary subspace,
\begin{equation}
  T_{ij\sigma,0}(\theta) = t_{ij\sigma,0}\left( \cos(2\theta \lambda_{ij\sigma,0}) - \theta{\rm sinc}(2\theta\lambda_{ij\sigma,0})\right),
\end{equation}
\begin{align}
  T_{ij\sigma,1}(\theta)  = & \frac{W_1^{(0)}}{\lambda_{ij\sigma,2}(1+\beta_{ij\sigma}^2)}\left(\cos(2\theta\alpha)- \theta{\rm sinc}(2\theta\alpha)\right) \nonumber\\
  &  + \frac{W_2^{(0)}}{\lambda_{ij\sigma,1}(1+1/\beta_{ij\sigma}^2)}\left(\cos(4\theta\alpha) - \theta{\rm sinc}(4\theta\alpha)\right),
\end{align}
\begin{align}
  T_{ij\sigma,2}(\theta)  = & \frac{W_1^{(0)}}{\lambda_{ij\sigma,1}(1+1/\beta_{ij\sigma}^2)}\left(\cos(2\theta\alpha)- \theta{\rm sinc}(2\theta\alpha)\right) \nonumber\\
  &  - \frac{W_2^{(0)}}{\lambda_{ij\sigma,2}(1+\beta_{ij\sigma}^2)}\left(\cos(4\theta\alpha) - \theta{\rm sinc}(4\theta\alpha)\right),
\end{align}
\begin{equation}
T_{ij\sigma,3}(\theta)  =  t_{ij\sigma,3}\left( \cos(2\theta\lambda_{ij\sigma,3}) -\theta {\rm sinc}(2\theta\lambda_{ij\sigma,3})\right),
\end{equation}
the electronic integrals $K(\theta)$ associated to operators that acts diagonally on each subspace,
\begin{equation}
K_{ij\sigma,0}(\theta)   =  t_{ij\sigma,0}\left( \sin(2\theta\lambda_{ij\sigma,0}) - \lambda_{ij\sigma,0}\theta^2\,{\rm sinc}^2(\theta\lambda_{ij\sigma,0})\right),
\end{equation}
  \begin{align}
  K_{ij\sigma,1}(\theta) = & \frac{W_1^{(0)}\theta}{\beta_{ij\sigma} + 1/\beta_{ij\sigma}} \left( 2\,{\rm sinc}(2\theta \alpha) -  \theta{\rm sinc}^2(\theta \alpha)\right) \nonumber\\
  &\frac{W_2^{(0)}\theta}{1 + 1/\beta_{ij\sigma}^2}\left(2\,{\rm sinc}(4\theta \alpha) -  \theta{\rm sinc}^2(2\theta \alpha)\right)
\end{align}
  \begin{align}
  K_{ij\sigma,2}(\theta) = & \frac{W_1^{(0)}\theta}{\beta_{ij\sigma} + 1/\beta_{ij\sigma}} \left( 2\,{\rm sinc}(2\theta \alpha) -  \theta{\rm sinc}^2(\theta \alpha)\right) \nonumber\\
  &-\frac{W_2^{(0)}\theta}{1 + \beta_{ij\sigma}^2}\left(2\,{\rm sinc}(4\theta \alpha) -  \theta{\rm sinc}^2(2\theta \alpha)\right)
\end{align}
\begin{equation}
K_{ij\sigma,3}(\theta)  =  t_{ij\sigma,3}\left( \sin(2\theta\lambda_{ij\sigma,3}) - \lambda_{ij\sigma,3}\theta^2\,{\rm sinc}^2(\theta\lambda_{ij\sigma,3})\right),
\end{equation}
and finally, the electronic integrals $J(\theta)$ and $L(\theta)$ associated to spin-spin operators and doubly occupied-empty sites operators,
\begin{equation}
J_{ij\sigma}(\theta)  =  (1/2)\left(K_{ij\sigma,1}(\theta)  - K_{ij\sigma,2}(\theta) \right),
\end{equation}
and
\begin{equation}
L_{ij\sigma}(\theta)  =  (1/2)\left(K_{ij\sigma,1}(\theta)/\beta_{ij\sigma}  - \beta_{ij\sigma} K_{ij\sigma,2}(\theta) \right),
\end{equation}
respectively.

    %
    %
    %
% Proof VV %
In the following we show that for the homogenous case and at the saddle point, the variational SW transformation with generator $\hat{G} = \theta\hat{S}$ fulfills the VV condition in Eq.~(\ref{eq:VV_MBPT}).
  More precisely, in this case, Eq.~(\ref{eq:VV_MBPT}) reads
  \begin{align}
    [\theta \hat{S}, \hat{H}^0] = -\hat{V} - \sum_{n=1}^{\infty}c_n \theta^{2n}\mathcal{S}^{2n}(\hat{V}), \nonumber \\
    \theta \hat{V} = \hat{V} + \sum_{n=1}^{\infty}c_n \theta^{2n}\mathcal{S}^{2n}(\hat{V}). \label{eq:VV1}
  \end{align}
  Inserting Eq.~(\ref{eq:S2n}) on the right handside of Eq.~(\ref{eq:VV1}) with $T_{ij\sigma,0}^{(2n)} = T_{ij\sigma,3}^{(2n)} = 0$ and $T_{ij\sigma,1}^{(2n)} = -T_{ij\sigma,2}^{(2n)} = (-1)^n 4t/U$ for the homogeneous case
  [see Eqs.~(\ref{eq:tau_ij_0}), (\ref{eq:tau_ij_3}), (\ref{eq:tau_ij_1-2}) and (\ref{eq:tau_ij_2-2})] leads to
  \begin{align}
    \theta \hat{V} =  \left(1 +\sum_{n=1}^{\infty}c_n (-1)^n \left(\dfrac{4 t \theta}{U}\right)^{2n}\right) \hat{V} 
     %\theta =  \left(1 +\sum_{n=0}^{\infty}c_n (-1)^n (4 t \theta/U)^{2n}\right) 
  \end{align}
Using $c_n = B_{2n}2^{2n}/(2n)!$ $ = (-1)^{n-1}|B_{2n}|2^{2n}/(2n)!$,
we obtain
  \begin{align}
    \theta =  1 - \sum_{n=1}^{\infty} \dfrac{|B_{2n}|2^{2n}}{(2n)!} \left(\dfrac{4 t \theta}{U}\right)^{2n} \\
     \theta = \left(\dfrac{4 t \theta}{U}\right) {\rm cotan}\left(\dfrac{4 t \theta}{U}\right)
     %\theta =  \left(1 +\sum_{n=0}^{\infty}c_n (-1)^n (4 t \theta/U)^{2n}\right) 
  \end{align}
  or, equivalently, $\theta = (U/4t)\tan^{-1}{(4t/U)}$,
  which is the analytical expression of the minimizing variational parameter in Eq.~(\ref{eq:theta_analytic}).

\section{Iterative generator}
\label{app:Itgen}

In this section, we detail the construction of the iterative transformation, in the spirit of the Foldy--Wouthuysen transformation~\cite{foldy_on_1950}:
\begin{equation}
  \label{eq:Ite_SWT_U_B1}
  U^{(N_s)} = \left(\prod_{s = 0}^{N_s-1}e^{\hat{S}^{(s)}}\right).
\end{equation}
At each iteration, the generator takes
the form
\begin{equation}
  \hat{S}^{(s)}  =\frac{1}{2}\sum_{i\neq j, \sigma} \sum_x \lambda_{ij\sigma, x}^{(s)} \hat{p}_{ij\bar{\sigma},x}\left(\hat{\gamma}_{ij\sigma} - \hat{\gamma}_{ji\sigma}\right),
\end{equation}
and
the iteration $s=0$ refers to the standard SW transformation at infinite order, for which  $\lambda_{ij\sigma, x}^{(0)}$ parameters are set to satisfy Eq.~(\ref{eq:1storder}), i.e., 
\begin{align}
  [\hat{H}^0, \hat{S}^{(0)}]    = \hat{V}_X. \label{eq:S_H0_B1}
\end{align}
This equation can be rewritten as follows,
\begin{align}
 \sum_{i\neq j, \sigma} \sum_{x=0}^3\left( f_{ij\sigma,x}^{(0)}(\lambda_{ij\sigma,x}^{(0)}) - t_{ij\sigma,x} \right )\hat{p}_{ij \bar{\sigma},x}\left(\hat{\gamma}_{ij\sigma} + \hat{\gamma}_{ji\sigma}\right)  = \hat{0}, \label{eq:S_H0_B1_2}
\end{align}
with 
\begin{eqnarray}
f_{ij\sigma,0}^{(0)} &=& \lambda_{ij\sigma,0}^{(0)}\Delta \mu_{ij}/2,\nonumber \\
f_{ij\sigma,1}^{(0)} &=& \lambda_{ij\sigma,1}^{(0)}(\Delta \mu_{ij} + U_i)/2, \nonumber \\
f_{ij\sigma,2}^{(0)} &=& \lambda_{ij\sigma,2}^{(0)}(\Delta \mu_{ij} - U_j)/2, \nonumber \\
f_{ij\sigma,3}^{(0)} &= &\lambda_{ij\sigma,x}^{(0)}(\Delta \mu_{ij} + \Delta U_{ij})/2
\end{eqnarray}
thus
leading to Eqs.~(\ref{eq:lamb0})--(\ref{eq:lamb3}).\\

Through the recursive relations, we established that $\bar{H}^{0} = e^{\hat{S}^{(0)}}\hat{H}e^{-\hat{S}^{(0)}}$ is given by Eq.~(\ref{eq:Htrans}) where $\bar{H}^{\rm cpl}$ is off-block-diagonal. 
The iterative process consists in repeating SW-type unitary transformation, 
providing that $\hat{H}^0 +  \bar{H}^{{\rm dia}(s)} + \bar{H}^{{\rm ex}(s)} + \bar{H}^{{\rm de}(s)}  \rightarrow \hat{H}^{0(s+1)}$ and $\bar{H}^{{\rm cpl}(s)} \rightarrow \hat{V}^{(s+1)}$. 
Eq.~(\ref{eq:S_H0_B1}) then defines the $(s+1)$ SW generator,
\begin{align}
  &[\hat{H}^{0(s+1)}, \hat{S}^{(s+1)}]   = \hat{V}^{(s+1)} , \nonumber \\
  &[\hat{H}^0 + \bar{H}^{{\rm dia}(s)} + \bar{H}^{{\rm ex}(s)} + \bar{H}^{{\rm de}(s)}, \hat{S}^{(s+1)}]   
  = \bar{H}^{{\rm cpl}(s)} ,
\end{align}
that, similarly to Eq.~(\ref{eq:S_H0_B1_2}), reduces a set of linearly coupled equations,
\begin{align}
f_{ij\sigma,x}^{(s+1)} - T^{(s)}_{ij\sigma,x}  = 0,
\end{align}
or, with more details,
\begin{align}
  f_{ij\sigma,0}^{(s+1)} = \lambda_{ij\sigma,0}^{(s+1)}\left( \Delta \mu_{ij} + 2K^{(s)}_{ij\sigma,0} \right),\nonumber
\end{align}
\begin{eqnarray}
   & f_{ij\sigma,1}^{(s+1)}& =  2 \lambda_{ij\sigma,2}^{(s+1)}L^{(s)}_{ij} +\lambda_{ij\sigma,1}^{(s+1)}\times \nonumber \\
   & &  \left( \Delta \mu_{ij} + U_i + 3K^{(s)}_{ij\sigma,1}  - K^{(s)}_{ij\sigma,2} +   2J^{(s)}_{ij} \right),\nonumber
\end{eqnarray}
\begin{eqnarray}
  &  f_{ij\sigma,2}^{(s+1)} &=  -2 \lambda_{ij\sigma,1}^{(s+1)}L^{(s)}_{ij\sigma} +\lambda_{ij\sigma,2}^{(s+1)} \times \nonumber\\
  & &  \left( \Delta \mu_{ij} - U_j + 3K^{(s)}_{ij\sigma,2}  - K^{(s)}_{ij\sigma,1} -   2J^{(s)}_{ij} \right),\nonumber
\end{eqnarray}
and, finally,
\begin{align}
  f_{ij\sigma,3}^{(s+1)} = \lambda_{ij\sigma,3}^{(s+1)}\left(  \Delta \mu_{ij} +  \Delta U_{ij}   + 2K^{(s)}_{ij\sigma,3}  \right).\nonumber
\end{align}
It follows straightforwardly that
\begin{align}
   \lambda_{ij\sigma,0}^{(s+1)} = T^{(s)}_{ij\sigma,0}/(\Delta \mu_{ij} + 2K^{(s)}_{ij\sigma,0}),
\end{align}
\begin{align}
   \lambda_{ij\sigma,1}^{(s+1)} = \frac{2T^{(s)}_{ij\sigma,2}L^{(s)}_{ij\sigma} + T^{(s)}_{ij\sigma,1} B_{2, ij\sigma}^{(s)}}{4L^{(s)}_{ij\sigma}L^{(s)}_{ij\sigma} + B_{1, ij\sigma}^{(s)}B_{2, ij\sigma}^{(s)}} ,
  \end{align}
\begin{align}
    \lambda_{ij\sigma,2}^{(s+1)}=  \frac{-2T^{(s)}_{ij\sigma,1}L^{(s)}_{ij\sigma} + T^{(s)}_{ij\sigma,2} B_{1, ij\sigma}^{(s)}}{4L^{(s)}_{ij\sigma}L^{(s)}_{ij\sigma} + B_{1, ij\sigma}^{(s)}B_{2, ij\sigma}^{(s)}},
  \end{align}
  and
\begin{align}
     \lambda_{ij\sigma,3}^{(s+1)}= T^{(s)}_{ij\sigma,3}/  (\Delta \mu_{ij} +  \Delta U_{ij}  + 2K^{(s)}_{ij\sigma,3})  ,
\end{align}
with $B_{1,ij\sigma}^{(s)} = ( \Delta \mu_{ij}+ U_i) + 3K^{(s)}_{ij\sigma,1}  - K^{(s)}_{ij\sigma,2} +  2J^{(s)}_{ij\sigma}$ and $B_{2, ij\sigma}^{(s)} =( \Delta \mu_{ij}- U_j) + 3K^{(s)}_{ij\sigma,2}  - K^{(s)}_{ij\sigma,1} -   2J^{(s)}_{ij\sigma}$ .

\section{Cost functions}
\label{app:min}

\begin{figure}[!h]
\resizebox{\columnwidth}{!}{
\includegraphics[scale=1]{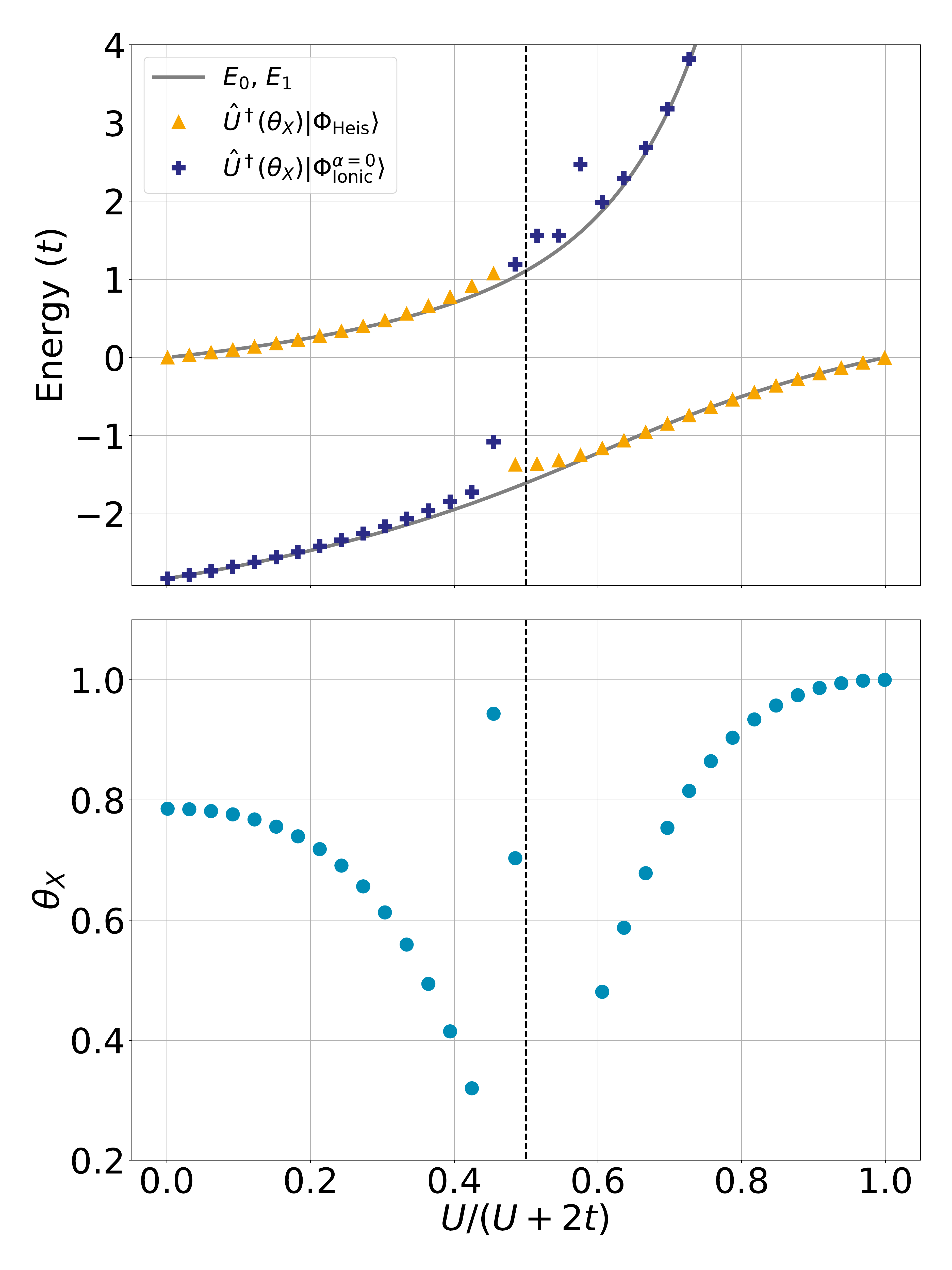}
}
\caption{Energies of the half-filled Hubbard dimer for $\Delta \mu/t = 2$ (top panel) with respect to the repulsion strength, using the variational MSW transformation method
to minimize the matrix elements of $|\bar{H}_{X}(\theta)|$. 
The minimizing parameter $\theta_{X}$ is shown in the bottom panel.
The vertical dotted line corresponds to $U = \Delta \mu$.}
\label{fig:min_cpl}
\end{figure}

\begin{figure}[!h]
\resizebox{\columnwidth}{!}{
\includegraphics[scale=1]{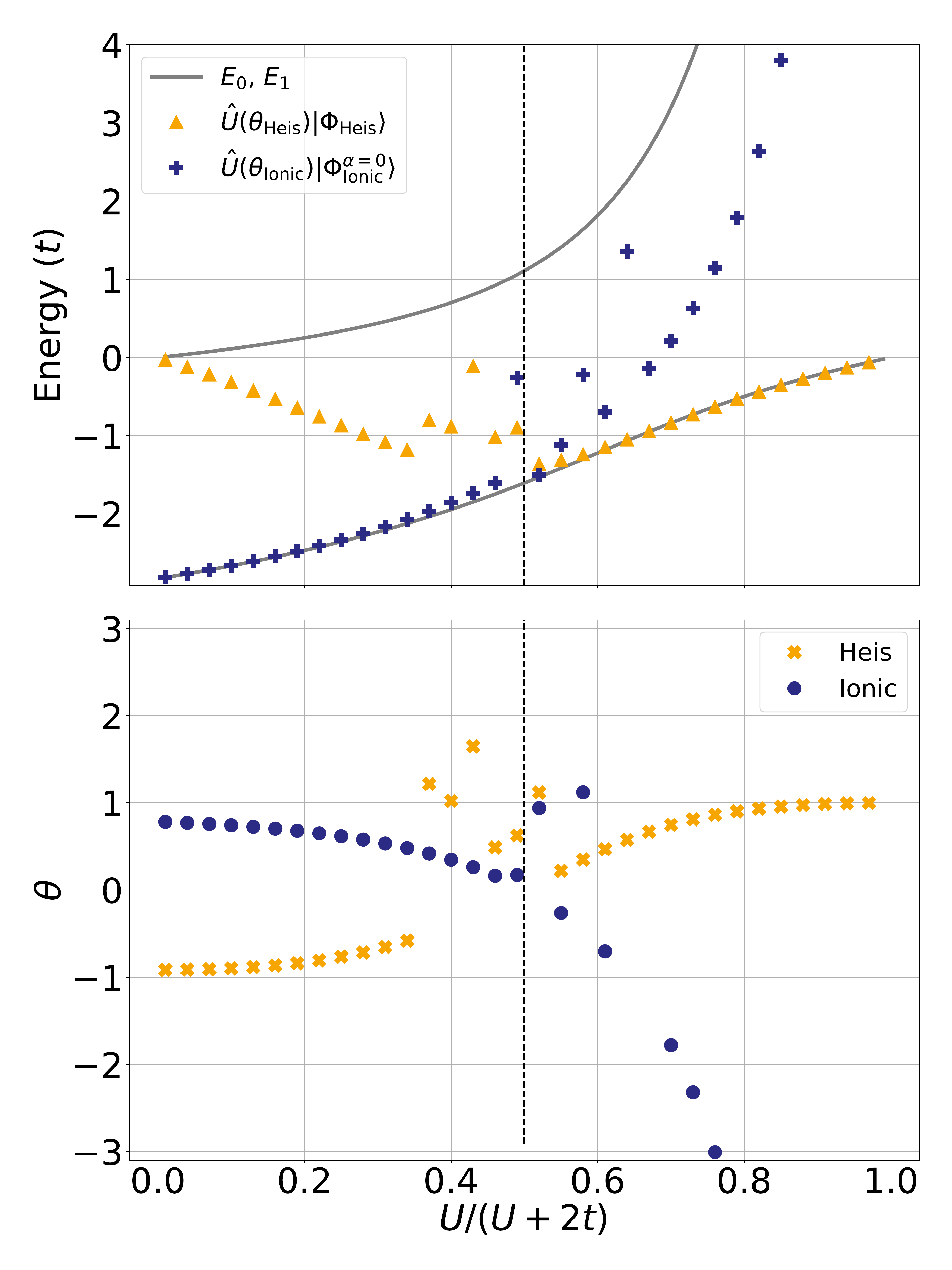}
}
\caption{Top panel: Energies of the half-filled Hubbard dimer for $\Delta \mu/t = 2$ with respect to the repulsion strength, using the variational SW method
to minimize the energy $\bra{\Phi_{\rm Heis}}\bar{H}(\theta)\ket{\Phi_{\rm Heis}}$ (orange triangles) or $\langle  \Phi_{\rm Ionic}^{\alpha = 0}\vert\bar{H}(\theta)\vert \Phi_{\rm Ionic}^{\alpha = 0}\rangle$ (blue crosses). 
Bottom panel: Minimizing parameter $\theta_{\rm Heis}$ (yellow crosses) and $\theta_{\rm Ionic}$ (blue dots).
The vertical dotted line corresponds to $U = \Delta \mu$.}
\label{fig:min_energy}
\end{figure}

%The variational approach described in Sec.~\ref{sec:variational} consists in finding the optimal parameter $\theta_{X}$ which minimizes the matrix elements of $\bar{H}^{\rm cpl}(\theta_{\rm cpl})$, thus enforcing the block-diagonalization of $\bar{H}(\theta_{\rm cpl})$.
%However, in practice it requires the direct measure of $\bra{\Phi_i} \bar{H} \ket{\Phi_j}$ on the quantum computer, which is generally non trivial compared to the evaluation of the energy in VQE, for instance, although one can note some recent improvements made in Refs.~\onlinecite{huggins2020non,stair2021simulating}.
In the variational approach described in Sec.~\ref{sec:variational},
one can optimize the scaling parameter $\theta$ to
minimize the contributions from the coupling operator $\bar{H}_X(\theta)$ ,
or to minimize the energy of $\bar{H}(\theta)$ restricted to the Heisenberg subspace.
In this section, we investigate the difference of the
two strategies.

The first one requires to minimize the Frobenius norm $||\bar{H}_X(\theta)||_F$, for which the saddle point $\theta_X$ gives the transformation that maximally decouples the Heisenberg subspace $\Omega$ 
from the complementary subspace, 
without any warranty that the low-energy ground state belongs to $\Omega$.
On a quantum computer, it requires the estimation of 
off-diagonal elements $\bra{\Phi_i} \bar{H}(\theta) \ket{\Phi_j}$, $i$ belonging to $\Omega$ and $j$ to the complementary subspace, which is generally non trivial, although one can note some recent improvements made in Refs.~[\onlinecite{huggins2020non}] and~[\onlinecite{stair2021simulating}].
Zhang and coworkers~\cite{zhang_quantum_2022} used a similar cost function, evaluated using only the estimation of off-diagonal terms over states belonging to $\Omega$.
In Fig.~\ref{fig:min_cpl},
we plot the minimizing $\theta_X$ and the energies associated to the rotated states $U^\dagger(\theta_X)\ket{\Phi_{\rm Heis}}$ and $U^\dagger(\theta_X)\ket{\Phi_{\rm Ionic}}$.

The other method consists in minimizing the energy $\bra{\Phi_{\rm Heis}} U(\theta) H U^\dagger(\theta)\ket{\Phi_{\rm Heis}}$ ($\bra{\Phi_{\rm Ionic}} U(\theta) H U^\dagger(\theta)\ket{\Phi_{\rm Ionic}}$) with respect to $\theta$ and is analog to the VQE algorithm. 
The saddle point obtained is denoted as $\theta_{\rm Heis}$ ($\theta_{\rm Ionic}$) and gives the transformation that maximally overlaps the rotated Heisenberg (Ionic) state with the exact ground state.
Results are shown in Fig.~\ref{fig:min_energy}

Comparing Fig.~\ref{fig:min_cpl} to Fig.~\ref{fig:min_energy}, one can
directly see that minimizing the energy with respect to the Heisenberg state
does work for $U \gg \Delta \mu$ only, where $\theta_{\rm Heis} \simeq \theta_X$, while minimizing the energy
with respect to the ionic state works only for $U \ll \Delta \mu$, where $\theta_{\rm ionic} \simeq \theta_X$.
Note that in virtue of the variational principle, the energies of the states $U^\dagger(\theta_{\rm Heis})\ket{\Phi_{\rm Heis}}$ and $U^\dagger(\theta_{\rm Ionic})\ket{\Phi_{\rm Ionic}}$ are always below $U^\dagger(\theta_{X})\ket{\Phi_{\rm Heis}}$  and $U^\dagger(\theta_{X})\ket{\Phi_{\rm Ionic}}$, respectively, for the entire range of correlation regime.
Therefore, minimizing $||\bar{H}_X(\theta)||_F$ gives the rotation that maximally satisfies the VV conditions [Eq.~(\ref{eq:VV_MBPT})] but, in contrast to the VQE-like algorithm, it does not ensure that the Heisenberg space 
$\Omega$ maximally overlaps with the low-effective sub-space.
%In contrast to Fig.~\ref{fig:var_deltav2} where the energy was minimized to obtain both $\theta_{\rm Heis}$ and $\theta_{\rm Ionic}$, a single minimization is performed to get $\theta_{\rm cpl}$ which block-diagonalizes $\bar{H}^{\rm cpl}(\theta_{\rm cpl})$ in the entire range of interaction strength.
%In the range of interaction where the obtained $\bar{H}(\theta)$ is not block-diagonal, we see that the energy of the Heisenberg (ionic) state in Fig.~\ref{fig:min_energy} are below the one of Fig.~\ref{fig:min_cpl}, as the ground state does not belong to the Heisenberg (ionic) subspace and because the energy is minimized rather than the coupling terms, such that the state is not eigenstate of $\bar{H}(\theta_{\rm Heis})$ ($\bar{H}(\theta_{\rm Ionic})$) anymore.

\newcommand{\Aa}[0]{Aa}


\begin{thebibliography}{52}%
\makeatletter
\providecommand \@ifxundefined [1]{%
 \@ifx{#1\undefined}
}%
\providecommand \@ifnum [1]{%
 \ifnum #1\expandafter \@firstoftwo
 \else \expandafter \@secondoftwo
 \fi
}%
\providecommand \@ifx [1]{%
 \ifx #1\expandafter \@firstoftwo
 \else \expandafter \@secondoftwo
 \fi
}%
\providecommand \natexlab [1]{#1}%
\providecommand \enquote  [1]{``#1''}%
\providecommand \bibnamefont  [1]{#1}%
\providecommand \bibfnamefont [1]{#1}%
\providecommand \citenamefont [1]{#1}%
\providecommand \href@noop [0]{\@secondoftwo}%
\providecommand \href [0]{\begingroup \@sanitize@url \@href}%
\providecommand \@href[1]{\@@startlink{#1}\@@href}%
\providecommand \@@href[1]{\endgroup#1\@@endlink}%
\providecommand \@sanitize@url [0]{\catcode `\\12\catcode `\$12\catcode
  `\&12\catcode `\#12\catcode `\^12\catcode `\_12\catcode `\%12\relax}%
\providecommand \@@startlink[1]{}%
\providecommand \@@endlink[0]{}%
\providecommand \url  [0]{\begingroup\@sanitize@url \@url }%
\providecommand \@url [1]{\endgroup\@href {#1}{\urlprefix }}%
\providecommand \urlprefix  [0]{URL }%
\providecommand \Eprint [0]{\href }%
\providecommand \doibase [0]{http://dx.doi.org/}%
\providecommand \selectlanguage [0]{\@gobble}%
\providecommand \bibinfo  [0]{\@secondoftwo}%
\providecommand \bibfield  [0]{\@secondoftwo}%
\providecommand \translation [1]{[#1]}%
\providecommand \BibitemOpen [0]{}%
\providecommand \bibitemStop [0]{}%
\providecommand \bibitemNoStop [0]{.\EOS\space}%
\providecommand \EOS [0]{\spacefactor3000\relax}%
\providecommand \BibitemShut  [1]{\csname bibitem#1\endcsname}%
\let\auto@bib@innerbib\@empty
%</preamble>
\bibitem [{\citenamefont {Hubbard}(1963)}]{hubbard_electron_1963}%
  \BibitemOpen
  \bibfield  {author} {\bibinfo {author} {\bibfnamefont {J.}~\bibnamefont
  {Hubbard}},\ }\href {https://doi.org/10.1098/rspa.1963.0204} {\bibfield
  {journal} {\bibinfo  {journal} {Proc. Math. Phys. Eng. Sci. P ROY SOC A-MATH
  PHY}\ }\textbf {\bibinfo {volume} {276}},\ \bibinfo {pages} {238} (\bibinfo
  {year} {1963})}\BibitemShut {NoStop}%
\bibitem [{\citenamefont {Nagaoka}(1965)}]{nagaoka_ground_1965}%
  \BibitemOpen
  \bibfield  {author} {\bibinfo {author} {\bibfnamefont {Y.}~\bibnamefont
  {Nagaoka}},\ }\href {https://doi.org/10.1016/0038-1098(65)90266-8} {\bibfield
   {journal} {\bibinfo  {journal} {Solid State Commun.}\ }\textbf {\bibinfo
  {volume} {3}},\ \bibinfo {pages} {409} (\bibinfo {year} {1965})}\BibitemShut
  {NoStop}%
\bibitem [{\citenamefont {Lieb}\ and\ \citenamefont
  {Wu}(1968)}]{lieb_absence_1968}%
  \BibitemOpen
  \bibfield  {author} {\bibinfo {author} {\bibfnamefont {E.~H.}\ \bibnamefont
  {Lieb}}\ and\ \bibinfo {author} {\bibfnamefont {F.~Y.}\ \bibnamefont {Wu}},\
  }\href {https://doi.org/10.1103/PhysRevLett.20.1445} {\bibfield  {journal}
  {\bibinfo  {journal} {Phys. Rev. Lett.}\ }\textbf {\bibinfo {volume} {20}},\
  \bibinfo {pages} {1445} (\bibinfo {year} {1968})}\BibitemShut {NoStop}%
\bibitem [{\citenamefont {López-Sandoval}\ and\ \citenamefont
  {Pastor}(2002)}]{lopez-sandoval_density-matrix_2002}%
  \BibitemOpen
  \bibfield  {author} {\bibinfo {author} {\bibfnamefont {R.}~\bibnamefont
  {López-Sandoval}}\ and\ \bibinfo {author} {\bibfnamefont {G.~M.}\
  \bibnamefont {Pastor}},\ }\href {https://doi.org/10.1103/PhysRevB.66.155118}
  {\bibfield  {journal} {\bibinfo  {journal} {Phys. Rev. B}\ }\textbf {\bibinfo
  {volume} {66}} (\bibinfo {year} {2002})}\BibitemShut {NoStop}%
\bibitem [{\citenamefont {Lima}\ \emph {et~al.}(2003)\citenamefont {Lima},
  \citenamefont {Silva}, \citenamefont {Oliveira},\ and\ \citenamefont
  {Capelle}}]{lima_density_2003}%
  \BibitemOpen
  \bibfield  {author} {\bibinfo {author} {\bibfnamefont {N.~A.}\ \bibnamefont
  {Lima}}, \bibinfo {author} {\bibfnamefont {M.~F.}\ \bibnamefont {Silva}},
  \bibinfo {author} {\bibfnamefont {L.~N.}\ \bibnamefont {Oliveira}}, \ and\
  \bibinfo {author} {\bibfnamefont {K.}~\bibnamefont {Capelle}},\ }\href
  {https://doi.org/10.1103/PhysRevLett.90.146402} {\bibfield  {journal}
  {\bibinfo  {journal} {Phys. Rev. Lett.}\ }\textbf {\bibinfo {volume} {90}}
  (\bibinfo {year} {2003})}\BibitemShut {NoStop}%
\bibitem [{\citenamefont {Georges}\ \emph {et~al.}(1996)\citenamefont
  {Georges}, \citenamefont {Kotliar}, \citenamefont {Krauth},\ and\
  \citenamefont {Rozenberg}}]{georges_dynamical_1996}%
  \BibitemOpen
  \bibfield  {author} {\bibinfo {author} {\bibfnamefont {A.}~\bibnamefont
  {Georges}}, \bibinfo {author} {\bibfnamefont {G.}~\bibnamefont {Kotliar}},
  \bibinfo {author} {\bibfnamefont {W.}~\bibnamefont {Krauth}}, \ and\ \bibinfo
  {author} {\bibfnamefont {M.~J.}\ \bibnamefont {Rozenberg}},\ }\href
  {https://doi.org/10.1103/RevModPhys.68.13} {\bibfield  {journal} {\bibinfo
  {journal} {Rev. Mod. Phys.}\ }\textbf {\bibinfo {volume} {68}},\ \bibinfo
  {pages} {13} (\bibinfo {year} {1996})}\BibitemShut {NoStop}%
\bibitem [{\citenamefont {Sénéchal}\ \emph {et~al.}(2000)\citenamefont
  {Sénéchal}, \citenamefont {Perez},\ and\ \citenamefont
  {Pioro-Ladriere}}]{senechal_spectral_2000}%
  \BibitemOpen
  \bibfield  {author} {\bibinfo {author} {\bibfnamefont {D.}~\bibnamefont
  {Sénéchal}}, \bibinfo {author} {\bibfnamefont {D.}~\bibnamefont {Perez}}, \
  and\ \bibinfo {author} {\bibfnamefont {M.}~\bibnamefont {Pioro-Ladriere}},\
  }\href {https://doi.org/10.1103/PhysRevLett.84.522} {\bibfield  {journal}
  {\bibinfo  {journal} {Phys. Rev. Lett.}\ }\textbf {\bibinfo {volume} {84}},\
  \bibinfo {pages} {522} (\bibinfo {year} {2000})}\BibitemShut {NoStop}%
\bibitem [{\citenamefont
  {Potthoff}(2003)}]{potthoff_self-energy-functional_2003}%
  \BibitemOpen
  \bibfield  {author} {\bibinfo {author} {\bibfnamefont {M.}~\bibnamefont
  {Potthoff}},\ }\href {https://doi.org/10.1140/epjb/e2003-00121-8} {\bibfield
  {journal} {\bibinfo  {journal} {Eur. Phys. J. B}\ }\textbf {\bibinfo {volume}
  {32}},\ \bibinfo {pages} {429} (\bibinfo {year} {2003})}\BibitemShut
  {NoStop}%
\bibitem [{\citenamefont {White}(1992)}]{white_density_1992}%
  \BibitemOpen
  \bibfield  {author} {\bibinfo {author} {\bibfnamefont {S.~R.}\ \bibnamefont
  {White}},\ }\href {https://doi.org/10.1103/PhysRevLett.69.2863} {\bibfield
  {journal} {\bibinfo  {journal} {Phys. Rev. Lett.}\ }\textbf {\bibinfo
  {volume} {69}},\ \bibinfo {pages} {2863} (\bibinfo {year}
  {1992})}\BibitemShut {NoStop}%
\bibitem [{\citenamefont {Knizia}\ and\ \citenamefont
  {Chan}(2012)}]{knizia_density_2012}%
  \BibitemOpen
  \bibfield  {author} {\bibinfo {author} {\bibfnamefont {G.}~\bibnamefont
  {Knizia}}\ and\ \bibinfo {author} {\bibfnamefont {G.~K.-L.}\ \bibnamefont
  {Chan}},\ }\href {https://doi.org/10.1103/PhysRevLett.109.186404} {\bibfield
  {journal} {\bibinfo  {journal} {Phys. Rev. Lett.}\ }\textbf {\bibinfo
  {volume} {109}} (\bibinfo {year} {2012})}\BibitemShut {NoStop}%
\bibitem [{\citenamefont {Sekaran}\ \emph {et~al.}(2021)\citenamefont
  {Sekaran}, \citenamefont {Tsuchiizu}, \citenamefont {Saubanère},\ and\
  \citenamefont {Fromager}}]{sekaran_householder_2021}%
  \BibitemOpen
  \bibfield  {author} {\bibinfo {author} {\bibfnamefont {S.}~\bibnamefont
  {Sekaran}}, \bibinfo {author} {\bibfnamefont {M.}~\bibnamefont {Tsuchiizu}},
  \bibinfo {author} {\bibfnamefont {M.}~\bibnamefont {Saubanère}}, \ and\
  \bibinfo {author} {\bibfnamefont {E.}~\bibnamefont {Fromager}},\ }\href
  {https://doi.org/10.1103/PhysRevB.104.035121} {\enquote {\bibinfo {title}
  {Householder transformed density matrix functional embedding theory},}\ }
  (\bibinfo {year} {2021})\BibitemShut {NoStop}%
\bibitem [{\citenamefont {Wecker}\ \emph
  {et~al.}(2015{\natexlab{a}})\citenamefont {Wecker}, \citenamefont {Hastings},
  \citenamefont {Wiebe}, \citenamefont {Clark}, \citenamefont {Nayak},\ and\
  \citenamefont {Troyer}}]{wecker_solving_2015}%
  \BibitemOpen
  \bibfield  {author} {\bibinfo {author} {\bibfnamefont {D.}~\bibnamefont
  {Wecker}}, \bibinfo {author} {\bibfnamefont {M.~B.}\ \bibnamefont
  {Hastings}}, \bibinfo {author} {\bibfnamefont {N.}~\bibnamefont {Wiebe}},
  \bibinfo {author} {\bibfnamefont {B.~K.}\ \bibnamefont {Clark}}, \bibinfo
  {author} {\bibfnamefont {C.}~\bibnamefont {Nayak}}, \ and\ \bibinfo {author}
  {\bibfnamefont {M.}~\bibnamefont {Troyer}},\ }\href
  {https://doi.org/10.1103/PhysRevA.92.062318} {\bibfield  {journal} {\bibinfo
  {journal} {Phys. Rev. A}\ }\textbf {\bibinfo {volume} {92}},\ \bibinfo
  {pages} {062318} (\bibinfo {year} {2015}{\natexlab{a}})}\BibitemShut
  {NoStop}%
\bibitem [{\citenamefont {Wecker}\ \emph
  {et~al.}(2015{\natexlab{b}})\citenamefont {Wecker}, \citenamefont
  {Hastings},\ and\ \citenamefont {Troyer}}]{wecker_towards_2015}%
  \BibitemOpen
  \bibfield  {author} {\bibinfo {author} {\bibfnamefont {D.}~\bibnamefont
  {Wecker}}, \bibinfo {author} {\bibfnamefont {M.~B.}\ \bibnamefont
  {Hastings}}, \ and\ \bibinfo {author} {\bibfnamefont {M.}~\bibnamefont
  {Troyer}},\ }\href {https://doi.org/10.1103/PhysRevA.92.042303} {\bibfield
  {journal} {\bibinfo  {journal} {Phys. Rev. A}\ }\textbf {\bibinfo {volume}
  {92}},\ \bibinfo {pages} {042303} (\bibinfo {year}
  {2015}{\natexlab{b}})}\BibitemShut {NoStop}%
\bibitem [{\citenamefont {Kivlichan}\ \emph {et~al.}(2018)\citenamefont
  {Kivlichan}, \citenamefont {McClean}, \citenamefont {Wiebe}, \citenamefont
  {Gidney}, \citenamefont {Aspuru-Guzik}, \citenamefont {Chan},\ and\
  \citenamefont {Babbush}}]{kivlichan_quantum_2018}%
  \BibitemOpen
  \bibfield  {author} {\bibinfo {author} {\bibfnamefont {I.~D.}\ \bibnamefont
  {Kivlichan}}, \bibinfo {author} {\bibfnamefont {J.}~\bibnamefont {McClean}},
  \bibinfo {author} {\bibfnamefont {N.}~\bibnamefont {Wiebe}}, \bibinfo
  {author} {\bibfnamefont {C.}~\bibnamefont {Gidney}}, \bibinfo {author}
  {\bibfnamefont {A.}~\bibnamefont {Aspuru-Guzik}}, \bibinfo {author}
  {\bibfnamefont {G.~K.-L.}\ \bibnamefont {Chan}}, \ and\ \bibinfo {author}
  {\bibfnamefont {R.}~\bibnamefont {Babbush}},\ }\href
  {https://doi.org/10.1103/PhysRevLett.120.110501} {\bibfield  {journal}
  {\bibinfo  {journal} {Phys. Rev. Lett.}\ }\textbf {\bibinfo {volume} {120}},\
  \bibinfo {pages} {110501} (\bibinfo {year} {2018})}\BibitemShut {NoStop}%
\bibitem [{\citenamefont {Reiner}\ \emph {et~al.}(2019)\citenamefont {Reiner},
  \citenamefont {Wilhelm-Mauch}, \citenamefont {Schön},\ and\ \citenamefont
  {Marthaler}}]{reiner_finding_2019}%
  \BibitemOpen
  \bibfield  {author} {\bibinfo {author} {\bibfnamefont {J.-M.}\ \bibnamefont
  {Reiner}}, \bibinfo {author} {\bibfnamefont {F.}~\bibnamefont
  {Wilhelm-Mauch}}, \bibinfo {author} {\bibfnamefont {G.}~\bibnamefont
  {Schön}}, \ and\ \bibinfo {author} {\bibfnamefont {M.}~\bibnamefont
  {Marthaler}},\ }\href {https://doi.org/10.1088/2058-9565/ab1e85} {\bibfield
  {journal} {\bibinfo  {journal} {Quantum Sci. Technol.}\ }\textbf {\bibinfo
  {volume} {4}},\ \bibinfo {pages} {035005} (\bibinfo {year}
  {2019})}\BibitemShut {NoStop}%
\bibitem [{\citenamefont {Montanaro}\ and\ \citenamefont
  {Stanisic}(2020)}]{montanaro_compressed_2020}%
  \BibitemOpen
  \bibfield  {author} {\bibinfo {author} {\bibfnamefont {A.}~\bibnamefont
  {Montanaro}}\ and\ \bibinfo {author} {\bibfnamefont {S.}~\bibnamefont
  {Stanisic}},\ }\href {http://arxiv.org/abs/2006.01179} {\enquote {\bibinfo
  {title} {Compressed variational quantum eigensolver for the {Fermi}-{Hubbard}
  model},}\ } (\bibinfo {year} {2020})\BibitemShut {NoStop}%
\bibitem [{\citenamefont {Cai}(2020)}]{cai_resource_2020}%
  \BibitemOpen
  \bibfield  {author} {\bibinfo {author} {\bibfnamefont {Z.}~\bibnamefont
  {Cai}},\ }\href {https://doi.org/10.1103/PhysRevApplied.14.014059} {\bibfield
   {journal} {\bibinfo  {journal} {Phys. Rev. Applied}\ }\textbf {\bibinfo
  {volume} {14}},\ \bibinfo {pages} {014059} (\bibinfo {year}
  {2020})}\BibitemShut {NoStop}%
\bibitem [{\citenamefont {Cade}\ \emph {et~al.}(2020)\citenamefont {Cade},
  \citenamefont {Mineh}, \citenamefont {Montanaro},\ and\ \citenamefont
  {Stanisic}}]{cade_strategies_2020}%
  \BibitemOpen
  \bibfield  {author} {\bibinfo {author} {\bibfnamefont {C.}~\bibnamefont
  {Cade}}, \bibinfo {author} {\bibfnamefont {L.}~\bibnamefont {Mineh}},
  \bibinfo {author} {\bibfnamefont {A.}~\bibnamefont {Montanaro}}, \ and\
  \bibinfo {author} {\bibfnamefont {S.}~\bibnamefont {Stanisic}},\ }\href
  {https://doi.org/10.1103/PhysRevB.102.235122} {\bibfield  {journal} {\bibinfo
   {journal} {Phys. Rev. B}\ }\textbf {\bibinfo {volume} {102}},\ \bibinfo
  {pages} {235122} (\bibinfo {year} {2020})}\BibitemShut {NoStop}%
\bibitem [{\citenamefont {Mineh}\ and\ \citenamefont
  {Montanaro}(2022)}]{mineh_solving_2022}%
  \BibitemOpen
  \bibfield  {author} {\bibinfo {author} {\bibfnamefont {L.}~\bibnamefont
  {Mineh}}\ and\ \bibinfo {author} {\bibfnamefont {A.}~\bibnamefont
  {Montanaro}},\ }\href {https://doi.org/10.1103/PhysRevB.105.125117}
  {\bibfield  {journal} {\bibinfo  {journal} {Phys. Rev. B}\ }\textbf {\bibinfo
  {volume} {105}},\ \bibinfo {pages} {125117} (\bibinfo {year}
  {2022})}\BibitemShut {NoStop}%
\bibitem [{\citenamefont {Martin}\ \emph {et~al.}(2022)\citenamefont {Martin},
  \citenamefont {Simon},\ and\ \citenamefont
  {Rančić}}]{martin_simulating_2022}%
  \BibitemOpen
  \bibfield  {author} {\bibinfo {author} {\bibfnamefont {B.~A.}\ \bibnamefont
  {Martin}}, \bibinfo {author} {\bibfnamefont {P.}~\bibnamefont {Simon}}, \
  and\ \bibinfo {author} {\bibfnamefont {M.~J.}\ \bibnamefont {Rančić}},\
  }\href {https://doi.org/10.1103/PhysRevResearch.4.023190} {\bibfield
  {journal} {\bibinfo  {journal} {Phys. Rev. Research}\ }\textbf {\bibinfo
  {volume} {4}},\ \bibinfo {pages} {023190} (\bibinfo {year}
  {2022})}\BibitemShut {NoStop}%
\bibitem [{\citenamefont {Stanisic}\ \emph {et~al.}(2022)\citenamefont
  {Stanisic}, \citenamefont {Bosse}, \citenamefont {Gambetta}, \citenamefont
  {Santos}, \citenamefont {Mruczkiewicz}, \citenamefont {O'Brien},
  \citenamefont {Ostby},\ and\ \citenamefont
  {Montanaro}}]{stanisic_observing_2022}%
  \BibitemOpen
  \bibfield  {author} {\bibinfo {author} {\bibfnamefont {S.}~\bibnamefont
  {Stanisic}}, \bibinfo {author} {\bibfnamefont {J.~L.}\ \bibnamefont {Bosse}},
  \bibinfo {author} {\bibfnamefont {F.~M.}\ \bibnamefont {Gambetta}}, \bibinfo
  {author} {\bibfnamefont {R.~A.}\ \bibnamefont {Santos}}, \bibinfo {author}
  {\bibfnamefont {W.}~\bibnamefont {Mruczkiewicz}}, \bibinfo {author}
  {\bibfnamefont {T.~E.}\ \bibnamefont {O'Brien}}, \bibinfo {author}
  {\bibfnamefont {E.}~\bibnamefont {Ostby}}, \ and\ \bibinfo {author}
  {\bibfnamefont {A.}~\bibnamefont {Montanaro}},\ }\href
  {https://doi.org/10.1038/s41467-022-33335-4} {\bibfield  {journal} {\bibinfo
  {journal} {Nat. Commun.}\ }\textbf {\bibinfo {volume} {13}},\ \bibinfo
  {pages} {5743} (\bibinfo {year} {2022})}\BibitemShut {NoStop}%
\bibitem [{\citenamefont {Dallaire-Demers}\ \emph {et~al.}(2019)\citenamefont
  {Dallaire-Demers}, \citenamefont {Romero}, \citenamefont {Veis},
  \citenamefont {Sim},\ and\ \citenamefont
  {Aspuru-Guzik}}]{dallaire-demers_low-depth_2018}%
  \BibitemOpen
  \bibfield  {author} {\bibinfo {author} {\bibfnamefont {P.-L.}\ \bibnamefont
  {Dallaire-Demers}}, \bibinfo {author} {\bibfnamefont {J.}~\bibnamefont
  {Romero}}, \bibinfo {author} {\bibfnamefont {L.}~\bibnamefont {Veis}},
  \bibinfo {author} {\bibfnamefont {S.}~\bibnamefont {Sim}}, \ and\ \bibinfo
  {author} {\bibfnamefont {A.}~\bibnamefont {Aspuru-Guzik}},\ }\href
  {https://doi.org/10.1088/2058-9565/ab3951} {\bibfield  {journal} {\bibinfo
  {journal} {Quantum Sci. Technol.}\ }\textbf {\bibinfo {volume} {4}},\
  \bibinfo {pages} {045005} (\bibinfo {year} {2019})}\BibitemShut {NoStop}%
\bibitem [{\citenamefont {Dallaire-Demers}\ \emph {et~al.}(2020)\citenamefont
  {Dallaire-Demers}, \citenamefont {St\c{e}ch{\l}y}, \citenamefont {Gonthier},
  \citenamefont {Bashige}, \citenamefont {Romero},\ and\ \citenamefont
  {Cao}}]{dallaire-demers_application_2020}%
  \BibitemOpen
  \bibfield  {author} {\bibinfo {author} {\bibfnamefont {P.-L.}\ \bibnamefont
  {Dallaire-Demers}}, \bibinfo {author} {\bibfnamefont {M.}~\bibnamefont
  {St\c{e}ch{\l}y}}, \bibinfo {author} {\bibfnamefont {J.~F.}\ \bibnamefont
  {Gonthier}}, \bibinfo {author} {\bibfnamefont {N.~T.}\ \bibnamefont
  {Bashige}}, \bibinfo {author} {\bibfnamefont {J.}~\bibnamefont {Romero}}, \
  and\ \bibinfo {author} {\bibfnamefont {Y.}~\bibnamefont {Cao}},\ }\href
  {http://arxiv.org/abs/2003.01862} {\bibfield  {journal} {\bibinfo  {journal}
  {arXiv:2003.01862}\ } (\bibinfo {year} {2020})}\BibitemShut {NoStop}%
\bibitem [{\citenamefont {Suchsland}\ \emph {et~al.}(2022)\citenamefont
  {Suchsland}, \citenamefont {Barkoutsos}, \citenamefont {Tavernelli},
  \citenamefont {Fischer},\ and\ \citenamefont
  {Neupert}}]{suchsland_simulating_2022}%
  \BibitemOpen
  \bibfield  {author} {\bibinfo {author} {\bibfnamefont {P.}~\bibnamefont
  {Suchsland}}, \bibinfo {author} {\bibfnamefont {P.~K.}\ \bibnamefont
  {Barkoutsos}}, \bibinfo {author} {\bibfnamefont {I.}~\bibnamefont
  {Tavernelli}}, \bibinfo {author} {\bibfnamefont {M.~H.}\ \bibnamefont
  {Fischer}}, \ and\ \bibinfo {author} {\bibfnamefont {T.}~\bibnamefont
  {Neupert}},\ }\href {https://doi.org/10.1103/PhysRevResearch.4.013165}
  {\bibfield  {journal} {\bibinfo  {journal} {Phys. Rev. Research}\ }\textbf
  {\bibinfo {volume} {4}},\ \bibinfo {pages} {013165} (\bibinfo {year}
  {2022})}\BibitemShut {NoStop}%
\bibitem [{\citenamefont {Gard}\ and\ \citenamefont
  {Meier}(2022)}]{gard_classically_2022}%
  \BibitemOpen
  \bibfield  {author} {\bibinfo {author} {\bibfnamefont {B.~T.}\ \bibnamefont
  {Gard}}\ and\ \bibinfo {author} {\bibfnamefont {A.~M.}\ \bibnamefont
  {Meier}},\ }\href {https://doi.org/10.1103/PhysRevA.105.042602} {\bibfield
  {journal} {\bibinfo  {journal} {Phys. Rev. A}\ }\textbf {\bibinfo {volume}
  {105}},\ \bibinfo {pages} {042602} (\bibinfo {year} {2022})}\BibitemShut
  {NoStop}%
\bibitem [{\citenamefont {Kivlichan}\ \emph {et~al.}(2020)\citenamefont
  {Kivlichan}, \citenamefont {Gidney}, \citenamefont {Berry}, \citenamefont
  {Wiebe}, \citenamefont {McClean}, \citenamefont {Sun}, \citenamefont {Jiang},
  \citenamefont {Rubin}, \citenamefont {Fowler}, \citenamefont {Aspuru-Guzik},
  \citenamefont {Neven},\ and\ \citenamefont
  {Babbush}}]{kivlichan_improved_2020}%
  \BibitemOpen
  \bibfield  {author} {\bibinfo {author} {\bibfnamefont {I.~D.}\ \bibnamefont
  {Kivlichan}}, \bibinfo {author} {\bibfnamefont {C.}~\bibnamefont {Gidney}},
  \bibinfo {author} {\bibfnamefont {D.~W.}\ \bibnamefont {Berry}}, \bibinfo
  {author} {\bibfnamefont {N.}~\bibnamefont {Wiebe}}, \bibinfo {author}
  {\bibfnamefont {J.}~\bibnamefont {McClean}}, \bibinfo {author} {\bibfnamefont
  {W.}~\bibnamefont {Sun}}, \bibinfo {author} {\bibfnamefont {Z.}~\bibnamefont
  {Jiang}}, \bibinfo {author} {\bibfnamefont {N.}~\bibnamefont {Rubin}},
  \bibinfo {author} {\bibfnamefont {A.}~\bibnamefont {Fowler}}, \bibinfo
  {author} {\bibfnamefont {A.}~\bibnamefont {Aspuru-Guzik}}, \bibinfo {author}
  {\bibfnamefont {H.}~\bibnamefont {Neven}}, \ and\ \bibinfo {author}
  {\bibfnamefont {R.}~\bibnamefont {Babbush}},\ }\href
  {https://doi.org/10.22331/q-2020-07-16-296} {\bibfield  {journal} {\bibinfo
  {journal} {Quantum}\ }\textbf {\bibinfo {volume} {4}},\ \bibinfo {pages}
  {296} (\bibinfo {year} {2020})}\BibitemShut {NoStop}%
\bibitem [{\citenamefont {Campbell}(2022)}]{campbell_early_2022}%
  \BibitemOpen
  \bibfield  {author} {\bibinfo {author} {\bibfnamefont {E.~T.}\ \bibnamefont
  {Campbell}},\ }\href {https://doi.org/10.1088/2058-9565/ac3110} {\bibfield
  {journal} {\bibinfo  {journal} {Quantum Sci. Technol.}\ }\textbf {\bibinfo
  {volume} {7}},\ \bibinfo {pages} {015007} (\bibinfo {year}
  {2022})}\BibitemShut {NoStop}%
\bibitem [{\citenamefont {Clinton}\ \emph {et~al.}(2021)\citenamefont
  {Clinton}, \citenamefont {Bausch},\ and\ \citenamefont
  {Cubitt}}]{clinton_hamiltonian_2021}%
  \BibitemOpen
  \bibfield  {author} {\bibinfo {author} {\bibfnamefont {L.}~\bibnamefont
  {Clinton}}, \bibinfo {author} {\bibfnamefont {J.}~\bibnamefont {Bausch}}, \
  and\ \bibinfo {author} {\bibfnamefont {T.}~\bibnamefont {Cubitt}},\ }\href
  {https://doi.org/10.1038/s41467-021-25196-0} {\bibfield  {journal} {\bibinfo
  {journal} {Nat. Commun.}\ }\textbf {\bibinfo {volume} {12}},\ \bibinfo
  {pages} {4989} (\bibinfo {year} {2021})}\BibitemShut {NoStop}%
\bibitem [{\citenamefont {Peruzzo}\ \emph {et~al.}(2014)\citenamefont
  {Peruzzo}, \citenamefont {McClean}, \citenamefont {Shadbolt}, \citenamefont
  {Yung}, \citenamefont {Zhou}, \citenamefont {Love}, \citenamefont
  {Aspuru-Guzik},\ and\ \citenamefont {O’Brien}}]{peruzzo_variational_2014}%
  \BibitemOpen
  \bibfield  {author} {\bibinfo {author} {\bibfnamefont {A.}~\bibnamefont
  {Peruzzo}}, \bibinfo {author} {\bibfnamefont {J.}~\bibnamefont {McClean}},
  \bibinfo {author} {\bibfnamefont {P.}~\bibnamefont {Shadbolt}}, \bibinfo
  {author} {\bibfnamefont {M.-H.}\ \bibnamefont {Yung}}, \bibinfo {author}
  {\bibfnamefont {X.-Q.}\ \bibnamefont {Zhou}}, \bibinfo {author}
  {\bibfnamefont {P.~J.}\ \bibnamefont {Love}}, \bibinfo {author}
  {\bibfnamefont {A.}~\bibnamefont {Aspuru-Guzik}}, \ and\ \bibinfo {author}
  {\bibfnamefont {J.~L.}\ \bibnamefont {O’Brien}},\ }\href
  {https://doi.org/10.1038/ncomms5213} {\bibfield  {journal} {\bibinfo
  {journal} {Nat. Commun.}\ }\textbf {\bibinfo {volume} {5}},\ \bibinfo {pages}
  {4213} (\bibinfo {year} {2014})}\BibitemShut {NoStop}%
\bibitem [{\citenamefont {Bharti}\ \emph {et~al.}(2022)\citenamefont {Bharti},
  \citenamefont {Cervera-Lierta}, \citenamefont {Kyaw}, \citenamefont {Haug},
  \citenamefont {Alperin-Lea}, \citenamefont {Anand}, \citenamefont {Degroote},
  \citenamefont {Heimonen}, \citenamefont {Kottmann}, \citenamefont {Menke},
  \citenamefont {Mok}, \citenamefont {Sim}, \citenamefont {Kwek},\ and\
  \citenamefont {Aspuru-Guzik}}]{bharti_noisy_2022}%
  \BibitemOpen
  \bibfield  {author} {\bibinfo {author} {\bibfnamefont {K.}~\bibnamefont
  {Bharti}}, \bibinfo {author} {\bibfnamefont {A.}~\bibnamefont
  {Cervera-Lierta}}, \bibinfo {author} {\bibfnamefont {T.~H.}\ \bibnamefont
  {Kyaw}}, \bibinfo {author} {\bibfnamefont {T.}~\bibnamefont {Haug}}, \bibinfo
  {author} {\bibfnamefont {S.}~\bibnamefont {Alperin-Lea}}, \bibinfo {author}
  {\bibfnamefont {A.}~\bibnamefont {Anand}}, \bibinfo {author} {\bibfnamefont
  {M.}~\bibnamefont {Degroote}}, \bibinfo {author} {\bibfnamefont
  {H.}~\bibnamefont {Heimonen}}, \bibinfo {author} {\bibfnamefont {J.~S.}\
  \bibnamefont {Kottmann}}, \bibinfo {author} {\bibfnamefont {T.}~\bibnamefont
  {Menke}}, \bibinfo {author} {\bibfnamefont {W.-K.}\ \bibnamefont {Mok}},
  \bibinfo {author} {\bibfnamefont {S.}~\bibnamefont {Sim}}, \bibinfo {author}
  {\bibfnamefont {L.-C.}\ \bibnamefont {Kwek}}, \ and\ \bibinfo {author}
  {\bibfnamefont {A.}~\bibnamefont {Aspuru-Guzik}},\ }\href
  {https://doi.org/10.1103/RevModPhys.94.015004} {\bibfield  {journal}
  {\bibinfo  {journal} {Rev. Mod. Phys.}\ }\textbf {\bibinfo {volume} {94}},\
  \bibinfo {pages} {015004} (\bibinfo {year} {2022})}\BibitemShut {NoStop}%
\bibitem [{\citenamefont {Van~Vleck}(1929)}]{van_vleck__1929}%
  \BibitemOpen
  \bibfield  {author} {\bibinfo {author} {\bibfnamefont {J.~H.}\ \bibnamefont
  {Van~Vleck}},\ }\href {https://doi.org/10.1103/PhysRev.33.467} {\bibfield
  {journal} {\bibinfo  {journal} {Phys. Rev.}\ }\textbf {\bibinfo {volume}
  {33}},\ \bibinfo {pages} {467} (\bibinfo {year} {1929})}\BibitemShut
  {NoStop}%
\bibitem [{\citenamefont {Jordahl}(1934)}]{jordahl_effect_1934}%
  \BibitemOpen
  \bibfield  {author} {\bibinfo {author} {\bibfnamefont {O.~M.}\ \bibnamefont
  {Jordahl}},\ }\href {https://doi.org/10.1103/PhysRev.45.87} {\bibfield
  {journal} {\bibinfo  {journal} {Phys. Rev.}\ }\textbf {\bibinfo {volume}
  {45}},\ \bibinfo {pages} {87} (\bibinfo {year} {1934})}\BibitemShut {NoStop}%
\bibitem [{\citenamefont {Foldy}\ and\ \citenamefont
  {Wouthuysen}(1950)}]{foldy_on_1950}%
  \BibitemOpen
  \bibfield  {author} {\bibinfo {author} {\bibfnamefont {L.~L.}\ \bibnamefont
  {Foldy}}\ and\ \bibinfo {author} {\bibfnamefont {S.~A.}\ \bibnamefont
  {Wouthuysen}},\ }\href {https://doi.org/10.1103/PhysRev.78.29} {\bibfield
  {journal} {\bibinfo  {journal} {Phys. Rev.}\ }\textbf {\bibinfo {volume}
  {78}},\ \bibinfo {pages} {29} (\bibinfo {year} {1950})}\BibitemShut {NoStop}%
\bibitem [{\citenamefont {Primas}(1963)}]{primas_generalized_1963}%
  \BibitemOpen
  \bibfield  {author} {\bibinfo {author} {\bibfnamefont {H.}~\bibnamefont
  {Primas}},\ }\href {https://doi.org/10.1103/RevModPhys.35.710} {\bibfield
  {journal} {\bibinfo  {journal} {Rev. Mod. Phys.}\ }\textbf {\bibinfo {volume}
  {35}},\ \bibinfo {pages} {710} (\bibinfo {year} {1963})}\BibitemShut
  {NoStop}%
\bibitem [{\citenamefont {Brandow}(1979)}]{brandow_formal_1979}%
  \BibitemOpen
  \bibfield  {author} {\bibinfo {author} {\bibfnamefont {B.~H.}\ \bibnamefont
  {Brandow}},\ }\href {https://doi.org/10.1002/qua.560150207} {\bibfield
  {journal} {\bibinfo  {journal} {Int. J. Quantum Chem.}\ }\textbf {\bibinfo
  {volume} {15}},\ \bibinfo {pages} {207} (\bibinfo {year} {1979})}\BibitemShut
  {NoStop}%
\bibitem [{\citenamefont {Shavitt}\ and\ \citenamefont
  {Redmon}(1980)}]{shavitt_quasidegenerate_1980}%
  \BibitemOpen
  \bibfield  {author} {\bibinfo {author} {\bibfnamefont {I.}~\bibnamefont
  {Shavitt}}\ and\ \bibinfo {author} {\bibfnamefont {L.~T.}\ \bibnamefont
  {Redmon}},\ }\href {https://doi.org/10.1063/1.440050} {\bibfield  {journal}
  {\bibinfo  {journal} {J. Chem. Phys.}\ }\textbf {\bibinfo {volume} {73}},\
  \bibinfo {pages} {5711} (\bibinfo {year} {1980})}\BibitemShut {NoStop}%
\bibitem [{\citenamefont {Bravyi}\ \emph {et~al.}(2011)\citenamefont {Bravyi},
  \citenamefont {DiVincenzo},\ and\ \citenamefont
  {Loss}}]{bravyi_schriefferwolff_2011}%
  \BibitemOpen
  \bibfield  {author} {\bibinfo {author} {\bibfnamefont {S.}~\bibnamefont
  {Bravyi}}, \bibinfo {author} {\bibfnamefont {D.~P.}\ \bibnamefont
  {DiVincenzo}}, \ and\ \bibinfo {author} {\bibfnamefont {D.}~\bibnamefont
  {Loss}},\ }\href {https://doi.org/10.1016/j.aop.2011.06.004} {\bibfield
  {journal} {\bibinfo  {journal} {Ann. Physik.}\ }\textbf {\bibinfo {volume}
  {326}},\ \bibinfo {pages} {2793} (\bibinfo {year} {2011})}\BibitemShut
  {NoStop}%
\bibitem [{\citenamefont {Shavitt}\ and\ \citenamefont
  {Bartlett}(2009)}]{shavitt_many-body_2009}%
  \BibitemOpen
  \bibfield  {author} {\bibinfo {author} {\bibfnamefont {I.}~\bibnamefont
  {Shavitt}}\ and\ \bibinfo {author} {\bibfnamefont {R.~J.}\ \bibnamefont
  {Bartlett}},\ }\href {https://doi.org/10.1017/CBO9780511596834} {\emph
  {\bibinfo {title} {Many-{Body} {Methods} in {Chemistry} and {Physics}: {MBPT}
  and {Coupled}-{Cluster} {Theory}}}},\ \bibinfo {edition} {1st}\ ed.\
  (\bibinfo  {publisher} {Cambridge University Press},\ \bibinfo {year}
  {2009})\BibitemShut {NoStop}%
\bibitem [{\citenamefont {Schrieffer}\ and\ \citenamefont
  {Wolff}(1966)}]{schrieffer_relation_1966}%
  \BibitemOpen
  \bibfield  {author} {\bibinfo {author} {\bibfnamefont {J.~R.}\ \bibnamefont
  {Schrieffer}}\ and\ \bibinfo {author} {\bibfnamefont {P.~A.}\ \bibnamefont
  {Wolff}},\ }\href {https://doi.org/10.1103/PhysRev.149.491} {\bibfield
  {journal} {\bibinfo  {journal} {Phys. Rev.}\ }\textbf {\bibinfo {volume}
  {149}},\ \bibinfo {pages} {491} (\bibinfo {year} {1966})}\BibitemShut
  {NoStop}%
\bibitem [{\citenamefont {Harris}\ and\ \citenamefont
  {Lange}(1967)}]{harris_single-particle_nodate}%
  \BibitemOpen
  \bibfield  {author} {\bibinfo {author} {\bibfnamefont {A.~B.}\ \bibnamefont
  {Harris}}\ and\ \bibinfo {author} {\bibfnamefont {R.~V.}\ \bibnamefont
  {Lange}},\ }\href {https://doi.org/10.1103/PhysRev.157.295} {\bibfield
  {journal} {\bibinfo  {journal} {Phys. Rev.}\ }\textbf {\bibinfo {volume}
  {157}},\ \bibinfo {pages} {295} (\bibinfo {year} {1967})}\BibitemShut
  {NoStop}%
\bibitem [{\citenamefont {Chao}\ \emph {et~al.}(1977)\citenamefont {Chao},
  \citenamefont {Spalek},\ and\ \citenamefont {Oles}}]{chao_kinetic_1977}%
  \BibitemOpen
  \bibfield  {author} {\bibinfo {author} {\bibfnamefont {K.~A.}\ \bibnamefont
  {Chao}}, \bibinfo {author} {\bibfnamefont {J.}~\bibnamefont {Spalek}}, \ and\
  \bibinfo {author} {\bibfnamefont {A.~M.}\ \bibnamefont {Oles}},\ }\href
  {https://doi.org/10.1088/0022-3719/10/10/002} {\bibfield  {journal} {\bibinfo
   {journal} {J. Phys. C: Solid State Phys.}\ }\textbf {\bibinfo {volume}
  {10}},\ \bibinfo {pages} {L271} (\bibinfo {year} {1977})}\BibitemShut
  {NoStop}%
\bibitem [{\citenamefont {Zhang}\ \emph {et~al.}(2022)\citenamefont {Zhang},
  \citenamefont {Yang}, \citenamefont {Xu},\ and\ \citenamefont
  {Li}}]{zhang_quantum_2022}%
  \BibitemOpen
  \bibfield  {author} {\bibinfo {author} {\bibfnamefont {Z.}~\bibnamefont
  {Zhang}}, \bibinfo {author} {\bibfnamefont {Y.}~\bibnamefont {Yang}},
  \bibinfo {author} {\bibfnamefont {X.}~\bibnamefont {Xu}}, \ and\ \bibinfo
  {author} {\bibfnamefont {Y.}~\bibnamefont {Li}},\ }\href
  {https://doi.org/10.1103/PhysRevResearch.4.043023} {\bibfield  {journal}
  {\bibinfo  {journal} {Phys. Rev. Research}\ }\textbf {\bibinfo {volume}
  {4}},\ \bibinfo {pages} {043023} (\bibinfo {year} {2022})}\BibitemShut
  {NoStop}%
\bibitem [{\citenamefont {Romero}\ \emph {et~al.}(2018)\citenamefont {Romero},
  \citenamefont {Babbush}, \citenamefont {McClean}, \citenamefont {Hempel},
  \citenamefont {Love},\ and\ \citenamefont
  {Aspuru-Guzik}}]{romero2018strategies}%
  \BibitemOpen
  \bibfield  {author} {\bibinfo {author} {\bibfnamefont {J.}~\bibnamefont
  {Romero}}, \bibinfo {author} {\bibfnamefont {R.}~\bibnamefont {Babbush}},
  \bibinfo {author} {\bibfnamefont {J.~R.}\ \bibnamefont {McClean}}, \bibinfo
  {author} {\bibfnamefont {C.}~\bibnamefont {Hempel}}, \bibinfo {author}
  {\bibfnamefont {P.~J.}\ \bibnamefont {Love}}, \ and\ \bibinfo {author}
  {\bibfnamefont {A.}~\bibnamefont {Aspuru-Guzik}},\ }\href
  {https://doi.org/10.1088/2058-9565/aad3e4} {\bibfield  {journal} {\bibinfo
  {journal} {Quantum Sci. Technol.}\ }\textbf {\bibinfo {volume} {4}},\
  \bibinfo {pages} {014008} (\bibinfo {year} {2018})}\BibitemShut {NoStop}%
\bibitem [{\citenamefont {tA~v}\ \emph {et~al.}(2021)\citenamefont {tA~v},
  \citenamefont {ANIS}, \citenamefont {Abby-Mitchell}, \citenamefont {Abraham},
  \citenamefont {AduOffei}, \citenamefont {Agarwal}, \citenamefont {Agliardi},
  \citenamefont {Aharoni}, \citenamefont {Ajith}, \citenamefont {Akhalwaya},
  \citenamefont {Aleksandrowicz}, \citenamefont {Alexander}, \citenamefont
  {Amy}, \citenamefont {Anagolum}, \citenamefont {Anthony-Gandon},
  \citenamefont {Araujo}, \citenamefont {Arbel}, \citenamefont {Asfaw},
  \citenamefont {Ashimine}, \citenamefont {Athalye}, \citenamefont {Avkhadiev},
  \citenamefont {Azaustre}, \citenamefont {BHOLE}, \citenamefont {Bajpe},
  \citenamefont {Banerjee}, \citenamefont {Banerjee}, \citenamefont {Bang},
  \citenamefont {Bansal}, \citenamefont {Barkoutsos}, \citenamefont {Barnawal},
  \citenamefont {Barron}, \citenamefont {Barron}, \citenamefont {Bello},
  \citenamefont {Ben-Haim}, \citenamefont {Bennett}, \citenamefont {Bevenius},
  \citenamefont {Bhatnagar}, \citenamefont {Bhatnagar}, \citenamefont {Bhobe},
  \citenamefont {Bianchini}, \citenamefont {Bishop}, \citenamefont {Blank},
  \citenamefont {Bolos}, \citenamefont {Bopardikar}, \citenamefont {Bosch},
  \citenamefont {Brandhofer}, \citenamefont {Brandon}, \citenamefont {Bravyi},
  \citenamefont {Bryce-Fuller}, \citenamefont {Bucher}, \citenamefont
  {Burgholzer}, \citenamefont {Burov}, \citenamefont {Cabrera}, \citenamefont
  {Calpin}, \citenamefont {Capelluto}, \citenamefont {Carballo}, \citenamefont
  {Carrascal}, \citenamefont {Carriker}, \citenamefont {Carvalho},
  \citenamefont {Chakrabarti}, \citenamefont {Chen}, \citenamefont {Chen},
  \citenamefont {Chen}, \citenamefont {Chen}, \citenamefont {Chen},
  \citenamefont {Chevallier}, \citenamefont {Chinda}, \citenamefont
  {Cholarajan}, \citenamefont {Chow}, \citenamefont {Churchill}, \citenamefont
  {CisterMoke}, \citenamefont {Claus}, \citenamefont {Clauss}, \citenamefont
  {Clothier}, \citenamefont {Cocking}, \citenamefont {Cocuzzo}, \citenamefont
  {Connor}, \citenamefont {Correa}, \citenamefont {Crockett}, \citenamefont
  {Cross}, \citenamefont {Cross}, \citenamefont {Cross}, \citenamefont
  {Cruz-Benito}, \citenamefont {Culver}, \citenamefont {C{\'o}rcoles-Gonzales},
  \citenamefont {D}, \citenamefont {Dague}, \citenamefont {Dandachi},
  \citenamefont {Dangwal}, \citenamefont {Daniel}, \citenamefont {DanielAja},
  \citenamefont {Daniels}, \citenamefont {Dartiailh}, \citenamefont {Davila},
  \citenamefont {Debouni}, \citenamefont {Dekusar}, \citenamefont {Deshmukh},
  \citenamefont {Deshpande}, \citenamefont {Ding}, \citenamefont {Doi},
  \citenamefont {Dow}, \citenamefont {Downing}, \citenamefont {Drechsler},
  \citenamefont {Drudis}, \citenamefont {Dumitrescu}, \citenamefont {Dumon},
  \citenamefont {Duran}, \citenamefont {EL-Safty}, \citenamefont {Eastman},
  \citenamefont {Eberle}, \citenamefont {Ebrahimi}, \citenamefont {Eendebak},
  \citenamefont {Egger}, \citenamefont {EgrettaThula}, \citenamefont {ElePT},
  \citenamefont {Elsayed}, \citenamefont {Emilio}, \citenamefont {Espiricueta},
  \citenamefont {Everitt}, \citenamefont {Facoetti}, \citenamefont {Farida},
  \citenamefont {Fern{\'a}ndez}, \citenamefont {Ferracin}, \citenamefont
  {Ferrari}, \citenamefont {Ferrera}, \citenamefont {Fouilland}, \citenamefont
  {Frisch}, \citenamefont {Fuhrer}, \citenamefont {Fuller}, \citenamefont
  {GEORGE}, \citenamefont {Gacon}, \citenamefont {Gago}, \citenamefont
  {Gambella}, \citenamefont {Gambetta}, \citenamefont {Gammanpila},
  \citenamefont {Garcia}, \citenamefont {Garg}, \citenamefont {Garion},
  \citenamefont {Garrison}, \citenamefont {Garrison}, \citenamefont {Gates},
  \citenamefont {Gavrielov}, \citenamefont {Gentinetta}, \citenamefont
  {Georgiev}, \citenamefont {Gil}, \citenamefont {Gilliam}, \citenamefont
  {Giridharan}, \citenamefont {Glen}, \citenamefont {Gomez-Mosquera},
  \citenamefont {Gonzalo}, \citenamefont {de~la Puente~Gonz{\'a}lez},
  \citenamefont {Gorzinski}, \citenamefont {Gould}, \citenamefont {Greenberg},
  \citenamefont {Grinko}, \citenamefont {Guan}, \citenamefont {Guijo},
  \citenamefont {Guillermo-Mijares-Vilarino}, \citenamefont {Gunnels},
  \citenamefont {Gupta}, \citenamefont {Gupta}, \citenamefont {G{\"u}nther},
  \citenamefont {Haglund}, \citenamefont {Haide}, \citenamefont {Hamamura},
  \citenamefont {Hamido}, \citenamefont {Harkins}, \citenamefont {Hartman},
  \citenamefont {Hasan}, \citenamefont {Havlicek}, \citenamefont {Hellmers},
  \citenamefont {Herok}, \citenamefont {Hill}, \citenamefont {Hillmich},
  \citenamefont {Hincks}, \citenamefont {Hong}, \citenamefont {Horii},
  \citenamefont {Howington}, \citenamefont {Hu}, \citenamefont {Hu},
  \citenamefont {Huang}, \citenamefont {Huang}, \citenamefont {Huisman},
  \citenamefont {Imai}, \citenamefont {Imamichi}, \citenamefont {Ishizaki},
  \citenamefont {Ishwor}, \citenamefont {Iten}, \citenamefont {Itoko},
  \citenamefont {Ivrii}, \citenamefont {Javadi}, \citenamefont {Javadi-Abhari},
  \citenamefont {Javed}, \citenamefont {Jianhua}, \citenamefont {Jivrajani},
  \citenamefont {Johns}, \citenamefont {Johnstun}, \citenamefont
  {Jonathan-Shoemaker}, \citenamefont {JosDenmark}, \citenamefont {JoshDumo},
  \citenamefont {Judge}, \citenamefont {Kachmann}, \citenamefont {Kale},
  \citenamefont {Kanazawa}, \citenamefont {Kane}, \citenamefont {Kang-Bae},
  \citenamefont {Kapila}, \citenamefont {Karazeev}, \citenamefont {Kassebaum},
  \citenamefont {Kato}, \citenamefont {Kehrer}, \citenamefont {Kelso},
  \citenamefont {Kelso}, \citenamefont {van Kemenade}, \citenamefont
  {Khanderao}, \citenamefont {King}, \citenamefont {Kobayashi}, \citenamefont
  {Kovi11Day}, \citenamefont {Kovyrshin}, \citenamefont {Krishna},
  \citenamefont {Krishnakumar}, \citenamefont {Krishnamurthy}, \citenamefont
  {Krishnan}, \citenamefont {Krsulich}, \citenamefont {Kumkar}, \citenamefont
  {Kus}, \citenamefont {LNoorl}, \citenamefont {LaRose}, \citenamefont {Lacal},
  \citenamefont {Lambert}, \citenamefont {Landa}, \citenamefont {Lapeyre},
  \citenamefont {Lasecki}, \citenamefont {Latone}, \citenamefont {Lawrence},
  \citenamefont {Lee}, \citenamefont {Li}, \citenamefont {Liang}, \citenamefont
  {Lishman}, \citenamefont {Liu}, \citenamefont {Liu}, \citenamefont {Lolcroc},
  \citenamefont {M}, \citenamefont {Madden}, \citenamefont {Maeng},
  \citenamefont {Maheshkar}, \citenamefont {Majmudar}, \citenamefont
  {Malyshev}, \citenamefont {Mandouh}, \citenamefont {Manela}, \citenamefont
  {Manjula}, \citenamefont {Marecek}, \citenamefont {Marques}, \citenamefont
  {Marwaha}, \citenamefont {Maslov}, \citenamefont {Maszota}, \citenamefont
  {Mathews}, \citenamefont {Matsuo}, \citenamefont {Mazhandu}, \citenamefont
  {McClure}, \citenamefont {McElaney}, \citenamefont {McElroy}, \citenamefont
  {McGarry}, \citenamefont {McKay}, \citenamefont {McPherson}, \citenamefont
  {Meesala}, \citenamefont {Meirom}, \citenamefont {Mendell}, \citenamefont
  {Metcalfe}, \citenamefont {Mevissen}, \citenamefont {Meyer}, \citenamefont
  {Mezzacapo}, \citenamefont {Midha}, \citenamefont {Millar}, \citenamefont
  {Miller}, \citenamefont {Miller}, \citenamefont {Minev}, \citenamefont
  {Mitchell}, \citenamefont {Mohammad}, \citenamefont {Moll}, \citenamefont
  {Montanez}, \citenamefont {Monteiro}, \citenamefont {Mooring}, \citenamefont
  {Morales}, \citenamefont {Moran}, \citenamefont {Morcuende}, \citenamefont
  {Mostafa}, \citenamefont {Motta}, \citenamefont {Moyard}, \citenamefont
  {Murali}, \citenamefont {Murata}, \citenamefont {M{\"u}ggenburg},
  \citenamefont {NEMOZ}, \citenamefont {Nadlinger}, \citenamefont {Nakanishi},
  \citenamefont {Nannicini}, \citenamefont {Nation}, \citenamefont {Navarro},
  \citenamefont {Naveh}, \citenamefont {Neagle}, \citenamefont {Neuweiler},
  \citenamefont {Ngoueya}, \citenamefont {Nguyen}, \citenamefont {Nicander},
  \citenamefont {Nick-Singstock}, \citenamefont {Niroula}, \citenamefont
  {Norlen}, \citenamefont {NuoWenLei}, \citenamefont {O'Riordan}, \citenamefont
  {Ogunbayo}, \citenamefont {Ollitrault}, \citenamefont {Onodera},
  \citenamefont {Otaolea}, \citenamefont {Oud}, \citenamefont {Padilha},
  \citenamefont {Paik}, \citenamefont {Pal}, \citenamefont {Pang},
  \citenamefont {Panigrahi}, \citenamefont {Pascuzzi}, \citenamefont
  {Perriello}, \citenamefont {Peterson}, \citenamefont {Phan}, \citenamefont
  {Pilch}, \citenamefont {Piro}, \citenamefont {Pistoia}, \citenamefont
  {Piveteau}, \citenamefont {Plewa}, \citenamefont {Pocreau}, \citenamefont
  {Possel}, \citenamefont {Pozas-Kerstjens}, \citenamefont {Pracht},
  \citenamefont {Prokop}, \citenamefont {Prutyanov}, \citenamefont {Puri},
  \citenamefont {Puzzuoli}, \citenamefont {Pythonix}, \citenamefont
  {P{\'e}rez}, \citenamefont {Quant02}, \citenamefont {Quintiii}, \citenamefont
  {Rahman}, \citenamefont {Raja}, \citenamefont {Rajeev}, \citenamefont
  {Rajput}, \citenamefont {Ramagiri}, \citenamefont {Rao}, \citenamefont
  {Raymond}, \citenamefont {Reardon-Smith}, \citenamefont {Redondo},
  \citenamefont {Reuter}, \citenamefont {Rice}, \citenamefont {Riedemann},
  \citenamefont {Rietesh}, \citenamefont {Risinger}, \citenamefont {Rivero},
  \citenamefont {Rocca}, \citenamefont {Rodr{\'\i}guez}, \citenamefont
  {RohithKarur}, \citenamefont {Rosand}, \citenamefont {Rossmannek},
  \citenamefont {Ryu}, \citenamefont {SAPV}, \citenamefont {Sa}, \citenamefont
  {Saha}, \citenamefont {Ash-Saki}, \citenamefont {Salman}, \citenamefont
  {Sanand}, \citenamefont {Sandberg}, \citenamefont {Sandesara}, \citenamefont
  {Sapra}, \citenamefont {Sargsyan}, \citenamefont {Sarkar}, \citenamefont
  {Sathaye}, \citenamefont {Savola}, \citenamefont {Schmitt}, \citenamefont
  {Schnabel}, \citenamefont {Schoenfeld}, \citenamefont {Scholten},
  \citenamefont {Schoute}, \citenamefont {Schuhmacher}, \citenamefont
  {Schulterbrandt}, \citenamefont {Schwarm}, \citenamefont {Schweigert},
  \citenamefont {Seaward}, \citenamefont {Sergi}, \citenamefont {Serrano},
  \citenamefont {Sertage}, \citenamefont {Setia}, \citenamefont {Shah},
  \citenamefont {Shammah}, \citenamefont {Shanks}, \citenamefont {Sharma},
  \citenamefont {Shaw}, \citenamefont {Shi}, \citenamefont {Shoemaker},
  \citenamefont {Silva}, \citenamefont {Simonetto}, \citenamefont {Singh},
  \citenamefont {Singh}, \citenamefont {Singh}, \citenamefont {Singkanipa},
  \citenamefont {Siraichi}, \citenamefont {Siri}, \citenamefont {Sistos},
  \citenamefont {Sistos}, \citenamefont {Sitdikov}, \citenamefont {Sivarajah},
  \citenamefont {Slavikmew}, \citenamefont {Sletfjerding}, \citenamefont
  {Smolin}, \citenamefont {Soeken}, \citenamefont {Sokolov}, \citenamefont
  {Sokolov}, \citenamefont {Soloviev}, \citenamefont {SooluThomas},
  \citenamefont {Starfish}, \citenamefont {Steenken}, \citenamefont
  {Stypulkoski}, \citenamefont {Suau}, \citenamefont {Sun}, \citenamefont
  {Sung}, \citenamefont {Suwama}, \citenamefont {S{\l}owik}, \citenamefont
  {Taeja}, \citenamefont {Takahashi}, \citenamefont {Takawale}, \citenamefont
  {Tavernelli}, \citenamefont {Taylor}, \citenamefont {Taylour}, \citenamefont
  {Thomas}, \citenamefont {Tian}, \citenamefont {Tillet}, \citenamefont {Tod},
  \citenamefont {Tomasik}, \citenamefont {Tornow}, \citenamefont {de~la Torre},
  \citenamefont {Toural}, \citenamefont {Trabing}, \citenamefont {Treinish},
  \citenamefont {Trenev}, \citenamefont {TrishaPe}, \citenamefont {Truger},
  \citenamefont {TsafrirA}, \citenamefont {Tsilimigkounakis}, \citenamefont
  {Tsuoka}, \citenamefont {Tulsi}, \citenamefont {Tuna}, \citenamefont
  {Turner}, \citenamefont {Vaknin}, \citenamefont {Valcarce}, \citenamefont
  {Varchon}, \citenamefont {Vartak}, \citenamefont {Vazquez}, \citenamefont
  {Vijaywargiya}, \citenamefont {Villar}, \citenamefont {Vishnu}, \citenamefont
  {Vogt-Lee}, \citenamefont {Vuillot}, \citenamefont {WQ}, \citenamefont
  {Weaver}, \citenamefont {Weidenfeller}, \citenamefont {Wieczorek},
  \citenamefont {Wildstrom}, \citenamefont {Wilson}, \citenamefont {Winston},
  \citenamefont {WinterSoldier}, \citenamefont {Woehr}, \citenamefont
  {Woerner}, \citenamefont {Woo}, \citenamefont {Wood}, \citenamefont {Wood},
  \citenamefont {Wood}, \citenamefont {Wootton}, \citenamefont {Wright},
  \citenamefont {Xing}, \citenamefont {YU}, \citenamefont {Yaiza},
  \citenamefont {Yang}, \citenamefont {Yang}, \citenamefont {Yao},
  \citenamefont {Yeralin}, \citenamefont {Yonekura}, \citenamefont
  {Yonge-Mallo}, \citenamefont {Yoshida}, \citenamefont {Young}, \citenamefont
  {Yu}, \citenamefont {Yu}, \citenamefont {Yuma-Nakamura}, \citenamefont
  {Zachow}, \citenamefont {Zdanski}, \citenamefont {Zhang}, \citenamefont
  {Zheltonozhskii}, \citenamefont {Zidaru}, \citenamefont {Zimmermann},
  \citenamefont {Zindorf}, \citenamefont {Zoufal}, \citenamefont {a~matsuo},
  \citenamefont {aeddins ibm}, \citenamefont {alexzhang13}, \citenamefont
  {b63}, \citenamefont {bartek bartlomiej}, \citenamefont {bcamorrison},
  \citenamefont {brandhsn}, \citenamefont {nick bronn}, \citenamefont
  {chetmurthy}, \citenamefont {choerst ibm}, \citenamefont {comet},
  \citenamefont {dalin27}, \citenamefont {deeplokhande}, \citenamefont
  {dekel.meirom}, \citenamefont {derwind}, \citenamefont {dime10},
  \citenamefont {ehchen}, \citenamefont {ewinston}, \citenamefont
  {fanizzamarco}, \citenamefont {fs1132429}, \citenamefont {gadial},
  \citenamefont {galeinston}, \citenamefont {georgezhou20}, \citenamefont
  {georgios ts}, \citenamefont {gruu}, \citenamefont {hhorii}, \citenamefont
  {hhyap}, \citenamefont {hykavitha}, \citenamefont {itoko}, \citenamefont
  {jeppevinkel}, \citenamefont {jessica angel7}, \citenamefont {jezerjojo14},
  \citenamefont {jliu45}, \citenamefont {johannesgreiner}, \citenamefont
  {jscott2}, \citenamefont {kUmezawa}, \citenamefont {klinvill}, \citenamefont
  {krutik2966}, \citenamefont {ma5x}, \citenamefont {michelle4654},
  \citenamefont {msuwama}, \citenamefont {nico lgrs}, \citenamefont
  {nrhawkins}, \citenamefont {ntgiwsvp}, \citenamefont {ordmoj}, \citenamefont
  {sagar pahwa}, \citenamefont {pritamsinha2304}, \citenamefont {rithikaadiga},
  \citenamefont {ryancocuzzo}, \citenamefont {saktar unr}, \citenamefont
  {saswati qiskit}, \citenamefont {sebastian mair}, \citenamefont {septembrr},
  \citenamefont {sethmerkel}, \citenamefont {sg495}, \citenamefont {shaashwat},
  \citenamefont {smturro2}, \citenamefont {sternparky}, \citenamefont
  {strickroman}, \citenamefont {tigerjack}, \citenamefont {tsura crisaldo},
  \citenamefont {upsideon}, \citenamefont {vadebayo49}, \citenamefont {welien},
  \citenamefont {willhbang}, \citenamefont {wmurphy collabstar}, \citenamefont
  {yang.luh}, \citenamefont {yuri@FreeBSD},\ and\ \citenamefont
  {{\v{C}}epulkovskis}}]{Qiskit}%
  \BibitemOpen
  \bibfield  {author} {\bibinfo {author} {\bibfnamefont {A.}~\bibnamefont
  {tA~v}}, \bibinfo {author} {\bibfnamefont {M.~S.}\ \bibnamefont {ANIS}},
  \bibinfo {author} {\bibnamefont {Abby-Mitchell}}, \bibinfo {author}
  {\bibfnamefont {H.}~\bibnamefont {Abraham}}, \bibinfo {author} {\bibnamefont
  {AduOffei}}, \bibinfo {author} {\bibfnamefont {R.}~\bibnamefont {Agarwal}},
  \bibinfo {author} {\bibfnamefont {G.}~\bibnamefont {Agliardi}}, \bibinfo
  {author} {\bibfnamefont {M.}~\bibnamefont {Aharoni}}, \bibinfo {author}
  {\bibfnamefont {V.}~\bibnamefont {Ajith}}, \bibinfo {author} {\bibfnamefont
  {I.~Y.}\ \bibnamefont {Akhalwaya}}, \bibinfo {author} {\bibfnamefont
  {G.}~\bibnamefont {Aleksandrowicz}}, \bibinfo {author} {\bibfnamefont
  {T.}~\bibnamefont {Alexander}}, \bibinfo {author} {\bibfnamefont
  {M.}~\bibnamefont {Amy}}, \bibinfo {author} {\bibfnamefont {S.}~\bibnamefont
  {Anagolum}}, \bibinfo {author} {\bibnamefont {Anthony-Gandon}}, \bibinfo
  {author} {\bibfnamefont {I.~F.}\ \bibnamefont {Araujo}}, \bibinfo {author}
  {\bibfnamefont {E.}~\bibnamefont {Arbel}}, \bibinfo {author} {\bibfnamefont
  {A.}~\bibnamefont {Asfaw}}, \bibinfo {author} {\bibfnamefont {I.~E.}\
  \bibnamefont {Ashimine}}, \bibinfo {author} {\bibfnamefont {A.}~\bibnamefont
  {Athalye}}, \bibinfo {author} {\bibfnamefont {A.}~\bibnamefont {Avkhadiev}},
  \bibinfo {author} {\bibfnamefont {C.}~\bibnamefont {Azaustre}}, \bibinfo
  {author} {\bibfnamefont {P.}~\bibnamefont {BHOLE}}, \bibinfo {author}
  {\bibfnamefont {V.}~\bibnamefont {Bajpe}}, \bibinfo {author} {\bibfnamefont
  {A.}~\bibnamefont {Banerjee}}, \bibinfo {author} {\bibfnamefont
  {S.}~\bibnamefont {Banerjee}}, \bibinfo {author} {\bibfnamefont
  {W.}~\bibnamefont {Bang}}, \bibinfo {author} {\bibfnamefont {A.}~\bibnamefont
  {Bansal}}, \bibinfo {author} {\bibfnamefont {P.}~\bibnamefont {Barkoutsos}},
  \bibinfo {author} {\bibfnamefont {A.}~\bibnamefont {Barnawal}}, \bibinfo
  {author} {\bibfnamefont {G.}~\bibnamefont {Barron}}, \bibinfo {author}
  {\bibfnamefont {G.~S.}\ \bibnamefont {Barron}}, \bibinfo {author}
  {\bibfnamefont {L.}~\bibnamefont {Bello}}, \bibinfo {author} {\bibfnamefont
  {Y.}~\bibnamefont {Ben-Haim}}, \bibinfo {author} {\bibfnamefont {M.~C.}\
  \bibnamefont {Bennett}}, \bibinfo {author} {\bibfnamefont {D.}~\bibnamefont
  {Bevenius}}, \bibinfo {author} {\bibfnamefont {D.}~\bibnamefont {Bhatnagar}},
  \bibinfo {author} {\bibfnamefont {P.}~\bibnamefont {Bhatnagar}}, \bibinfo
  {author} {\bibfnamefont {A.}~\bibnamefont {Bhobe}}, \bibinfo {author}
  {\bibfnamefont {P.}~\bibnamefont {Bianchini}}, \bibinfo {author}
  {\bibfnamefont {L.~S.}\ \bibnamefont {Bishop}}, \bibinfo {author}
  {\bibfnamefont {C.}~\bibnamefont {Blank}}, \bibinfo {author} {\bibfnamefont
  {S.}~\bibnamefont {Bolos}}, \bibinfo {author} {\bibfnamefont
  {S.}~\bibnamefont {Bopardikar}}, \bibinfo {author} {\bibfnamefont
  {S.}~\bibnamefont {Bosch}}, \bibinfo {author} {\bibfnamefont
  {S.}~\bibnamefont {Brandhofer}}, \bibinfo {author} {\bibnamefont {Brandon}},
  \bibinfo {author} {\bibfnamefont {S.}~\bibnamefont {Bravyi}}, \bibinfo
  {author} {\bibnamefont {Bryce-Fuller}}, \bibinfo {author} {\bibfnamefont
  {D.}~\bibnamefont {Bucher}}, \bibinfo {author} {\bibfnamefont
  {L.}~\bibnamefont {Burgholzer}}, \bibinfo {author} {\bibfnamefont
  {A.}~\bibnamefont {Burov}}, \bibinfo {author} {\bibfnamefont
  {F.}~\bibnamefont {Cabrera}}, \bibinfo {author} {\bibfnamefont
  {P.}~\bibnamefont {Calpin}}, \bibinfo {author} {\bibfnamefont
  {L.}~\bibnamefont {Capelluto}}, \bibinfo {author} {\bibfnamefont
  {J.}~\bibnamefont {Carballo}}, \bibinfo {author} {\bibfnamefont
  {G.}~\bibnamefont {Carrascal}}, \bibinfo {author} {\bibfnamefont
  {A.}~\bibnamefont {Carriker}}, \bibinfo {author} {\bibfnamefont
  {I.}~\bibnamefont {Carvalho}}, \bibinfo {author} {\bibfnamefont
  {R.}~\bibnamefont {Chakrabarti}}, \bibinfo {author} {\bibfnamefont
  {A.}~\bibnamefont {Chen}}, \bibinfo {author} {\bibfnamefont {C.-F.}\
  \bibnamefont {Chen}}, \bibinfo {author} {\bibfnamefont {E.}~\bibnamefont
  {Chen}}, \bibinfo {author} {\bibfnamefont {J.~C.}\ \bibnamefont {Chen}},
  \bibinfo {author} {\bibfnamefont {R.}~\bibnamefont {Chen}}, \bibinfo {author}
  {\bibfnamefont {F.}~\bibnamefont {Chevallier}}, \bibinfo {author}
  {\bibfnamefont {K.}~\bibnamefont {Chinda}}, \bibinfo {author} {\bibfnamefont
  {R.}~\bibnamefont {Cholarajan}}, \bibinfo {author} {\bibfnamefont {J.~M.}\
  \bibnamefont {Chow}}, \bibinfo {author} {\bibfnamefont {S.}~\bibnamefont
  {Churchill}}, \bibinfo {author} {\bibnamefont {CisterMoke}}, \bibinfo
  {author} {\bibfnamefont {C.}~\bibnamefont {Claus}}, \bibinfo {author}
  {\bibfnamefont {C.}~\bibnamefont {Clauss}}, \bibinfo {author} {\bibfnamefont
  {C.}~\bibnamefont {Clothier}}, \bibinfo {author} {\bibfnamefont
  {R.}~\bibnamefont {Cocking}}, \bibinfo {author} {\bibfnamefont
  {R.}~\bibnamefont {Cocuzzo}}, \bibinfo {author} {\bibfnamefont
  {J.}~\bibnamefont {Connor}}, \bibinfo {author} {\bibfnamefont
  {F.}~\bibnamefont {Correa}}, \bibinfo {author} {\bibfnamefont
  {Z.}~\bibnamefont {Crockett}}, \bibinfo {author} {\bibfnamefont {A.~J.}\
  \bibnamefont {Cross}}, \bibinfo {author} {\bibfnamefont {A.~W.}\ \bibnamefont
  {Cross}}, \bibinfo {author} {\bibfnamefont {S.}~\bibnamefont {Cross}},
  \bibinfo {author} {\bibfnamefont {J.}~\bibnamefont {Cruz-Benito}}, \bibinfo
  {author} {\bibfnamefont {C.}~\bibnamefont {Culver}}, \bibinfo {author}
  {\bibfnamefont {A.~D.}\ \bibnamefont {C{\'o}rcoles-Gonzales}}, \bibinfo
  {author} {\bibfnamefont {N.}~\bibnamefont {D}}, \bibinfo {author}
  {\bibfnamefont {S.}~\bibnamefont {Dague}}, \bibinfo {author} {\bibfnamefont
  {T.~E.}\ \bibnamefont {Dandachi}}, \bibinfo {author} {\bibfnamefont {A.~N.}\
  \bibnamefont {Dangwal}}, \bibinfo {author} {\bibfnamefont {J.}~\bibnamefont
  {Daniel}}, \bibinfo {author} {\bibnamefont {DanielAja}}, \bibinfo {author}
  {\bibfnamefont {M.}~\bibnamefont {Daniels}}, \bibinfo {author} {\bibfnamefont
  {M.}~\bibnamefont {Dartiailh}}, \bibinfo {author} {\bibfnamefont {A.~R.}\
  \bibnamefont {Davila}}, \bibinfo {author} {\bibfnamefont {F.}~\bibnamefont
  {Debouni}}, \bibinfo {author} {\bibfnamefont {A.}~\bibnamefont {Dekusar}},
  \bibinfo {author} {\bibfnamefont {A.}~\bibnamefont {Deshmukh}}, \bibinfo
  {author} {\bibfnamefont {M.}~\bibnamefont {Deshpande}}, \bibinfo {author}
  {\bibfnamefont {D.}~\bibnamefont {Ding}}, \bibinfo {author} {\bibfnamefont
  {J.}~\bibnamefont {Doi}}, \bibinfo {author} {\bibfnamefont {E.~M.}\
  \bibnamefont {Dow}}, \bibinfo {author} {\bibfnamefont {P.}~\bibnamefont
  {Downing}}, \bibinfo {author} {\bibfnamefont {E.}~\bibnamefont {Drechsler}},
  \bibinfo {author} {\bibfnamefont {M.~S.}\ \bibnamefont {Drudis}}, \bibinfo
  {author} {\bibfnamefont {E.}~\bibnamefont {Dumitrescu}}, \bibinfo {author}
  {\bibfnamefont {K.}~\bibnamefont {Dumon}}, \bibinfo {author} {\bibfnamefont
  {I.}~\bibnamefont {Duran}}, \bibinfo {author} {\bibfnamefont
  {K.}~\bibnamefont {EL-Safty}}, \bibinfo {author} {\bibfnamefont
  {E.}~\bibnamefont {Eastman}}, \bibinfo {author} {\bibfnamefont
  {G.}~\bibnamefont {Eberle}}, \bibinfo {author} {\bibfnamefont
  {A.}~\bibnamefont {Ebrahimi}}, \bibinfo {author} {\bibfnamefont
  {P.}~\bibnamefont {Eendebak}}, \bibinfo {author} {\bibfnamefont
  {D.}~\bibnamefont {Egger}}, \bibinfo {author} {\bibnamefont {EgrettaThula}},
  \bibinfo {author} {\bibnamefont {ElePT}}, \bibinfo {author} {\bibfnamefont
  {I.}~\bibnamefont {Elsayed}}, \bibinfo {author} {\bibnamefont {Emilio}},
  \bibinfo {author} {\bibfnamefont {A.}~\bibnamefont {Espiricueta}}, \bibinfo
  {author} {\bibfnamefont {M.}~\bibnamefont {Everitt}}, \bibinfo {author}
  {\bibfnamefont {D.}~\bibnamefont {Facoetti}}, \bibinfo {author} {\bibnamefont
  {Farida}}, \bibinfo {author} {\bibfnamefont {P.~M.}\ \bibnamefont
  {Fern{\'a}ndez}}, \bibinfo {author} {\bibfnamefont {S.}~\bibnamefont
  {Ferracin}}, \bibinfo {author} {\bibfnamefont {D.}~\bibnamefont {Ferrari}},
  \bibinfo {author} {\bibfnamefont {A.~H.}\ \bibnamefont {Ferrera}}, \bibinfo
  {author} {\bibfnamefont {R.}~\bibnamefont {Fouilland}}, \bibinfo {author}
  {\bibfnamefont {A.}~\bibnamefont {Frisch}}, \bibinfo {author} {\bibfnamefont
  {A.}~\bibnamefont {Fuhrer}}, \bibinfo {author} {\bibfnamefont
  {B.}~\bibnamefont {Fuller}}, \bibinfo {author} {\bibfnamefont
  {M.}~\bibnamefont {GEORGE}}, \bibinfo {author} {\bibfnamefont
  {J.}~\bibnamefont {Gacon}}, \bibinfo {author} {\bibfnamefont {B.~G.}\
  \bibnamefont {Gago}}, \bibinfo {author} {\bibfnamefont {C.}~\bibnamefont
  {Gambella}}, \bibinfo {author} {\bibfnamefont {J.~M.}\ \bibnamefont
  {Gambetta}}, \bibinfo {author} {\bibfnamefont {A.}~\bibnamefont
  {Gammanpila}}, \bibinfo {author} {\bibfnamefont {L.}~\bibnamefont {Garcia}},
  \bibinfo {author} {\bibfnamefont {T.}~\bibnamefont {Garg}}, \bibinfo {author}
  {\bibfnamefont {S.}~\bibnamefont {Garion}}, \bibinfo {author} {\bibfnamefont
  {J.~R.}\ \bibnamefont {Garrison}}, \bibinfo {author} {\bibfnamefont
  {J.}~\bibnamefont {Garrison}}, \bibinfo {author} {\bibfnamefont
  {T.}~\bibnamefont {Gates}}, \bibinfo {author} {\bibfnamefont
  {N.}~\bibnamefont {Gavrielov}}, \bibinfo {author} {\bibfnamefont
  {G.}~\bibnamefont {Gentinetta}}, \bibinfo {author} {\bibfnamefont
  {H.}~\bibnamefont {Georgiev}}, \bibinfo {author} {\bibfnamefont
  {L.}~\bibnamefont {Gil}}, \bibinfo {author} {\bibfnamefont {A.}~\bibnamefont
  {Gilliam}}, \bibinfo {author} {\bibfnamefont {A.}~\bibnamefont {Giridharan}},
  \bibinfo {author} {\bibnamefont {Glen}}, \bibinfo {author} {\bibfnamefont
  {J.}~\bibnamefont {Gomez-Mosquera}}, \bibinfo {author} {\bibnamefont
  {Gonzalo}}, \bibinfo {author} {\bibfnamefont {S.}~\bibnamefont {de~la
  Puente~Gonz{\'a}lez}}, \bibinfo {author} {\bibfnamefont {J.}~\bibnamefont
  {Gorzinski}}, \bibinfo {author} {\bibfnamefont {I.}~\bibnamefont {Gould}},
  \bibinfo {author} {\bibfnamefont {D.}~\bibnamefont {Greenberg}}, \bibinfo
  {author} {\bibfnamefont {D.}~\bibnamefont {Grinko}}, \bibinfo {author}
  {\bibfnamefont {W.}~\bibnamefont {Guan}}, \bibinfo {author} {\bibfnamefont
  {D.}~\bibnamefont {Guijo}}, \bibinfo {author} {\bibnamefont
  {Guillermo-Mijares-Vilarino}}, \bibinfo {author} {\bibfnamefont {J.~A.}\
  \bibnamefont {Gunnels}}, \bibinfo {author} {\bibfnamefont {H.}~\bibnamefont
  {Gupta}}, \bibinfo {author} {\bibfnamefont {N.}~\bibnamefont {Gupta}},
  \bibinfo {author} {\bibfnamefont {J.~M.}\ \bibnamefont {G{\"u}nther}},
  \bibinfo {author} {\bibfnamefont {M.}~\bibnamefont {Haglund}}, \bibinfo
  {author} {\bibfnamefont {I.}~\bibnamefont {Haide}}, \bibinfo {author}
  {\bibfnamefont {I.}~\bibnamefont {Hamamura}}, \bibinfo {author}
  {\bibfnamefont {O.~C.}\ \bibnamefont {Hamido}}, \bibinfo {author}
  {\bibfnamefont {F.}~\bibnamefont {Harkins}}, \bibinfo {author} {\bibfnamefont
  {K.}~\bibnamefont {Hartman}}, \bibinfo {author} {\bibfnamefont
  {A.}~\bibnamefont {Hasan}}, \bibinfo {author} {\bibfnamefont
  {V.}~\bibnamefont {Havlicek}}, \bibinfo {author} {\bibfnamefont
  {J.}~\bibnamefont {Hellmers}}, \bibinfo {author} {\bibfnamefont
  {{\L}.}~\bibnamefont {Herok}}, \bibinfo {author} {\bibfnamefont
  {R.}~\bibnamefont {Hill}}, \bibinfo {author} {\bibfnamefont {S.}~\bibnamefont
  {Hillmich}}, \bibinfo {author} {\bibfnamefont {I.}~\bibnamefont {Hincks}},
  \bibinfo {author} {\bibfnamefont {C.}~\bibnamefont {Hong}}, \bibinfo {author}
  {\bibfnamefont {H.}~\bibnamefont {Horii}}, \bibinfo {author} {\bibfnamefont
  {C.}~\bibnamefont {Howington}}, \bibinfo {author} {\bibfnamefont
  {S.}~\bibnamefont {Hu}}, \bibinfo {author} {\bibfnamefont {W.}~\bibnamefont
  {Hu}}, \bibinfo {author} {\bibfnamefont {C.-H.}\ \bibnamefont {Huang}},
  \bibinfo {author} {\bibfnamefont {J.}~\bibnamefont {Huang}}, \bibinfo
  {author} {\bibfnamefont {R.}~\bibnamefont {Huisman}}, \bibinfo {author}
  {\bibfnamefont {H.}~\bibnamefont {Imai}}, \bibinfo {author} {\bibfnamefont
  {T.}~\bibnamefont {Imamichi}}, \bibinfo {author} {\bibfnamefont
  {K.}~\bibnamefont {Ishizaki}}, \bibinfo {author} {\bibnamefont {Ishwor}},
  \bibinfo {author} {\bibfnamefont {R.}~\bibnamefont {Iten}}, \bibinfo {author}
  {\bibfnamefont {T.}~\bibnamefont {Itoko}}, \bibinfo {author} {\bibfnamefont
  {A.}~\bibnamefont {Ivrii}}, \bibinfo {author} {\bibfnamefont
  {A.}~\bibnamefont {Javadi}}, \bibinfo {author} {\bibfnamefont
  {A.}~\bibnamefont {Javadi-Abhari}}, \bibinfo {author} {\bibfnamefont
  {W.}~\bibnamefont {Javed}}, \bibinfo {author} {\bibfnamefont
  {Q.}~\bibnamefont {Jianhua}}, \bibinfo {author} {\bibfnamefont
  {M.}~\bibnamefont {Jivrajani}}, \bibinfo {author} {\bibfnamefont
  {K.}~\bibnamefont {Johns}}, \bibinfo {author} {\bibfnamefont
  {S.}~\bibnamefont {Johnstun}}, \bibinfo {author} {\bibnamefont
  {Jonathan-Shoemaker}}, \bibinfo {author} {\bibnamefont {JosDenmark}},
  \bibinfo {author} {\bibnamefont {JoshDumo}}, \bibinfo {author} {\bibfnamefont
  {J.}~\bibnamefont {Judge}}, \bibinfo {author} {\bibfnamefont
  {T.}~\bibnamefont {Kachmann}}, \bibinfo {author} {\bibfnamefont
  {A.}~\bibnamefont {Kale}}, \bibinfo {author} {\bibfnamefont {N.}~\bibnamefont
  {Kanazawa}}, \bibinfo {author} {\bibfnamefont {J.}~\bibnamefont {Kane}},
  \bibinfo {author} {\bibnamefont {Kang-Bae}}, \bibinfo {author} {\bibfnamefont
  {A.}~\bibnamefont {Kapila}}, \bibinfo {author} {\bibfnamefont
  {A.}~\bibnamefont {Karazeev}}, \bibinfo {author} {\bibfnamefont
  {P.}~\bibnamefont {Kassebaum}}, \bibinfo {author} {\bibfnamefont
  {T.}~\bibnamefont {Kato}}, \bibinfo {author} {\bibfnamefont {T.}~\bibnamefont
  {Kehrer}}, \bibinfo {author} {\bibfnamefont {J.}~\bibnamefont {Kelso}},
  \bibinfo {author} {\bibfnamefont {S.}~\bibnamefont {Kelso}}, \bibinfo
  {author} {\bibfnamefont {H.}~\bibnamefont {van Kemenade}}, \bibinfo {author}
  {\bibfnamefont {V.}~\bibnamefont {Khanderao}}, \bibinfo {author}
  {\bibfnamefont {S.}~\bibnamefont {King}}, \bibinfo {author} {\bibfnamefont
  {Y.}~\bibnamefont {Kobayashi}}, \bibinfo {author} {\bibnamefont {Kovi11Day}},
  \bibinfo {author} {\bibfnamefont {A.}~\bibnamefont {Kovyrshin}}, \bibinfo
  {author} {\bibfnamefont {J.}~\bibnamefont {Krishna}}, \bibinfo {author}
  {\bibfnamefont {R.}~\bibnamefont {Krishnakumar}}, \bibinfo {author}
  {\bibfnamefont {P.}~\bibnamefont {Krishnamurthy}}, \bibinfo {author}
  {\bibfnamefont {V.}~\bibnamefont {Krishnan}}, \bibinfo {author}
  {\bibfnamefont {K.}~\bibnamefont {Krsulich}}, \bibinfo {author}
  {\bibfnamefont {P.}~\bibnamefont {Kumkar}}, \bibinfo {author} {\bibfnamefont
  {G.}~\bibnamefont {Kus}}, \bibinfo {author} {\bibnamefont {LNoorl}}, \bibinfo
  {author} {\bibfnamefont {R.}~\bibnamefont {LaRose}}, \bibinfo {author}
  {\bibfnamefont {E.}~\bibnamefont {Lacal}}, \bibinfo {author} {\bibfnamefont
  {R.}~\bibnamefont {Lambert}}, \bibinfo {author} {\bibfnamefont
  {H.}~\bibnamefont {Landa}}, \bibinfo {author} {\bibfnamefont
  {J.}~\bibnamefont {Lapeyre}}, \bibinfo {author} {\bibfnamefont
  {D.}~\bibnamefont {Lasecki}}, \bibinfo {author} {\bibfnamefont
  {J.}~\bibnamefont {Latone}}, \bibinfo {author} {\bibfnamefont
  {S.}~\bibnamefont {Lawrence}}, \bibinfo {author} {\bibfnamefont
  {C.}~\bibnamefont {Lee}}, \bibinfo {author} {\bibfnamefont {G.}~\bibnamefont
  {Li}}, \bibinfo {author} {\bibfnamefont {T.~J.}\ \bibnamefont {Liang}},
  \bibinfo {author} {\bibfnamefont {J.}~\bibnamefont {Lishman}}, \bibinfo
  {author} {\bibfnamefont {D.}~\bibnamefont {Liu}}, \bibinfo {author}
  {\bibfnamefont {P.}~\bibnamefont {Liu}}, \bibinfo {author} {\bibnamefont
  {Lolcroc}}, \bibinfo {author} {\bibfnamefont {A.~K.}\ \bibnamefont {M}},
  \bibinfo {author} {\bibfnamefont {L.}~\bibnamefont {Madden}}, \bibinfo
  {author} {\bibfnamefont {Y.}~\bibnamefont {Maeng}}, \bibinfo {author}
  {\bibfnamefont {S.}~\bibnamefont {Maheshkar}}, \bibinfo {author}
  {\bibfnamefont {K.}~\bibnamefont {Majmudar}}, \bibinfo {author}
  {\bibfnamefont {A.}~\bibnamefont {Malyshev}}, \bibinfo {author}
  {\bibfnamefont {M.~E.}\ \bibnamefont {Mandouh}}, \bibinfo {author}
  {\bibfnamefont {J.}~\bibnamefont {Manela}}, \bibinfo {author} {\bibnamefont
  {Manjula}}, \bibinfo {author} {\bibfnamefont {J.}~\bibnamefont {Marecek}},
  \bibinfo {author} {\bibfnamefont {M.}~\bibnamefont {Marques}}, \bibinfo
  {author} {\bibfnamefont {K.}~\bibnamefont {Marwaha}}, \bibinfo {author}
  {\bibfnamefont {D.}~\bibnamefont {Maslov}}, \bibinfo {author} {\bibfnamefont
  {P.}~\bibnamefont {Maszota}}, \bibinfo {author} {\bibfnamefont
  {D.}~\bibnamefont {Mathews}}, \bibinfo {author} {\bibfnamefont
  {A.}~\bibnamefont {Matsuo}}, \bibinfo {author} {\bibfnamefont
  {F.}~\bibnamefont {Mazhandu}}, \bibinfo {author} {\bibfnamefont
  {D.}~\bibnamefont {McClure}}, \bibinfo {author} {\bibfnamefont
  {M.}~\bibnamefont {McElaney}}, \bibinfo {author} {\bibfnamefont
  {J.}~\bibnamefont {McElroy}}, \bibinfo {author} {\bibfnamefont
  {C.}~\bibnamefont {McGarry}}, \bibinfo {author} {\bibfnamefont
  {D.}~\bibnamefont {McKay}}, \bibinfo {author} {\bibfnamefont
  {D.}~\bibnamefont {McPherson}}, \bibinfo {author} {\bibfnamefont
  {S.}~\bibnamefont {Meesala}}, \bibinfo {author} {\bibfnamefont
  {D.}~\bibnamefont {Meirom}}, \bibinfo {author} {\bibfnamefont
  {C.}~\bibnamefont {Mendell}}, \bibinfo {author} {\bibfnamefont
  {T.}~\bibnamefont {Metcalfe}}, \bibinfo {author} {\bibfnamefont
  {M.}~\bibnamefont {Mevissen}}, \bibinfo {author} {\bibfnamefont
  {A.}~\bibnamefont {Meyer}}, \bibinfo {author} {\bibfnamefont
  {A.}~\bibnamefont {Mezzacapo}}, \bibinfo {author} {\bibfnamefont
  {R.}~\bibnamefont {Midha}}, \bibinfo {author} {\bibfnamefont
  {D.}~\bibnamefont {Millar}}, \bibinfo {author} {\bibfnamefont
  {D.}~\bibnamefont {Miller}}, \bibinfo {author} {\bibfnamefont
  {H.}~\bibnamefont {Miller}}, \bibinfo {author} {\bibfnamefont
  {Z.}~\bibnamefont {Minev}}, \bibinfo {author} {\bibfnamefont
  {A.}~\bibnamefont {Mitchell}}, \bibinfo {author} {\bibfnamefont
  {A.}~\bibnamefont {Mohammad}}, \bibinfo {author} {\bibfnamefont
  {N.}~\bibnamefont {Moll}}, \bibinfo {author} {\bibfnamefont {A.}~\bibnamefont
  {Montanez}}, \bibinfo {author} {\bibfnamefont {G.}~\bibnamefont {Monteiro}},
  \bibinfo {author} {\bibfnamefont {M.~D.}\ \bibnamefont {Mooring}}, \bibinfo
  {author} {\bibfnamefont {R.}~\bibnamefont {Morales}}, \bibinfo {author}
  {\bibfnamefont {N.}~\bibnamefont {Moran}}, \bibinfo {author} {\bibfnamefont
  {D.}~\bibnamefont {Morcuende}}, \bibinfo {author} {\bibfnamefont
  {S.}~\bibnamefont {Mostafa}}, \bibinfo {author} {\bibfnamefont
  {M.}~\bibnamefont {Motta}}, \bibinfo {author} {\bibfnamefont
  {R.}~\bibnamefont {Moyard}}, \bibinfo {author} {\bibfnamefont
  {P.}~\bibnamefont {Murali}}, \bibinfo {author} {\bibfnamefont
  {D.}~\bibnamefont {Murata}}, \bibinfo {author} {\bibfnamefont
  {J.}~\bibnamefont {M{\"u}ggenburg}}, \bibinfo {author} {\bibfnamefont
  {T.}~\bibnamefont {NEMOZ}}, \bibinfo {author} {\bibfnamefont
  {D.}~\bibnamefont {Nadlinger}}, \bibinfo {author} {\bibfnamefont
  {K.}~\bibnamefont {Nakanishi}}, \bibinfo {author} {\bibfnamefont
  {G.}~\bibnamefont {Nannicini}}, \bibinfo {author} {\bibfnamefont
  {P.}~\bibnamefont {Nation}}, \bibinfo {author} {\bibfnamefont
  {E.}~\bibnamefont {Navarro}}, \bibinfo {author} {\bibfnamefont
  {Y.}~\bibnamefont {Naveh}}, \bibinfo {author} {\bibfnamefont {S.~W.}\
  \bibnamefont {Neagle}}, \bibinfo {author} {\bibfnamefont {P.}~\bibnamefont
  {Neuweiler}}, \bibinfo {author} {\bibfnamefont {A.}~\bibnamefont {Ngoueya}},
  \bibinfo {author} {\bibfnamefont {T.}~\bibnamefont {Nguyen}}, \bibinfo
  {author} {\bibfnamefont {J.}~\bibnamefont {Nicander}}, \bibinfo {author}
  {\bibnamefont {Nick-Singstock}}, \bibinfo {author} {\bibfnamefont
  {P.}~\bibnamefont {Niroula}}, \bibinfo {author} {\bibfnamefont
  {H.}~\bibnamefont {Norlen}}, \bibinfo {author} {\bibnamefont {NuoWenLei}},
  \bibinfo {author} {\bibfnamefont {L.~J.}\ \bibnamefont {O'Riordan}}, \bibinfo
  {author} {\bibfnamefont {O.}~\bibnamefont {Ogunbayo}}, \bibinfo {author}
  {\bibfnamefont {P.}~\bibnamefont {Ollitrault}}, \bibinfo {author}
  {\bibfnamefont {T.}~\bibnamefont {Onodera}}, \bibinfo {author} {\bibfnamefont
  {R.}~\bibnamefont {Otaolea}}, \bibinfo {author} {\bibfnamefont
  {S.}~\bibnamefont {Oud}}, \bibinfo {author} {\bibfnamefont {D.}~\bibnamefont
  {Padilha}}, \bibinfo {author} {\bibfnamefont {H.}~\bibnamefont {Paik}},
  \bibinfo {author} {\bibfnamefont {S.}~\bibnamefont {Pal}}, \bibinfo {author}
  {\bibfnamefont {Y.}~\bibnamefont {Pang}}, \bibinfo {author} {\bibfnamefont
  {A.}~\bibnamefont {Panigrahi}}, \bibinfo {author} {\bibfnamefont {V.~R.}\
  \bibnamefont {Pascuzzi}}, \bibinfo {author} {\bibfnamefont {S.}~\bibnamefont
  {Perriello}}, \bibinfo {author} {\bibfnamefont {E.}~\bibnamefont {Peterson}},
  \bibinfo {author} {\bibfnamefont {A.}~\bibnamefont {Phan}}, \bibinfo {author}
  {\bibfnamefont {K.}~\bibnamefont {Pilch}}, \bibinfo {author} {\bibfnamefont
  {F.}~\bibnamefont {Piro}}, \bibinfo {author} {\bibfnamefont {M.}~\bibnamefont
  {Pistoia}}, \bibinfo {author} {\bibfnamefont {C.}~\bibnamefont {Piveteau}},
  \bibinfo {author} {\bibfnamefont {J.}~\bibnamefont {Plewa}}, \bibinfo
  {author} {\bibfnamefont {P.}~\bibnamefont {Pocreau}}, \bibinfo {author}
  {\bibfnamefont {C.}~\bibnamefont {Possel}}, \bibinfo {author} {\bibfnamefont
  {A.}~\bibnamefont {Pozas-Kerstjens}}, \bibinfo {author} {\bibfnamefont
  {R.}~\bibnamefont {Pracht}}, \bibinfo {author} {\bibfnamefont
  {M.}~\bibnamefont {Prokop}}, \bibinfo {author} {\bibfnamefont
  {V.}~\bibnamefont {Prutyanov}}, \bibinfo {author} {\bibfnamefont
  {S.}~\bibnamefont {Puri}}, \bibinfo {author} {\bibfnamefont {D.}~\bibnamefont
  {Puzzuoli}}, \bibinfo {author} {\bibnamefont {Pythonix}}, \bibinfo {author}
  {\bibfnamefont {J.}~\bibnamefont {P{\'e}rez}}, \bibinfo {author}
  {\bibnamefont {Quant02}}, \bibinfo {author} {\bibnamefont {Quintiii}},
  \bibinfo {author} {\bibfnamefont {R.~I.}\ \bibnamefont {Rahman}}, \bibinfo
  {author} {\bibfnamefont {A.}~\bibnamefont {Raja}}, \bibinfo {author}
  {\bibfnamefont {R.}~\bibnamefont {Rajeev}}, \bibinfo {author} {\bibfnamefont
  {I.}~\bibnamefont {Rajput}}, \bibinfo {author} {\bibfnamefont
  {N.}~\bibnamefont {Ramagiri}}, \bibinfo {author} {\bibfnamefont
  {A.}~\bibnamefont {Rao}}, \bibinfo {author} {\bibfnamefont {R.}~\bibnamefont
  {Raymond}}, \bibinfo {author} {\bibfnamefont {O.}~\bibnamefont
  {Reardon-Smith}}, \bibinfo {author} {\bibfnamefont {R.~M.-C.}\ \bibnamefont
  {Redondo}}, \bibinfo {author} {\bibfnamefont {M.}~\bibnamefont {Reuter}},
  \bibinfo {author} {\bibfnamefont {J.}~\bibnamefont {Rice}}, \bibinfo {author}
  {\bibfnamefont {M.}~\bibnamefont {Riedemann}}, \bibinfo {author}
  {\bibnamefont {Rietesh}}, \bibinfo {author} {\bibfnamefont {D.}~\bibnamefont
  {Risinger}}, \bibinfo {author} {\bibfnamefont {P.}~\bibnamefont {Rivero}},
  \bibinfo {author} {\bibfnamefont {M.~L.}\ \bibnamefont {Rocca}}, \bibinfo
  {author} {\bibfnamefont {D.~M.}\ \bibnamefont {Rodr{\'\i}guez}}, \bibinfo
  {author} {\bibnamefont {RohithKarur}}, \bibinfo {author} {\bibfnamefont
  {B.}~\bibnamefont {Rosand}}, \bibinfo {author} {\bibfnamefont
  {M.}~\bibnamefont {Rossmannek}}, \bibinfo {author} {\bibfnamefont
  {M.}~\bibnamefont {Ryu}}, \bibinfo {author} {\bibfnamefont {T.}~\bibnamefont
  {SAPV}}, \bibinfo {author} {\bibfnamefont {N.~R.~C.}\ \bibnamefont {Sa}},
  \bibinfo {author} {\bibfnamefont {A.}~\bibnamefont {Saha}}, \bibinfo {author}
  {\bibfnamefont {A.}~\bibnamefont {Ash-Saki}}, \bibinfo {author}
  {\bibfnamefont {A.}~\bibnamefont {Salman}}, \bibinfo {author} {\bibfnamefont
  {S.}~\bibnamefont {Sanand}}, \bibinfo {author} {\bibfnamefont
  {M.}~\bibnamefont {Sandberg}}, \bibinfo {author} {\bibfnamefont
  {H.}~\bibnamefont {Sandesara}}, \bibinfo {author} {\bibfnamefont
  {R.}~\bibnamefont {Sapra}}, \bibinfo {author} {\bibfnamefont
  {H.}~\bibnamefont {Sargsyan}}, \bibinfo {author} {\bibfnamefont
  {A.}~\bibnamefont {Sarkar}}, \bibinfo {author} {\bibfnamefont
  {N.}~\bibnamefont {Sathaye}}, \bibinfo {author} {\bibfnamefont
  {N.}~\bibnamefont {Savola}}, \bibinfo {author} {\bibfnamefont
  {B.}~\bibnamefont {Schmitt}}, \bibinfo {author} {\bibfnamefont
  {C.}~\bibnamefont {Schnabel}}, \bibinfo {author} {\bibfnamefont
  {Z.}~\bibnamefont {Schoenfeld}}, \bibinfo {author} {\bibfnamefont {T.~L.}\
  \bibnamefont {Scholten}}, \bibinfo {author} {\bibfnamefont {E.}~\bibnamefont
  {Schoute}}, \bibinfo {author} {\bibfnamefont {J.}~\bibnamefont
  {Schuhmacher}}, \bibinfo {author} {\bibfnamefont {M.}~\bibnamefont
  {Schulterbrandt}}, \bibinfo {author} {\bibfnamefont {J.}~\bibnamefont
  {Schwarm}}, \bibinfo {author} {\bibfnamefont {P.}~\bibnamefont {Schweigert}},
  \bibinfo {author} {\bibfnamefont {J.}~\bibnamefont {Seaward}}, \bibinfo
  {author} {\bibnamefont {Sergi}}, \bibinfo {author} {\bibfnamefont {D.~E.}\
  \bibnamefont {Serrano}}, \bibinfo {author} {\bibfnamefont {I.~F.}\
  \bibnamefont {Sertage}}, \bibinfo {author} {\bibfnamefont {K.}~\bibnamefont
  {Setia}}, \bibinfo {author} {\bibfnamefont {F.}~\bibnamefont {Shah}},
  \bibinfo {author} {\bibfnamefont {N.}~\bibnamefont {Shammah}}, \bibinfo
  {author} {\bibfnamefont {W.}~\bibnamefont {Shanks}}, \bibinfo {author}
  {\bibfnamefont {R.}~\bibnamefont {Sharma}}, \bibinfo {author} {\bibfnamefont
  {P.}~\bibnamefont {Shaw}}, \bibinfo {author} {\bibfnamefont {Y.}~\bibnamefont
  {Shi}}, \bibinfo {author} {\bibfnamefont {J.}~\bibnamefont {Shoemaker}},
  \bibinfo {author} {\bibfnamefont {A.}~\bibnamefont {Silva}}, \bibinfo
  {author} {\bibfnamefont {A.}~\bibnamefont {Simonetto}}, \bibinfo {author}
  {\bibfnamefont {D.}~\bibnamefont {Singh}}, \bibinfo {author} {\bibfnamefont
  {D.}~\bibnamefont {Singh}}, \bibinfo {author} {\bibfnamefont
  {P.}~\bibnamefont {Singh}}, \bibinfo {author} {\bibfnamefont
  {P.}~\bibnamefont {Singkanipa}}, \bibinfo {author} {\bibfnamefont
  {Y.}~\bibnamefont {Siraichi}}, \bibinfo {author} {\bibnamefont {Siri}},
  \bibinfo {author} {\bibfnamefont {J.}~\bibnamefont {Sistos}}, \bibinfo
  {author} {\bibfnamefont {J.}~\bibnamefont {Sistos}}, \bibinfo {author}
  {\bibfnamefont {I.}~\bibnamefont {Sitdikov}}, \bibinfo {author}
  {\bibfnamefont {S.}~\bibnamefont {Sivarajah}}, \bibinfo {author}
  {\bibnamefont {Slavikmew}}, \bibinfo {author} {\bibfnamefont {M.~B.}\
  \bibnamefont {Sletfjerding}}, \bibinfo {author} {\bibfnamefont {J.~A.}\
  \bibnamefont {Smolin}}, \bibinfo {author} {\bibfnamefont {M.}~\bibnamefont
  {Soeken}}, \bibinfo {author} {\bibfnamefont {I.~O.}\ \bibnamefont {Sokolov}},
  \bibinfo {author} {\bibfnamefont {I.}~\bibnamefont {Sokolov}}, \bibinfo
  {author} {\bibfnamefont {V.~P.}\ \bibnamefont {Soloviev}}, \bibinfo {author}
  {\bibnamefont {SooluThomas}}, \bibinfo {author} {\bibnamefont {Starfish}},
  \bibinfo {author} {\bibfnamefont {D.}~\bibnamefont {Steenken}}, \bibinfo
  {author} {\bibfnamefont {M.}~\bibnamefont {Stypulkoski}}, \bibinfo {author}
  {\bibfnamefont {A.}~\bibnamefont {Suau}}, \bibinfo {author} {\bibfnamefont
  {S.}~\bibnamefont {Sun}}, \bibinfo {author} {\bibfnamefont {K.~J.}\
  \bibnamefont {Sung}}, \bibinfo {author} {\bibfnamefont {M.}~\bibnamefont
  {Suwama}}, \bibinfo {author} {\bibfnamefont {O.}~\bibnamefont {S{\l}owik}},
  \bibinfo {author} {\bibfnamefont {R.}~\bibnamefont {Taeja}}, \bibinfo
  {author} {\bibfnamefont {H.}~\bibnamefont {Takahashi}}, \bibinfo {author}
  {\bibfnamefont {T.}~\bibnamefont {Takawale}}, \bibinfo {author}
  {\bibfnamefont {I.}~\bibnamefont {Tavernelli}}, \bibinfo {author}
  {\bibfnamefont {C.}~\bibnamefont {Taylor}}, \bibinfo {author} {\bibfnamefont
  {P.}~\bibnamefont {Taylour}}, \bibinfo {author} {\bibfnamefont
  {S.}~\bibnamefont {Thomas}}, \bibinfo {author} {\bibfnamefont
  {K.}~\bibnamefont {Tian}}, \bibinfo {author} {\bibfnamefont {M.}~\bibnamefont
  {Tillet}}, \bibinfo {author} {\bibfnamefont {M.}~\bibnamefont {Tod}},
  \bibinfo {author} {\bibfnamefont {M.}~\bibnamefont {Tomasik}}, \bibinfo
  {author} {\bibfnamefont {C.}~\bibnamefont {Tornow}}, \bibinfo {author}
  {\bibfnamefont {E.}~\bibnamefont {de~la Torre}}, \bibinfo {author}
  {\bibfnamefont {J.~L.~S.}\ \bibnamefont {Toural}}, \bibinfo {author}
  {\bibfnamefont {K.}~\bibnamefont {Trabing}}, \bibinfo {author} {\bibfnamefont
  {M.}~\bibnamefont {Treinish}}, \bibinfo {author} {\bibfnamefont
  {D.}~\bibnamefont {Trenev}}, \bibinfo {author} {\bibnamefont {TrishaPe}},
  \bibinfo {author} {\bibfnamefont {F.}~\bibnamefont {Truger}}, \bibinfo
  {author} {\bibnamefont {TsafrirA}}, \bibinfo {author} {\bibfnamefont
  {G.}~\bibnamefont {Tsilimigkounakis}}, \bibinfo {author} {\bibfnamefont
  {K.}~\bibnamefont {Tsuoka}}, \bibinfo {author} {\bibfnamefont
  {D.}~\bibnamefont {Tulsi}}, \bibinfo {author} {\bibfnamefont
  {D.}~\bibnamefont {Tuna}}, \bibinfo {author} {\bibfnamefont {W.}~\bibnamefont
  {Turner}}, \bibinfo {author} {\bibfnamefont {Y.}~\bibnamefont {Vaknin}},
  \bibinfo {author} {\bibfnamefont {C.~R.}\ \bibnamefont {Valcarce}}, \bibinfo
  {author} {\bibfnamefont {F.}~\bibnamefont {Varchon}}, \bibinfo {author}
  {\bibfnamefont {A.}~\bibnamefont {Vartak}}, \bibinfo {author} {\bibfnamefont
  {A.~C.}\ \bibnamefont {Vazquez}}, \bibinfo {author} {\bibfnamefont
  {P.}~\bibnamefont {Vijaywargiya}}, \bibinfo {author} {\bibfnamefont
  {V.}~\bibnamefont {Villar}}, \bibinfo {author} {\bibfnamefont
  {B.}~\bibnamefont {Vishnu}}, \bibinfo {author} {\bibfnamefont
  {D.}~\bibnamefont {Vogt-Lee}}, \bibinfo {author} {\bibfnamefont
  {C.}~\bibnamefont {Vuillot}}, \bibinfo {author} {\bibnamefont {WQ}}, \bibinfo
  {author} {\bibfnamefont {J.}~\bibnamefont {Weaver}}, \bibinfo {author}
  {\bibfnamefont {J.}~\bibnamefont {Weidenfeller}}, \bibinfo {author}
  {\bibfnamefont {R.}~\bibnamefont {Wieczorek}}, \bibinfo {author}
  {\bibfnamefont {J.~A.}\ \bibnamefont {Wildstrom}}, \bibinfo {author}
  {\bibfnamefont {J.}~\bibnamefont {Wilson}}, \bibinfo {author} {\bibfnamefont
  {E.}~\bibnamefont {Winston}}, \bibinfo {author} {\bibnamefont
  {WinterSoldier}}, \bibinfo {author} {\bibfnamefont {J.~J.}\ \bibnamefont
  {Woehr}}, \bibinfo {author} {\bibfnamefont {S.}~\bibnamefont {Woerner}},
  \bibinfo {author} {\bibfnamefont {R.}~\bibnamefont {Woo}}, \bibinfo {author}
  {\bibfnamefont {C.~J.}\ \bibnamefont {Wood}}, \bibinfo {author}
  {\bibfnamefont {R.}~\bibnamefont {Wood}}, \bibinfo {author} {\bibfnamefont
  {S.}~\bibnamefont {Wood}}, \bibinfo {author} {\bibfnamefont {J.}~\bibnamefont
  {Wootton}}, \bibinfo {author} {\bibfnamefont {M.}~\bibnamefont {Wright}},
  \bibinfo {author} {\bibfnamefont {L.}~\bibnamefont {Xing}}, \bibinfo {author}
  {\bibfnamefont {J.}~\bibnamefont {YU}}, \bibinfo {author} {\bibnamefont
  {Yaiza}}, \bibinfo {author} {\bibfnamefont {B.}~\bibnamefont {Yang}},
  \bibinfo {author} {\bibfnamefont {U.}~\bibnamefont {Yang}}, \bibinfo {author}
  {\bibfnamefont {J.}~\bibnamefont {Yao}}, \bibinfo {author} {\bibfnamefont
  {D.}~\bibnamefont {Yeralin}}, \bibinfo {author} {\bibfnamefont
  {R.}~\bibnamefont {Yonekura}}, \bibinfo {author} {\bibfnamefont
  {D.}~\bibnamefont {Yonge-Mallo}}, \bibinfo {author} {\bibfnamefont
  {R.}~\bibnamefont {Yoshida}}, \bibinfo {author} {\bibfnamefont
  {R.}~\bibnamefont {Young}}, \bibinfo {author} {\bibfnamefont
  {J.}~\bibnamefont {Yu}}, \bibinfo {author} {\bibfnamefont {L.}~\bibnamefont
  {Yu}}, \bibinfo {author} {\bibnamefont {Yuma-Nakamura}}, \bibinfo {author}
  {\bibfnamefont {C.}~\bibnamefont {Zachow}}, \bibinfo {author} {\bibfnamefont
  {L.}~\bibnamefont {Zdanski}}, \bibinfo {author} {\bibfnamefont
  {H.}~\bibnamefont {Zhang}}, \bibinfo {author} {\bibfnamefont
  {E.}~\bibnamefont {Zheltonozhskii}}, \bibinfo {author} {\bibfnamefont
  {I.}~\bibnamefont {Zidaru}}, \bibinfo {author} {\bibfnamefont
  {B.}~\bibnamefont {Zimmermann}}, \bibinfo {author} {\bibfnamefont
  {B.}~\bibnamefont {Zindorf}}, \bibinfo {author} {\bibfnamefont
  {C.}~\bibnamefont {Zoufal}}, \bibinfo {author} {\bibnamefont {a~matsuo}},
  \bibinfo {author} {\bibnamefont {aeddins ibm}}, \bibinfo {author}
  {\bibnamefont {alexzhang13}}, \bibinfo {author} {\bibnamefont {b63}},
  \bibinfo {author} {\bibnamefont {bartek bartlomiej}}, \bibinfo {author}
  {\bibnamefont {bcamorrison}}, \bibinfo {author} {\bibnamefont {brandhsn}},
  \bibinfo {author} {\bibnamefont {nick bronn}}, \bibinfo {author}
  {\bibnamefont {chetmurthy}}, \bibinfo {author} {\bibnamefont {choerst ibm}},
  \bibinfo {author} {\bibnamefont {comet}}, \bibinfo {author} {\bibnamefont
  {dalin27}}, \bibinfo {author} {\bibnamefont {deeplokhande}}, \bibinfo
  {author} {\bibnamefont {dekel.meirom}}, \bibinfo {author} {\bibnamefont
  {derwind}}, \bibinfo {author} {\bibnamefont {dime10}}, \bibinfo {author}
  {\bibnamefont {ehchen}}, \bibinfo {author} {\bibnamefont {ewinston}},
  \bibinfo {author} {\bibnamefont {fanizzamarco}}, \bibinfo {author}
  {\bibnamefont {fs1132429}}, \bibinfo {author} {\bibnamefont {gadial}},
  \bibinfo {author} {\bibnamefont {galeinston}}, \bibinfo {author}
  {\bibnamefont {georgezhou20}}, \bibinfo {author} {\bibnamefont {georgios
  ts}}, \bibinfo {author} {\bibnamefont {gruu}}, \bibinfo {author}
  {\bibnamefont {hhorii}}, \bibinfo {author} {\bibnamefont {hhyap}}, \bibinfo
  {author} {\bibnamefont {hykavitha}}, \bibinfo {author} {\bibnamefont
  {itoko}}, \bibinfo {author} {\bibnamefont {jeppevinkel}}, \bibinfo {author}
  {\bibnamefont {jessica angel7}}, \bibinfo {author} {\bibnamefont
  {jezerjojo14}}, \bibinfo {author} {\bibnamefont {jliu45}}, \bibinfo {author}
  {\bibnamefont {johannesgreiner}}, \bibinfo {author} {\bibnamefont {jscott2}},
  \bibinfo {author} {\bibnamefont {kUmezawa}}, \bibinfo {author} {\bibnamefont
  {klinvill}}, \bibinfo {author} {\bibnamefont {krutik2966}}, \bibinfo {author}
  {\bibnamefont {ma5x}}, \bibinfo {author} {\bibnamefont {michelle4654}},
  \bibinfo {author} {\bibnamefont {msuwama}}, \bibinfo {author} {\bibnamefont
  {nico lgrs}}, \bibinfo {author} {\bibnamefont {nrhawkins}}, \bibinfo {author}
  {\bibnamefont {ntgiwsvp}}, \bibinfo {author} {\bibnamefont {ordmoj}},
  \bibinfo {author} {\bibnamefont {sagar pahwa}}, \bibinfo {author}
  {\bibnamefont {pritamsinha2304}}, \bibinfo {author} {\bibnamefont
  {rithikaadiga}}, \bibinfo {author} {\bibnamefont {ryancocuzzo}}, \bibinfo
  {author} {\bibnamefont {saktar unr}}, \bibinfo {author} {\bibnamefont
  {saswati qiskit}}, \bibinfo {author} {\bibnamefont {sebastian mair}},
  \bibinfo {author} {\bibnamefont {septembrr}}, \bibinfo {author} {\bibnamefont
  {sethmerkel}}, \bibinfo {author} {\bibnamefont {sg495}}, \bibinfo {author}
  {\bibnamefont {shaashwat}}, \bibinfo {author} {\bibnamefont {smturro2}},
  \bibinfo {author} {\bibnamefont {sternparky}}, \bibinfo {author}
  {\bibnamefont {strickroman}}, \bibinfo {author} {\bibnamefont {tigerjack}},
  \bibinfo {author} {\bibnamefont {tsura crisaldo}}, \bibinfo {author}
  {\bibnamefont {upsideon}}, \bibinfo {author} {\bibnamefont {vadebayo49}},
  \bibinfo {author} {\bibnamefont {welien}}, \bibinfo {author} {\bibnamefont
  {willhbang}}, \bibinfo {author} {\bibnamefont {wmurphy collabstar}}, \bibinfo
  {author} {\bibnamefont {yang.luh}}, \bibinfo {author} {\bibnamefont
  {yuri@FreeBSD}}, \ and\ \bibinfo {author} {\bibfnamefont {M.}~\bibnamefont
  {{\v{C}}epulkovskis}},\ }\href {\doibase 10.5281/zenodo.2573505} {\enquote
  {\bibinfo {title} {Qiskit: An open-source framework for quantum computing},}\
  } (\bibinfo {year} {2021})\BibitemShut {NoStop}%
\bibitem [{\citenamefont {Jordan}\ and\ \citenamefont
  {Wigner}(1928)}]{jordan1928p}%
  \BibitemOpen
  \bibfield  {author} {\bibinfo {author} {\bibfnamefont {P.}~\bibnamefont
  {Jordan}}\ and\ \bibinfo {author} {\bibfnamefont {E.}~\bibnamefont
  {Wigner}},\ }\href {https://doi.org/10.1007/BF01331938} {\bibfield  {journal}
  {\bibinfo  {journal} {Z. Phys.}\ }\textbf {\bibinfo {volume} {47}},\ \bibinfo
  {pages} {631} (\bibinfo {year} {1928})}\BibitemShut {NoStop}%
\bibitem [{\citenamefont {Huggins}\ \emph {et~al.}(2020)\citenamefont
  {Huggins}, \citenamefont {Lee}, \citenamefont {Baek}, \citenamefont
  {O'Gorman},\ and\ \citenamefont {Whaley}}]{huggins2020non}%
  \BibitemOpen
  \bibfield  {author} {\bibinfo {author} {\bibfnamefont {W.~J.}\ \bibnamefont
  {Huggins}}, \bibinfo {author} {\bibfnamefont {J.}~\bibnamefont {Lee}},
  \bibinfo {author} {\bibfnamefont {U.}~\bibnamefont {Baek}}, \bibinfo {author}
  {\bibfnamefont {B.}~\bibnamefont {O'Gorman}}, \ and\ \bibinfo {author}
  {\bibfnamefont {K.~B.}\ \bibnamefont {Whaley}},\ }\href
  {https://doi.org/10.1088/1367-2630/ab867b} {\bibfield  {journal} {\bibinfo
  {journal} {New J. Phys.}\ } (\bibinfo {year} {2020})}\BibitemShut {NoStop}%
\bibitem [{\citenamefont {Stair}\ and\ \citenamefont
  {Evangelista}(2021)}]{stair2021simulating}%
  \BibitemOpen
  \bibfield  {author} {\bibinfo {author} {\bibfnamefont {N.~H.}\ \bibnamefont
  {Stair}}\ and\ \bibinfo {author} {\bibfnamefont {F.~A.}\ \bibnamefont
  {Evangelista}},\ }\href {https://doi.org/10.1103/PRXQuantum.2.030301}
  {\bibfield  {journal} {\bibinfo  {journal} {PRX Quantum}\ }\textbf {\bibinfo
  {volume} {2}},\ \bibinfo {pages} {030301} (\bibinfo {year}
  {2021})}\BibitemShut {NoStop}%
\bibitem [{\citenamefont {Cai}\ \emph {et~al.}(2022)\citenamefont {Cai},
  \citenamefont {Babbush}, \citenamefont {Benjamin}, \citenamefont {Endo},
  \citenamefont {Huggins}, \citenamefont {Li}, \citenamefont {McClean},\ and\
  \citenamefont {O'Brien}}]{cai2022quantum}%
  \BibitemOpen
  \bibfield  {author} {\bibinfo {author} {\bibfnamefont {Z.}~\bibnamefont
  {Cai}}, \bibinfo {author} {\bibfnamefont {R.}~\bibnamefont {Babbush}},
  \bibinfo {author} {\bibfnamefont {S.~C.}\ \bibnamefont {Benjamin}}, \bibinfo
  {author} {\bibfnamefont {S.}~\bibnamefont {Endo}}, \bibinfo {author}
  {\bibfnamefont {W.~J.}\ \bibnamefont {Huggins}}, \bibinfo {author}
  {\bibfnamefont {Y.}~\bibnamefont {Li}}, \bibinfo {author} {\bibfnamefont
  {J.~R.}\ \bibnamefont {McClean}}, \ and\ \bibinfo {author} {\bibfnamefont
  {T.~E.}\ \bibnamefont {O'Brien}},\ }\href
  {https://doi.org/10.48550/arXiv.2210.00921} {\bibfield  {journal} {\bibinfo
  {journal} {arXiv:2210.00921}\ } (\bibinfo {year} {2022})}\BibitemShut
  {NoStop}%
\bibitem [{\citenamefont {Grimsley}\ \emph {et~al.}(2019)\citenamefont
  {Grimsley}, \citenamefont {Claudino}, \citenamefont {Economou}, \citenamefont
  {Barnes},\ and\ \citenamefont {Mayhall}}]{grimsley2019trotterized}%
  \BibitemOpen
  \bibfield  {author} {\bibinfo {author} {\bibfnamefont {H.~R.}\ \bibnamefont
  {Grimsley}}, \bibinfo {author} {\bibfnamefont {D.}~\bibnamefont {Claudino}},
  \bibinfo {author} {\bibfnamefont {S.~E.}\ \bibnamefont {Economou}}, \bibinfo
  {author} {\bibfnamefont {E.}~\bibnamefont {Barnes}}, \ and\ \bibinfo {author}
  {\bibfnamefont {N.~J.}\ \bibnamefont {Mayhall}},\ }\href
  {https://doi.org/10.1021/acs.jctc.9b01083} {\bibfield  {journal} {\bibinfo
  {journal} {J. Comp. Theor. Chem.}\ }\textbf {\bibinfo {volume} {16}},\
  \bibinfo {pages} {1} (\bibinfo {year} {2019})}\BibitemShut {NoStop}%
\bibitem [{\citenamefont {McClean}\ \emph {et~al.}(2017)\citenamefont
  {McClean}, \citenamefont {Kimchi-Schwartz}, \citenamefont {Carter},\ and\
  \citenamefont {de~Jong}}]{mcclean2017hybrid}%
  \BibitemOpen
  \bibfield  {author} {\bibinfo {author} {\bibfnamefont {J.~R.}\ \bibnamefont
  {McClean}}, \bibinfo {author} {\bibfnamefont {M.~E.}\ \bibnamefont
  {Kimchi-Schwartz}}, \bibinfo {author} {\bibfnamefont {J.}~\bibnamefont
  {Carter}}, \ and\ \bibinfo {author} {\bibfnamefont {W.~A.}\ \bibnamefont
  {de~Jong}},\ }\href {https://doi.org/10.1103/PhysRevA.95.042308} {\bibfield
  {journal} {\bibinfo  {journal} {Phys. Rev. A}\ }\textbf {\bibinfo {volume}
  {95}},\ \bibinfo {pages} {042308} (\bibinfo {year} {2017})}\BibitemShut
  {NoStop}%
\bibitem [{\citenamefont {Motta}\ \emph {et~al.}(2020)\citenamefont {Motta},
  \citenamefont {Sun}, \citenamefont {Tan}, \citenamefont {O’Rourke},
  \citenamefont {Ye}, \citenamefont {Minnich}, \citenamefont {Brand{\~a}o},\
  and\ \citenamefont {Chan}}]{motta2020determining}%
  \BibitemOpen
  \bibfield  {author} {\bibinfo {author} {\bibfnamefont {M.}~\bibnamefont
  {Motta}}, \bibinfo {author} {\bibfnamefont {C.}~\bibnamefont {Sun}}, \bibinfo
  {author} {\bibfnamefont {A.~T.}\ \bibnamefont {Tan}}, \bibinfo {author}
  {\bibfnamefont {M.~J.}\ \bibnamefont {O’Rourke}}, \bibinfo {author}
  {\bibfnamefont {E.}~\bibnamefont {Ye}}, \bibinfo {author} {\bibfnamefont
  {A.~J.}\ \bibnamefont {Minnich}}, \bibinfo {author} {\bibfnamefont {F.~G.}\
  \bibnamefont {Brand{\~a}o}}, \ and\ \bibinfo {author} {\bibfnamefont
  {G.~K.-L.}\ \bibnamefont {Chan}},\ }\href
  {https://doi.org/10.1038/s41567-019-0704-4} {\bibfield  {journal} {\bibinfo
  {journal} {Nat. Phys.}\ }\textbf {\bibinfo {volume} {16}},\ \bibinfo {pages}
  {205} (\bibinfo {year} {2020})}\BibitemShut {NoStop}%
\bibitem [{\citenamefont {Stair}\ \emph {et~al.}(2020)\citenamefont {Stair},
  \citenamefont {Huang},\ and\ \citenamefont
  {Evangelista}}]{stair2020multireference}%
  \BibitemOpen
  \bibfield  {author} {\bibinfo {author} {\bibfnamefont {N.~H.}\ \bibnamefont
  {Stair}}, \bibinfo {author} {\bibfnamefont {R.}~\bibnamefont {Huang}}, \ and\
  \bibinfo {author} {\bibfnamefont {F.~A.}\ \bibnamefont {Evangelista}},\
  }\href {https://doi.org/10.1021/acs.jctc.9b01125} {\bibfield  {journal}
  {\bibinfo  {journal} {J. Chem. Theory Comput.}\ }\textbf {\bibinfo {volume}
  {16}},\ \bibinfo {pages} {2236} (\bibinfo {year} {2020})}\BibitemShut
  {NoStop}%
\end{thebibliography}
\end{document}